\documentclass[ reprint,
 amsmath,amssymb, aps,
prb,
floatfix,
nofootinbib,
showpacs ]{revtex4-1}

\usepackage{graphicx}
\usepackage{color} \usepackage{float} \usepackage{verbatim}
\usepackage{braket} \usepackage[mathscr]{euscript}
\usepackage{silence} 
\usepackage{helvet}
\usepackage{gensymb}
\usepackage{placeins}
\usepackage{afterpage}
\usepackage{xcolor}
\usepackage[normalem]{ulem}

 \newcommand{\multiref}[2]{\ref{#1}--\ref{#2}}

\usepackage[many]{tcolorbox}

\newtcolorbox{cross}{blank,parbox=false,
  overlay={\draw[red,line width=3pt] (interior.south west)--(interior.north east);
    \draw[red,line width=3pt] (interior.north west)--(interior.south east);}}

\usepackage{tikz}

\begin{document}

\newcommand{\kb}{$\mathbf{k}$}

\preprint{APS/123-QED}

\title{Effectiveness of smearing and tetrahedron methods: best practices in DFT codes}

\author{Jeremy J. Jorgensen, Gus L. W. Hart}
\affiliation{Department of Physics and Astronomy, Brigham Young University, Provo, Utah, 84602, USA} 


\date{\today}

\begin{abstract}

Density functional theory (DFT) codes are commonly treated as a ``black box" in high-throughput screening of materials, with users opting for the default values of the input parameters. Often, non-experts may not sufficiently consider the effect of these parameters on prediction quality. In this work, we attempt to identify a robust set of parameters related to smearing and tetrahedron methods that return numerically accurate and efficient results for a wide variety of metallic systems. The effects of smearing and tetrahedron methods on the total energy, number of self-consistent field cycles, and forces on atoms are studied in two popular DFT codes: the Vienna Ab initio Simulation Package (VASP) and Quantum Espresso (QE). From nearly 40,000 computations, it is apparent that the optimal smearing depends on the system, smearing method, smearing parameter, and \kb-point density. The benefit of smearing is a minor reduction in the number of self-consistent field cycles, which is independent of the smearing method or parameter. A large smearing parameter---what is considered large is system dependent---leads to inaccurate total energies and forces. Bl\"ochl's tetrahedron method leads to small improvements in total energies. When treating diverse systems with the same input parameters, we suggest using as little smearing as possible due to the system dependence of smearing and the risk of selecting a parameter that gives inaccurate energies and forces.

\end{abstract}

\maketitle

\section{Introduction}

Every year over the past decade tens of thousands of papers on density functional theory \cite{kohn1965self,hohenberg1964inhomogeneous} (DFT) have been published. Remarkably, 2 of the top 10, and 12 of the top 100 most cited papers relate to DFT \cite{van2014top}. The ubiquity of DFT stems from its ability to address a diverse portfolio of physical, material, chemical, and biological problems, including inorganic crystal structure \cite{woodley2008crystal,brandenburg2013dispersion,hautier2010finding}, band gaps \cite{xiao2011accurate,verma2017hle16,morales2017empirical}, charge transport \cite{cai2006density,lherbier2008charge,delgado2010tuning}, corrosion inhibitors \cite{xia2008molecular,obot2015density,verma2018density}, heterogeneous catalysis \cite{norskov2002universality,norskov2011density,bligaard2007ligand}, molecular properties and spectroscopy \cite{stephens1994ab,zhan2003ionization,neese2009prediction}, chemistry of transition metals \cite{connolly1983density,zhao2006new,cramer2009density}, phase transformations  \cite{connolly1983density,gracia2007characterization,vishnu2010phase}, surface structures and properties \cite{meyer2003density,norskov2011density,nolan2005density}, vibrational frequencies \cite{stephens1994ab,wong1996vibrational,scott1996harmonic}, molecular dynamics \cite{car1985unified,kresse1994ab,hafner2008ab} and chemical reactions in solutions and enzymes \cite{field2002simulating,mulholland2005modelling,hu2008free}. This broad applicability has enabled DFT simulations to (1) gain utility in academic research and industrial sectors \cite{eyert2018unravelling} where it has become an established tool in automotive, aerospace, energy, chemicals, electronics, oil and gas, metals and alloys, glass and ceramics, and polymer sectors, and (2) grow into the most popular electronic structure and quantum mechanical method \cite{van2014density,pribram2015dft}.

For many DFT practitioners, DFT codes are ``black boxes". This is understandable because the list of problems DFT can address continues to grow, the codes themselves are increasingly complex, and the ratio of experts to non-experts running DFT is shrinking. Often too little thought goes into the values of input parameters of DFT algorithms, with many opting for default values even though these may lead to inaccurate calculations. Mattsson et al.  \cite{mattsson2004designing} have demonstrated the sensitivity of numerically precise DFT results on \kb-point sampling, basis set cutoffs, and smearing. Mehl \cite{mehl2000occupation} draws attention to the fact that many use smearing with little justification or validation and drew attention to cases where smearing leads to inaccurate predictions of a material's properties.

The use case of interest in this paper is computational high-throughput screening of materials \cite{curtarolo2012aflow, jain2013commentary,saal2013materials}. An example of high-throughput screening is identifying stable alloys. To do this, a large number of DFT calculations are performed on a number of elements in varying concentrations and crystal structures. Those with the lowest formation enthalpy at a given concentration are most likely stable and qualify for further investigation. Tens of thousands of DFT calculations are performed without tailoring the parameters of the simulation for each calculation. A set of robust DFT parameters are desired that give numerically accurate results irrespective of the system.

We ran close to 40,000 DFT calculations to identify a smearing method and smearing parameter that would give improved computational performance for a wide variety of materials. We found that smearing reduces the number of self-consistency field (SCF) cycles to a small degree and large amounts of smearing leads to inaccurate DFT calculations. To complicate matters, the optimal smearing is dependent on the system, smearing method, smearing parameter, and \kb-point density. DFT manuals recommend default smearing parameters that are often too big and lead to inaccurate calculations.

We also tested the effect of tetrahedron methods and found they play a minor role in improved precision in DFT calculations, with uncorrected tetrahedron methods providing less numerical precision than the non-tetrahedron methods. Among tetrahedron methods, Bl\"{o}chl performs the best, but is only marginally better than integration without tetrahedra and without smearing.

\section{Background}

DFT codes calculate the ground state electronic density by means of a self-consistency field cycle. Smearing the electronic structure was proposed to deal with band sloshing \cite{marzari1996ab} and to reduce the number of \kb-points required by accelerating the convergence of the band energy calculation with respect to \kb-point density \cite{methfessel1989high}. Band sloshing typically occurs when there are many states near the Fermi level that alternately shift above and below the Fermi level in SCF cycles, resulting in the charge moving back and forth, never settling\cite{woods2018nature,tassone1994acceleration}. Tetrahedron methods were also developed to reduce the number of \kb-points in DFT calculations\cite{blochl1994improved}. 

The desired result of smearing or tetrahedron methods is more efficient DFT simulations from reduced SCF iterations or fewer \kb-points. This work shows that the optimal amount of smearing is usually much lower than what is recommended, and that there is no single, optimal parameter---the effectiveness of smearing is dependent on the metal, \kb-point density, and smearing method. In general, a smearing value smaller than the recommended value is safer and equally efficient.

\begin{figure}[htbp]
\includegraphics[width=\linewidth]{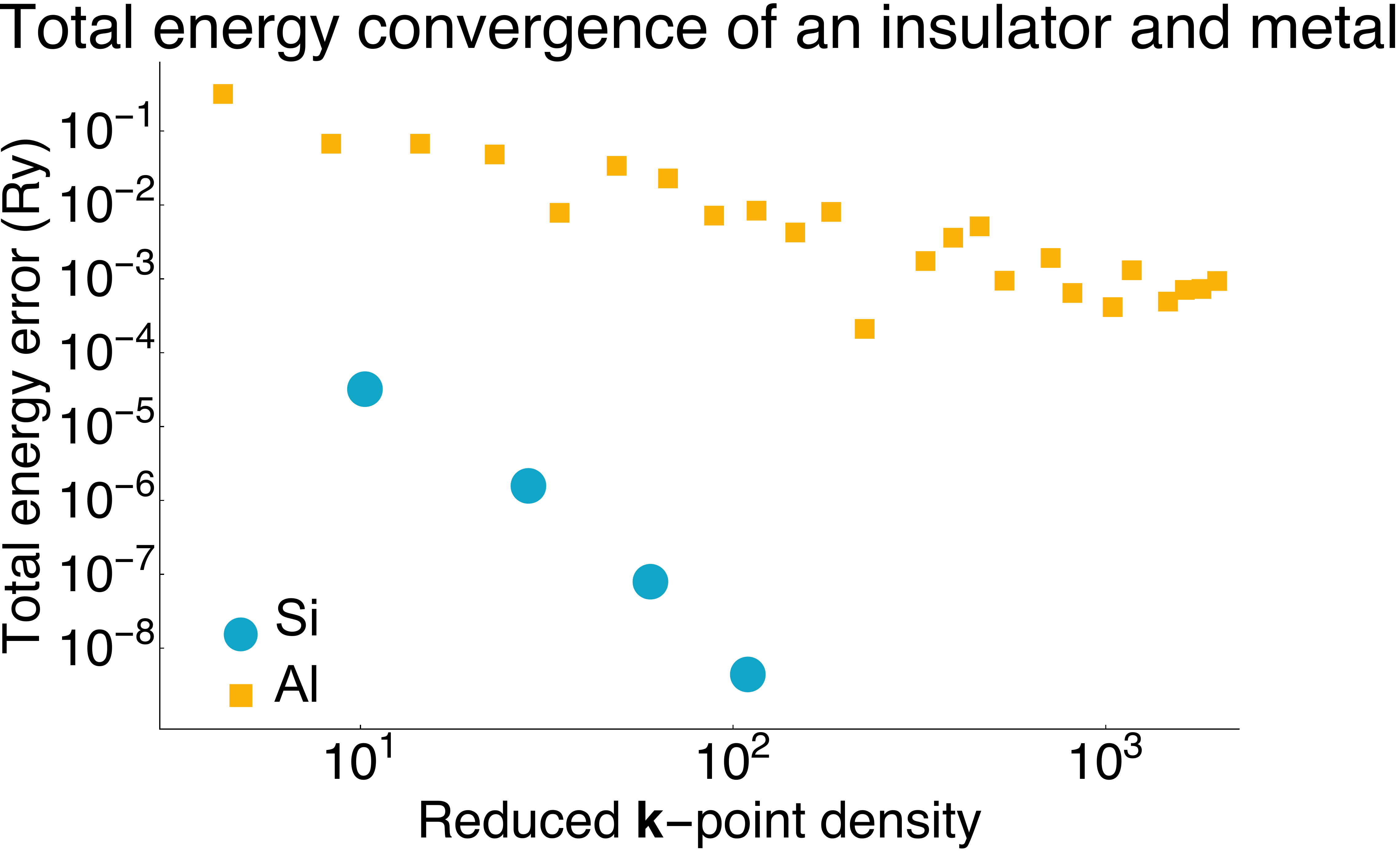}
\caption{A comparison of the convergence of the total energy of a metal, Al, and insulator, Si, with respect to \kb-point density. The rate of convergence of an insulator is exponential while that of a metal is linear. The differences in convergence rates and the erratic nature of the convergence of metals, leads to metals requiring far more computational resources, especially in circumstances such as high-throughput screening where high numerical accuracy is desired.}
\label{fig:metal_insulator}
\end{figure}

DFT calculations of metals take much more time than insulators often because they require much higher \kb-point densities to achieve the same numerical accuracy. The difficulty with metals stems from their Fermi surfaces. The Fermi surface is a surface separating the occupied regions of the electronic structure from the unoccupied regions. Fig. \ref{fig:metal_insulator} demonstrates the incredible convergence of an insulator, Si, compared to the considerably worse convergence of a metal, Al.

The band energy calculation for insulators converges quickly because the occupied bands are smooth and periodic. In Fig. \ref{fig:metal_insulator}, we used the rectangular integration method with a Monkorst-Pack grid (Monkhorst-Pack grids are \emph{regular grids} \cite{morgan2019generalized}) to calculate the band energy. The expected error convergence from integrating with the rectangular method decreases as $1/N^2$ in 1D where $N$ is the number of points, but there are classes of functions \cite{weideman2002numerical} that are smooth and periodic whose convergence rates are much faster, ranging from algebraic, but higher order ($1/N^4$, for example), geometric ($r^N, \, 0<r<1$), or exponential ($e^{-N}$). Insulators lack a Fermi surface and are cases where the rectangular method excels. 

\begin{figure}[htbp]
\includegraphics[width=200pt]{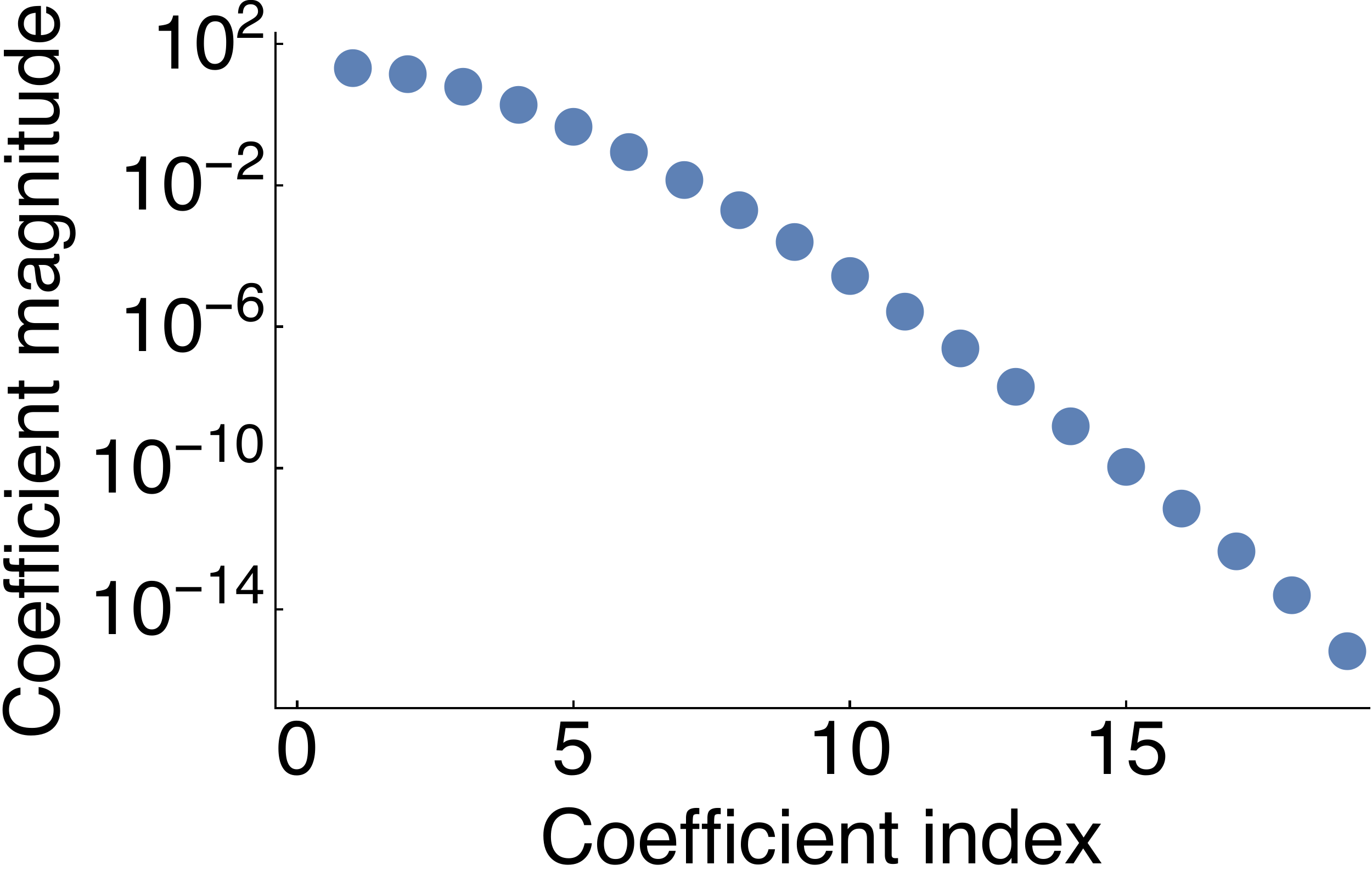}
\caption{The magnitude of the Fourier coefficients in a Fourier expansion of the function $y(x) = e^{2 \cos x}$. The magnitude of the 20th term in the expansion is around $10^{-15}$, which is proportional to the error in the integral of $y(x)$ over one period with the rectangular method with 20 integration points.}
\label{fig:fourier-coeff-convergence}
\end{figure}

\begin{figure}[htbp]
\includegraphics[width=200pt]{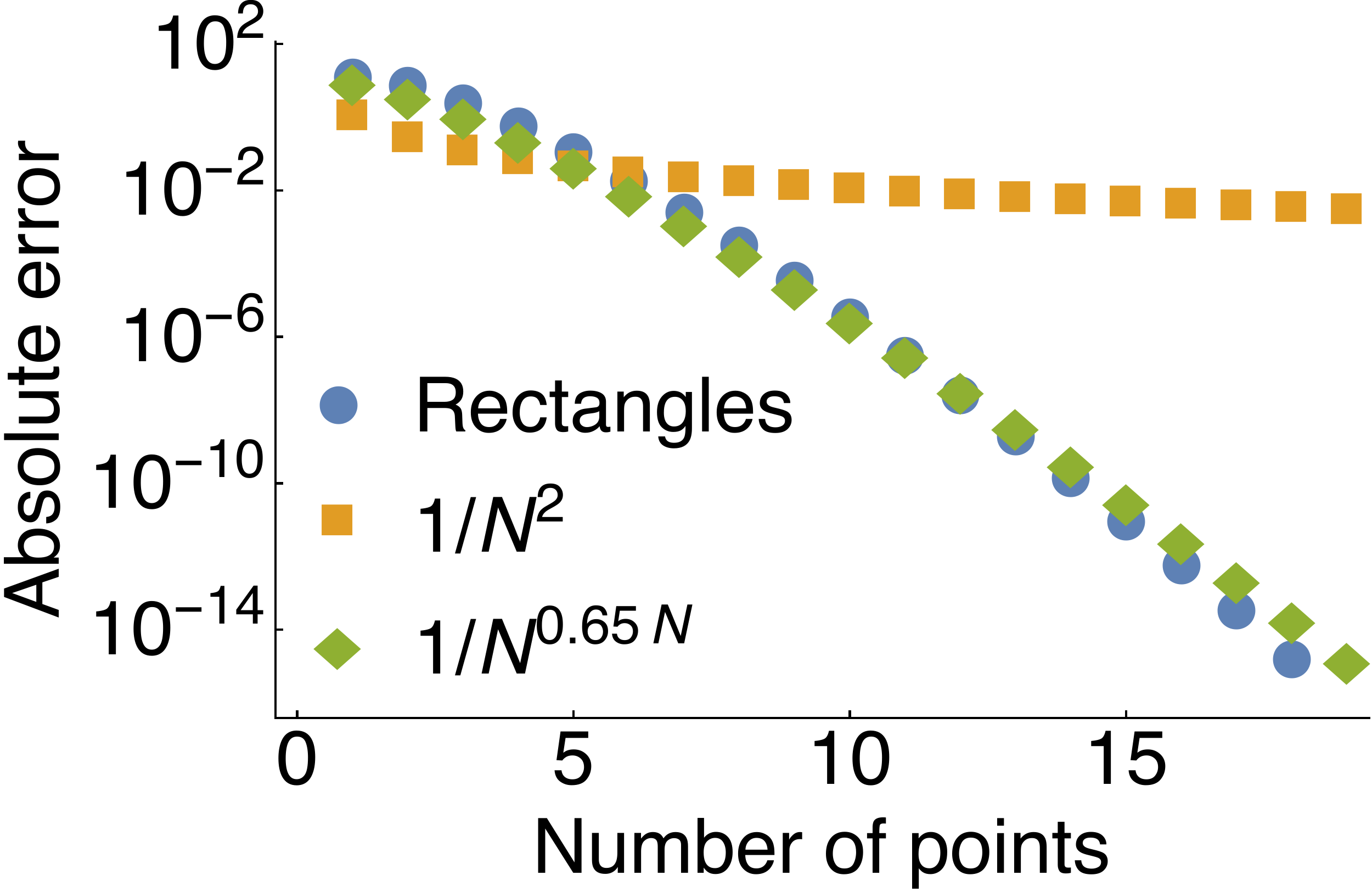}
\caption{A demonstration of the excellent convergence of the rectangular integration method on smooth periodic functions. We integrated the function $e^{2\cos{x}}$ over one period with varying numbers of integration points $N$. We show the error convergence with the rectangular method with blue circles, the expected algebraic convergence with orange squares, and a convergence fit with green diamonds.}
\label{fig:excellent-convergence}
\end{figure}

Integrals of smooth and periodic functions (functions whose Fourier expansions quickly drop to zero) with the rectangular method converge rapidly because lower order Fourier terms are integrated exactly with rectangles. This is illustrated in Fig. \ref{fig:fourier-coeff-convergence}. The rapid convergence of the Fourier coefficients should be compared to the error convergence with rectangular integration in Fig. \ref{fig:excellent-convergence}.

\begin{figure}[htbp]
\includegraphics[width=\linewidth]{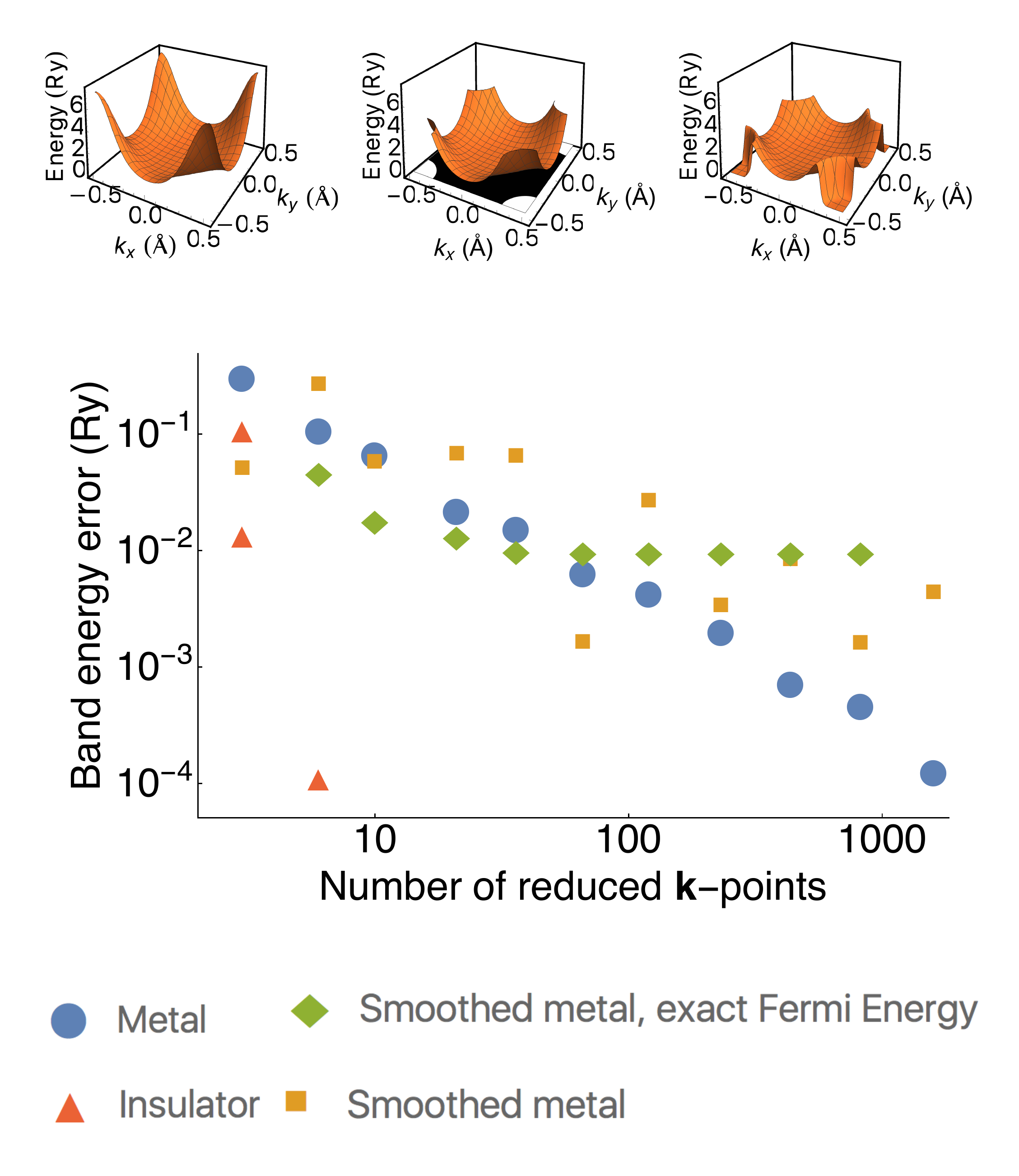}
\caption{A comparison of the band energy convergence of $w_1(\mathbf{k})=e^{-\textrm{cos}(2 \pi \mathbf{k}_x) - \textrm{cos}(2 \pi \mathbf{k}_y)}$ when treated as an insulator, a metal, a smoothed metal, and a metal with an exact Fermi level that has been smoothed. We show a single sheet of $w_1$ when treated as (a) an insulator, (b) a metal, and (c) a metal that has been smoothed. Below the band plots is shown their band energy convergences along with the convergence of a metal with an exact Fermi level that has had its discontinuity smoothed. Smoothing  shows no improvement in the band energy convergence. Smoothing with an exact Fermi level shows some improvement in band energy error, and eventually the band energy converges to the wrong value at high \kb-point densities.}
\label{fig:smoothed-convergence}
\end{figure}

The smoothness of the energy bands is removed for metals by the introduction of a Fermi level or highest occupied state that does not lie in a band gap (as it does for non-metals). Smearing methods attempt to restore the smoothness of the energy bands and the fast convergence of band energy calculations by smoothing out the discontinuities. See the top of Fig. \ref{fig:smoothed-convergence}

The band energy integral can take the following form
\begin{equation} \label{eq:band_energy}
\sum_n \int_{\mathbf{k} \in \mathbb{U}} \mathrm{d} \mathbf{k} \, E(\mathbf{k},n) \, \theta(E_f - E(\mathbf{k},n)),
\end{equation}
where $\mathbf{k}$ is a point in reciprocal space $\mathbb{R}^3$, $n$ is the band index, $\mathbb{U}$ is the reciprocal unit cell, $E(\mathbf{k},n)$ is the value of the electronic band structure at a given \kb-point and band, $E_f$ is the Fermi level, and $\theta$ is the step function. In this form, the integral is over the entire unit cell, and the integrand is discontinuous, resulting in the poor convergence of metals mentioned above. For insulators, the Fermi level lies in a band gap and the step function has no effect on the integrand.

\begin{figure}[htbp]
\begin{center}
\includegraphics[width=3in]{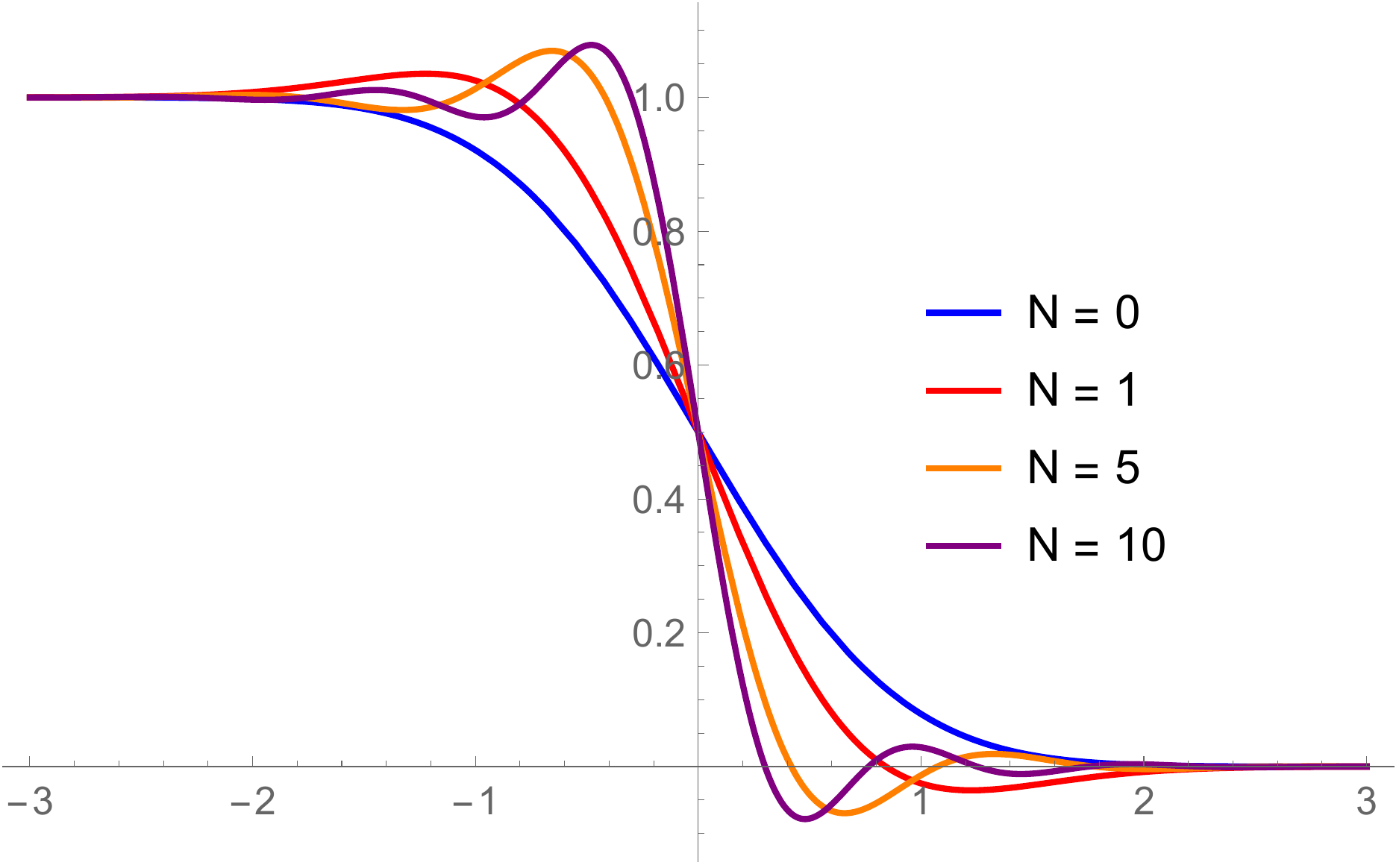}
\caption{Methfessel and Paxton \cite{methfessel1989high} used Hermite polynomial expansions to create smooth approximations to the step function. $N$ is the number of terms in the expansion.}
\label{fig:methfessel}
\end{center}
\end{figure}

The approach of Methfessel and Paxton (MP) is to replace the occupation step function with a continuous approximation of it so that the band structure remains smooth and continuous for metals \cite{methfessel1989high}. They expand the step function in Hermite polynomials; the expansion integrates polynomials of $N$-th order exactly. The $\delta$ function approximation is
\begin{equation}
  \delta(x) = \sum_{n = 0}^{N} A_n H_{2n}(x) e^{-x^2},
\end{equation}
where $A_n$ are expansion coefficients found from the orthogonality of Hermite polynomials, $H_{2n}(x)$ are Hermite polynomials, and $e^{-x^2}$ is a Gaussian weight function. Integrating the delta function they obtained approximations of the step function $S$:
\begin{align}
  S_0(x) &= \frac{1}{2}(1 - \text{erf}(x)) \\ S_N(x) &= S_0(x) +
  \sum_{n = 1}^{N} A_N H_{2n - 1}(x) e^{-x^2}.
\end{align}
These polynomial approximations are shown in Fig. \ref{fig:methfessel}. The zeroth order approximation of the step function is equivalent to Fermi-Dirac smearing; the higher order terms are corrections. In their paper, MP showed that Fermi-Dirac smearing is only accurate when the integrand is constant near the Fermi level. MP smearing is accurate when the integrand can be represented by a polynomial of degree $2N$ within an interval where the Gaussian weight function in the expansion is appreciably nonzero.

Marzari et al. draw attention to drawbacks to Methfessel-Paxton broadening \cite{marzari1996ab}: the thermal distribution loses the property of being monotonic, and the occupation numbers are no longer positive definite \cite{marzari1996ab}. As a consequence, the generalized entropy, the steepest descent directions, and the theorems for representing density matrices lose their explicit forms. Marzari's approximation of the delta function is
\begin{equation}\label{eq:marzari}
\tilde \delta (x) = \frac{2(2-\sqrt{2} x)}{\sqrt{\pi}} e^{-[x -1/\sqrt{2}]^2},
\end{equation}
where $x=\frac{\mu-\epsilon}{\sigma}$, $\mu$ is the Fermi level, $\epsilon$ is an energy variable, and $\sigma$ is the electronic temperature or smearing parameter. The benefit of Marzari-Vanderbilt smearing is occupations are positive definite. Lastly, Gaussian smearing is common in DFT codes and takes the form
\begin{equation}
g(\epsilon) = \frac{1}{\sigma \sqrt{2 \pi}} e^{-\frac{1}{2}((\epsilon - \mu)/\sigma)^2  },
\end{equation}
where all symbols are the same as those defined in Eq. \ref{eq:marzari}.

\onecolumngrid

\begin{figure}[htbp]
\begin{center}
\includegraphics[width=\linewidth]{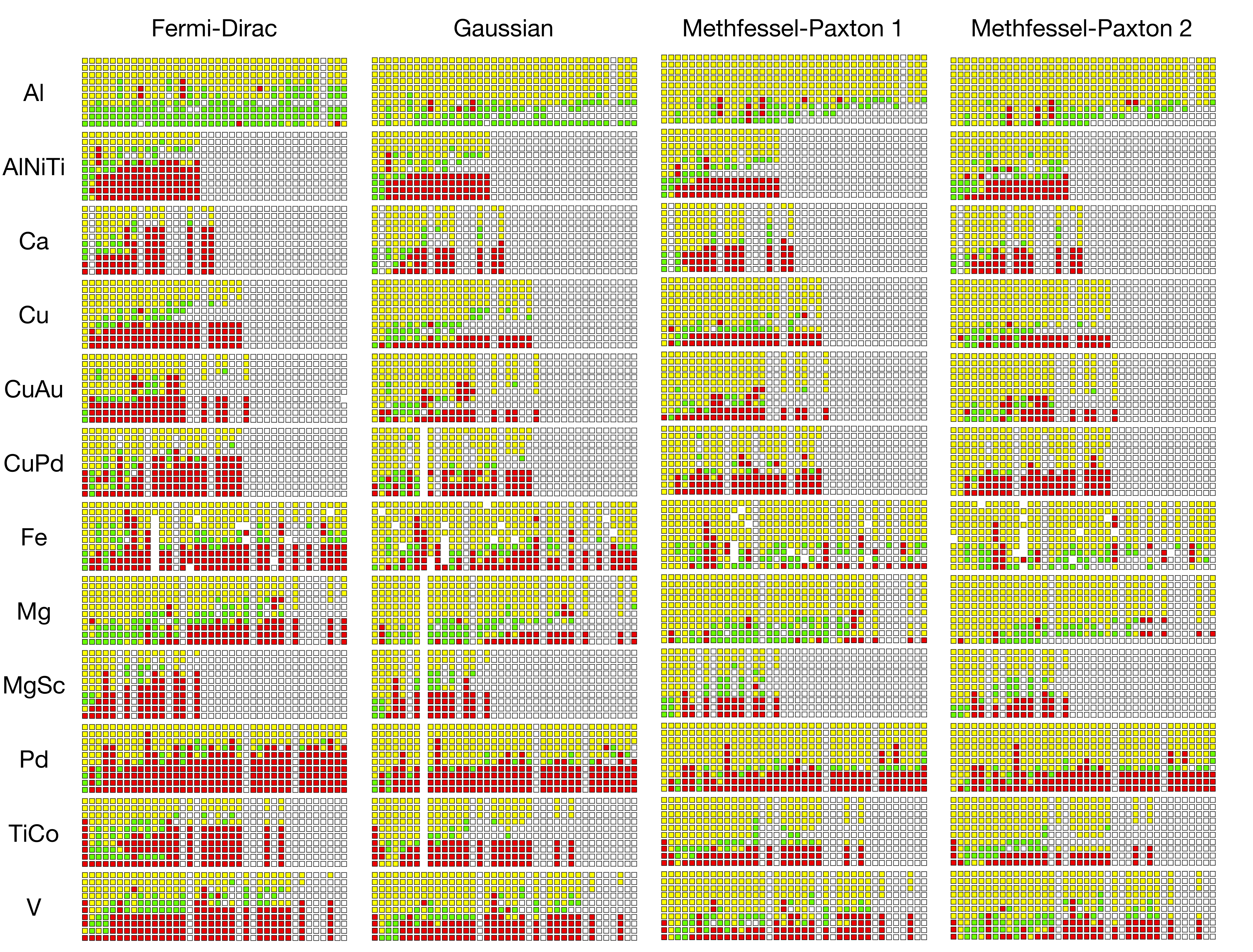}
\caption{Comparison of total energy errors in VASP with and without smearing. The value of the smearing parameter increases down the rows (values in eV are $1.0\times10^{-3}$, $1.0\times10^{-2}$, $2.5\times10^{-2}$, $5.0\times10^{-2}$, $7.5\times10^{-2}$, $1.0\times10^{-1}$, $2.5\times10^{-1}$, $5.0\times10^{-1}$, $7.5\times10^{-1}$, and $1.0\times10^{0}$). The \kb-point density gets larger with each column (number of \kb-points is $3^3$, $4^3$, \dots, $40^3$). Smearing in VASP usually does not improve total energy accuracy. Even in Al where smearing results are best, the most likely outcome of smearing is no improvement. The optimal smearing is dependent on the smearing parameter, smearing method, \kb-point density, and metal.}
\label{fig:tl-vasp-energy}
\end{center}
\end{figure}

\begin{figure}[htbp]
\begin{center}
\includegraphics[width=\linewidth]{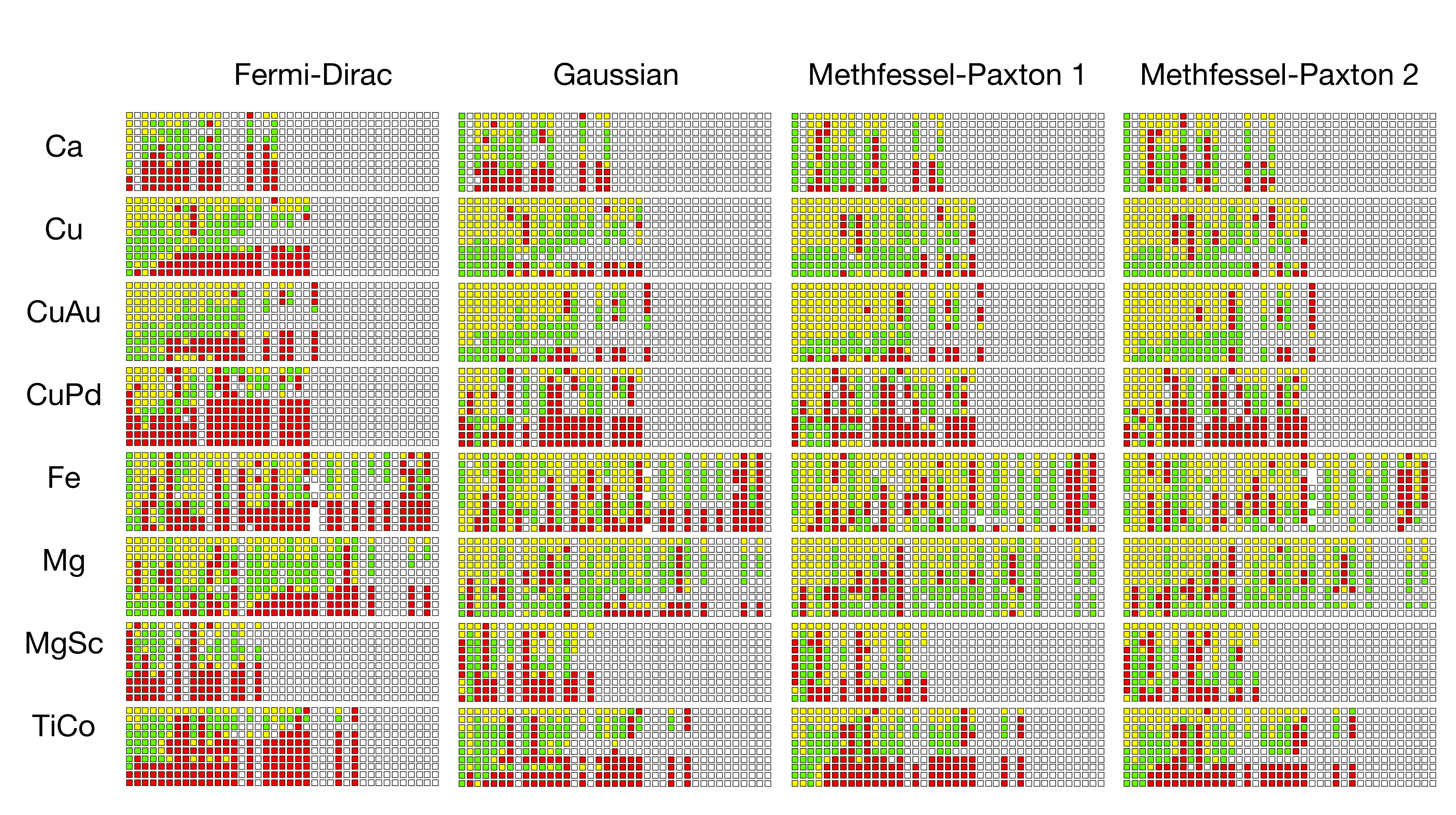}
\caption{Comparison of force errors in VASP with and without smearing. The value of the smearing parameter increasing down the rows (value in eV are $1.0\times10^{-3}$, $1.0\times10^{-2}$, $2.5\times10^{-2}$, $5.0\times10^{-2}$, $7.5\times10^{-2}$, $1.0\times10^{-1}$, $2.5\times10^{-1}$, $5.0\times10^{-1}$, $7.5\times10^{-1}$, and $1.0\times10^{0}$). The \kb-point density gets larger with each column (number of \kb-points is $3^3$, $4^3$, \dots, $40^3$). Smearing improves the accuracies of forces in VASP, but the optimal smearing depends on the smearing parameter, smearing method, \kb-point density, and metal.}
\label{fig:tl-vasp-forces}
\end{center}
\end{figure}

\onecolumngrid
\begin{figure}[htbp]
\begin{center}
\includegraphics[width=\linewidth]{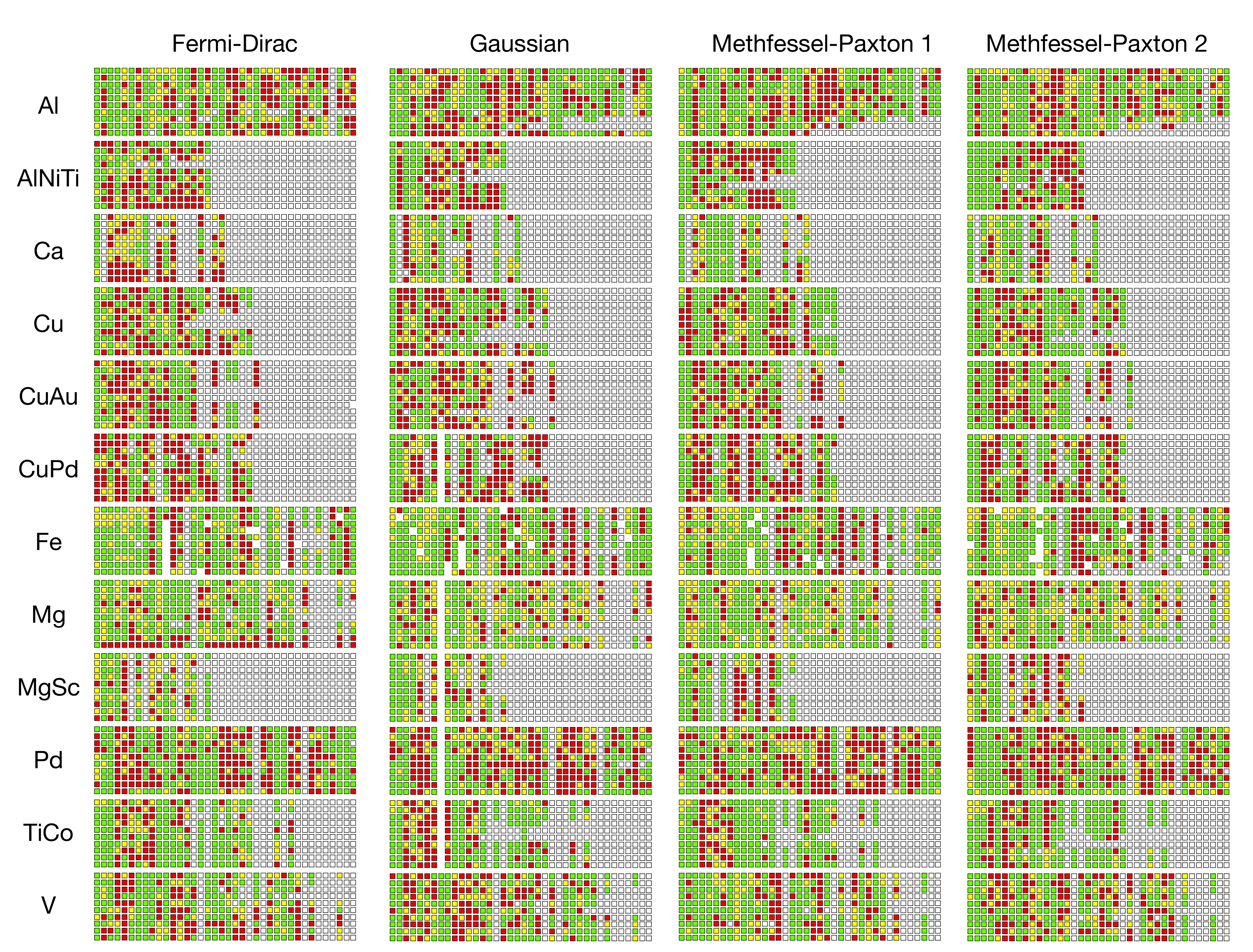}
\caption{Comparison of the number of SCF iterations in VASP with and without smearing. The value of the smearing parameter increasing down the rows (value in eV are $1.0\times10^{-3}$, $1.0\times10^{-2}$, $2.5\times10^{-2}$, $5.0\times10^{-2}$, $7.5\times10^{-2}$, $1.0\times10^{-1}$, $2.5\times10^{-1}$, $5.0\times10^{-1}$, $7.5\times10^{-1}$, and $1.0\times10^{0}$). The \kb-point density gets larger with each column (number of \kb-points is $3^3$, $4^3$, \dots, $40^3$). Smearing most often decreases the number of SCF iterations in VASP by at least 1 iteration; it less often but frequently increases the number of SCF iterations by 1.}
\label{fig:tl-vasp-iters}
\end{center}
\end{figure}

\begin{figure}[htbp]
\begin{center}
\includegraphics[width=\linewidth]{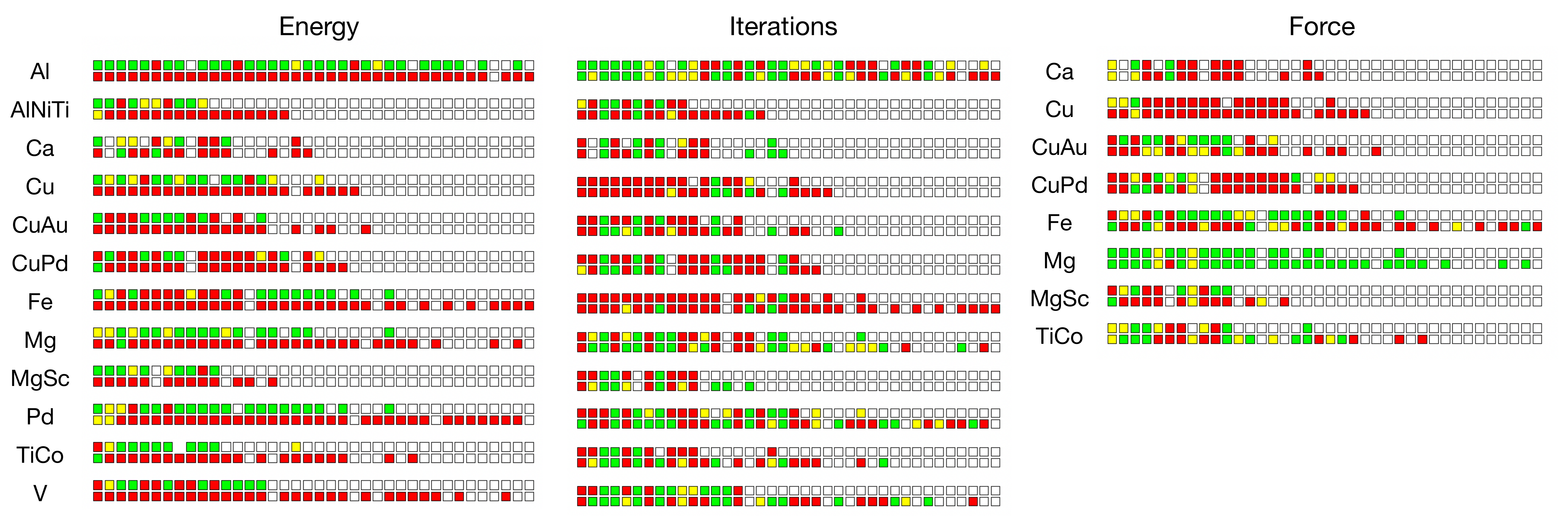}
\caption{The total energies, forces and number of SCF iterations in VASP with tetrahedron methods compared against the same values obtained without smearing or tetrahedra. The tetrahedron methods down the rows are Bl\"ochl's and linear tetrahedra. The \kb-point density gets larger with each column (number of \kb-points is $3^3$, $4^3$, \dots, $40^3$). Bl\"ochl's tetrahedron method most often improves the total energy accuracy. Tetrahedron methods may improve or reduce the number of SCF iterations and force accuracies, giving random and mixed results for both. Only for Mg did tetrahedra consistently improved the forces.}
\label{fig:tl-vasp-tet}
\end{center}
\end{figure}
\twocolumngrid

Tetrahedron methods are an alternative or additional approach to smearing that attempt to improve the poor convergence of metals. Tetrahedron methods split the unitcell or irreducible Brillouin zone into tetrahedra and interpolate the band structure with a linear polynomial within each tetrahedra \cite{lehmann1972numerical, jepson1971electronic}. Integrations are performed analytically within each tetrahedra, and the Fermi surface is approximated by a polygon. Bl\"{o}chl et al. improved upon the linear tetrahedron method by accounting for errors from missing terms in the polynomial expansion \cite{blochl1994improved}. Kawamura et al. generalized Bl\"{o}chl's corrections to more general Brillouin zone integrals, including phonon frequency and response functions \cite{kawamura2014improved}.

\section{Tests}

In order to test if smearing and tetrahedron methods reduce the number of SCF cycles, improve forces on atoms,  or reduce the \kb-point density required for a given accuracy, we ran close to 40,000 total DFT calculations in Quantum Espresso (QE) \cite{giannozzi2009quantum,giannozzi2017advanced} and VASP\cite{kresse1993ab,kresse1996efficiency}. VASP was chosen because of its popularity and QE because it is open source. In all calculations, the structural degrees of freedom were fixed.

In QE we tested 12 metallic systems (Al, AlNiTi, Ca, Cu, CuAu, CuPd, Fe, Mg, MgSc, Pd, TiCo, and V), and compared the total energy convergence, stress convergence, and number of SCF cycles for different smearing and tetrahedron methods. We used QE version 6.3 for all calculations. The smearing methods we tested were  Fermi-Dirac, Gaussian, Marzari-Vanderbilt, and Methfessel-Paxton smearing. Each smearing method was tested with 10 different smearing parameters, ranging from $10^{-10}$ Ry to $10^{-1}$ Ry. The tetrahedron methods tested in QE included linear, Bl\"{o}chl, and Kawamura. The pseudopotentials were obtained from \url{ http://www.quantum-espresso.org/pseudopotentials}. All pseudopotentials implemented the generalized gradient approximation \cite{perdew1996generalized}, had nonlinear core corrections, and were scalar relativistic. 

In VASP we tested the same 12 metallic systems tested in Quantum Espresso\cite{giannozzi2009quantum,giannozzi2017advanced}. We used a pre-release version of VASP version 6.0 for all calculations. The smearing methods we compared were Fermi-Dirac, Gaussian, and 1st and 2nd order Methfessel-Paxton. Each smearing method was tested with 11 different smearing values, ranging from $10^{-5}$ eV to $1$ eV. The tetrahedron methods included linear and Bl\"{o}chl's tetrahedron methods, and the amount of smearing for tetrahedra tests in VASP was $\sigma = 1\times10^{-5}$ eV with Gaussian smearing. All pseudopotentials for the tests in VASP used the projector-augmented wave method \cite{kresse1999ultrasoft,blochl1994projector,kresse1994norm}, the generalized-gradient approximation \cite{perdew1996generalized}, and augmentation charge corrections. 

Our results for smearing in VASP are shown in Figs. \multiref{fig:tl-vasp-energy}{fig:tl-vasp-iters}. Our results for tetrahedron methods in VASP are in Fig. \ref{fig:tl-vasp-tet}. We compare the convergence of the energy components of the total energy in VASP for a few metals in Fig. \ref{fig:vasp_Al-AlNiTi-Ca-1st-order-Methfessel-Paxton}. See the supplementary information for addition plots.

For both VASP and QE, the energy cutoffs were 2$\times$ the largest ENMAX or wfc\_cutoff in the system's pseudopotential file, respectively. Forces or stresses were not compared for 4 of the systems (Al, AlNiTi, Pd, and V) because forces were zero by symmetry.

To improve readability and because the results for VASP and QE are very similar, we only show results in the main text for VASP. Results for QE and $k$-point convergence plots for all calculations performed in this study can be found in the supplementary information.

\section{Discussion}

Our goal in running these tests was to find a smearing method and smearing parameter robust enough to accurately simulate many different metals, and identify input parameters that would make it possible to treat DFT codes as a black box. Another objective was to see if smearing or tetrahedra would increase the efficiency or accuracy of DFT simulations of metals. The quantity \emph{error ratio} is introduced to make it easier to compare total energies and forces with and without smearing. It is defined as
\begin{equation}
\epsilon_r = \frac{ \log(\frac{\epsilon_s}{\epsilon_n}) }{ \log(10) },
\end{equation}
where $\epsilon_s$ is the error with smearing and $\epsilon_n$ is the error without smearing, both of which are measured at the same \kb-point density. As an example, if the error ratio is $-1$ for the total energy error for a VASP or QE simulation, smearing is $10\times$ more accurate than the same simulation without smearing. If the error ratio were 1, no smearing would be $10\times$ more accurate than smearing. The error in the total energy for QE and VASP is simply the difference in the calculated energy and an energy ``answer" obtained without smearing at a large \kb-point density ($100\times100\times100$ Monkhorst-Pack grid). At the time of writing, one of the routines in VASP has a hard exit if the smearing parameter is less than $10^{-5}$ eV that prevents investigating smearing parameters of smaller values without adjusting internal parameters and recompiling. For this reason, the value of the smearing parameter for tests ``without smearing" is $10^{-5}$ eV. In QE, the value of the smearing parameter is $1 \times 10^{-10}$ Ry. The error for forces in VASP is 
\begin{equation}
\epsilon^{f} = \sum_\text{atoms} || \mathbf{F}_s - \mathbf{F}_a ||,
\end{equation}
or 2-norm where $\mathbf{F}_s$ is the force on one of the atoms in the atomic basis, $\mathbf{F}_a$ is the force ``answer", which we approximate by the force on the same atom without smearing at a very large \kb-point density. QE returns the stress tensor instead of the forces on atoms. The error for stresses is the difference in the determinants of the stress tensors. For the number of iterations, we do not use the error ratio but take the difference in the number of electronic iterations with and without smearing.

We include ``traffic light" plots to make it easier to see how smearing and no smearing compare for all the plots in the appendix. In traffic light plots for forces, stresses, and energies, there is a green box where smearing has 25\% less error than no smearing, a red box were smearing has 25\% more error, and a yellow box for in between the two. For the number of SCF iterations, there is a green box where smearing resulted in at least one fewer iteration than no smearing, a red box where smearing resulted in at least one more iteration, and a yellow box between. This information is also included in the legends of the figures. DFT codes have an accuracy up to around 1 meV/atom. We ignore simulations that resulted in total energy errors less than 0.1 meV/atom, and mark them with a white box. A missing box is placed where a DFT simulation failed. Traffic light plots are shown in Figs. \multiref{fig:tl-vasp-energy}{fig:tl-vasp-tet}.

\begin{figure}[htbp]
\includegraphics[width=\linewidth]{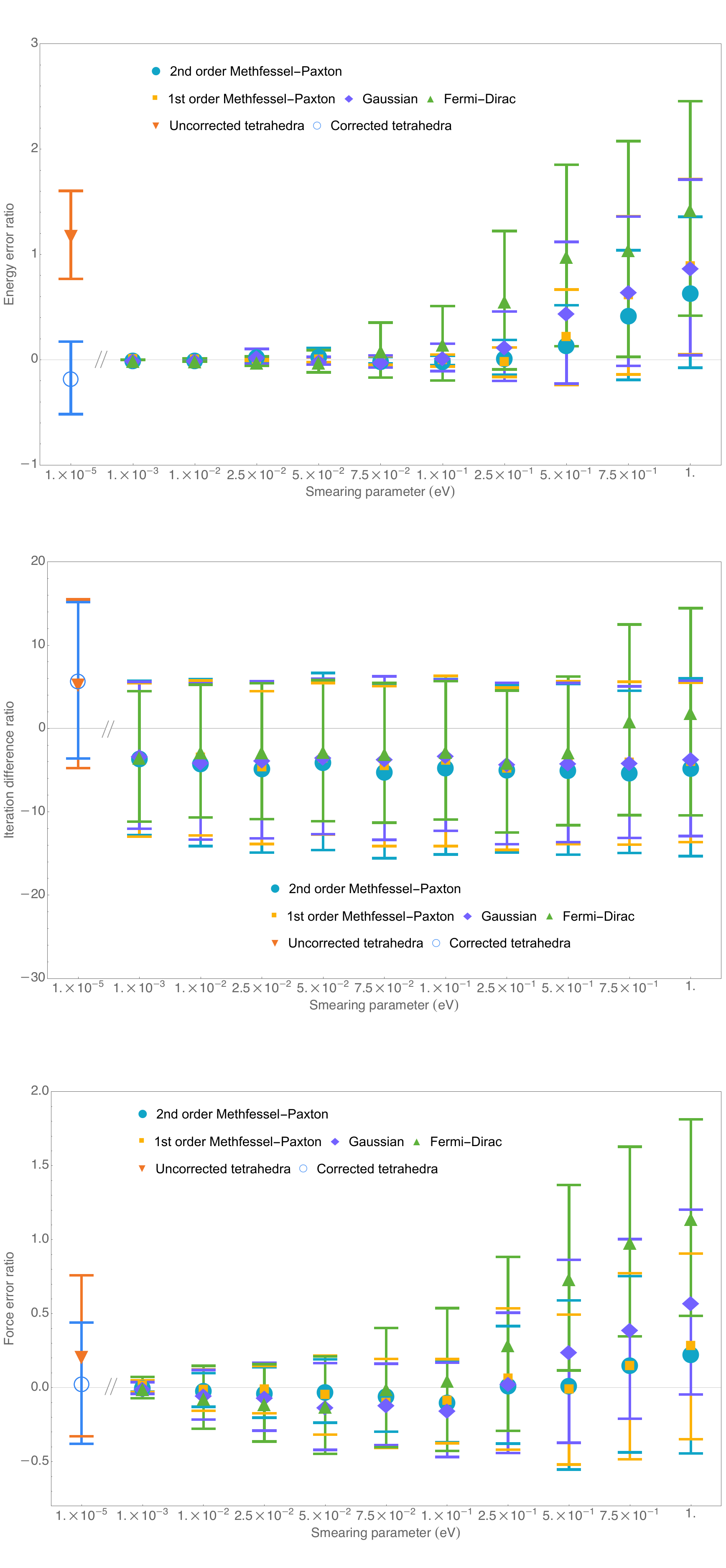}
\caption{Mean and mean deviation of energy ratios, iteration differences, and force ratios in VASP. Tetrahedron methods were only tested with the smallest smearing parameter.}
\label{fig:vasp-combined}
\end{figure}

To determine which set of parameters would work best in general, we took the mean and mean deviation of the energy error ratios, force or stress error ratios, and iteration differences for all 12 metals for a given smearing method and parameter or tetrahedron method. These are shown in Fig. \ref{fig:vasp-combined}. It is apparent from these plots that the smaller the smearing parameter, the more accurate the energies and forces. Smearing results in a reduction in the number of SCF iterations that is \emph{independent} of the value of the smearing parameter. The average reduction is 4--5 iterations regardless of the method of smearing or the smearing parameter. Smearing in VASP with a smearing parameter around $1 \times 10^{-1}$ eV to $1 \times 10^{-2}$ eV appears to improve forces by a small amount but have large mean deviations that often result in less accurate forces. The larger the smearing parameter, the more likely one will get large error in the energy or forces. Bl\"ochl's tetrahedron method in VASP resulted in more accurate total energies but a large mean deviation indicates it sometimes leads to less accurate energies.

\begin{figure}[htbp]
\includegraphics[width=\textwidth,height=\textheight,keepaspectratio]{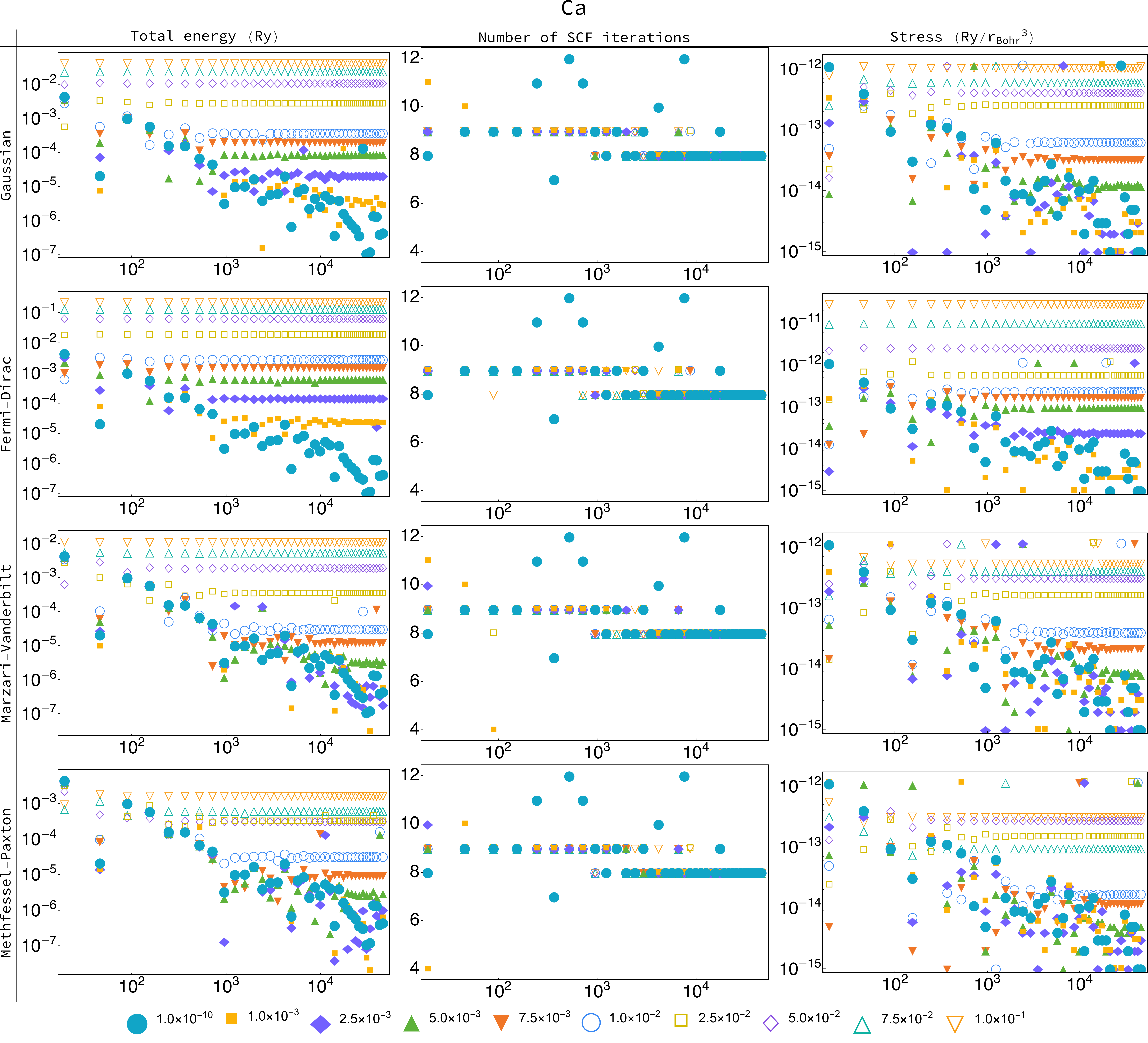}
\caption{The total energy convergence, number of SCF cycles, and force convergence for Ca in VASP. For all plots, the $x$-axis is the reduced \kb-point density in units of cubic Angstroms. The legend at the bottom gives the amount of smearing in electron volts.}
\label{fig:vasp_Ca-smooth}
\end{figure}

Fig. \ref{fig:vasp_Ca-smooth} shows a typical result of smearing on Ca as computed in VASP. Many more smearing tests can be found in the supplementary information. A major feature of the smearing tests is the leveling off of the convergence for large amounts of smearing. This is expected because the integral of the smeared band structure is different from the unsmeared, and the difference of the two integrals is the same as the error where the error convergence levels off. Smearing is expected to show improvement over no smearing for \kb-point densities just before the density where the convergence levels off. The lack of improvement could be related to uncertainty in the Fermi level or the energy value where smearing occurs (see Eq. \ref{eq:band_energy}). We demonstrate these ideas in Fig. \ref{fig:smoothed-convergence}. The DFT runs with practically no smearing ($\sigma = 1\times10^{-5}$ eV in VASP) often show the best error convergence.

\begin{figure}[htbp]
\begin{center}
\includegraphics[width=\linewidth,keepaspectratio]{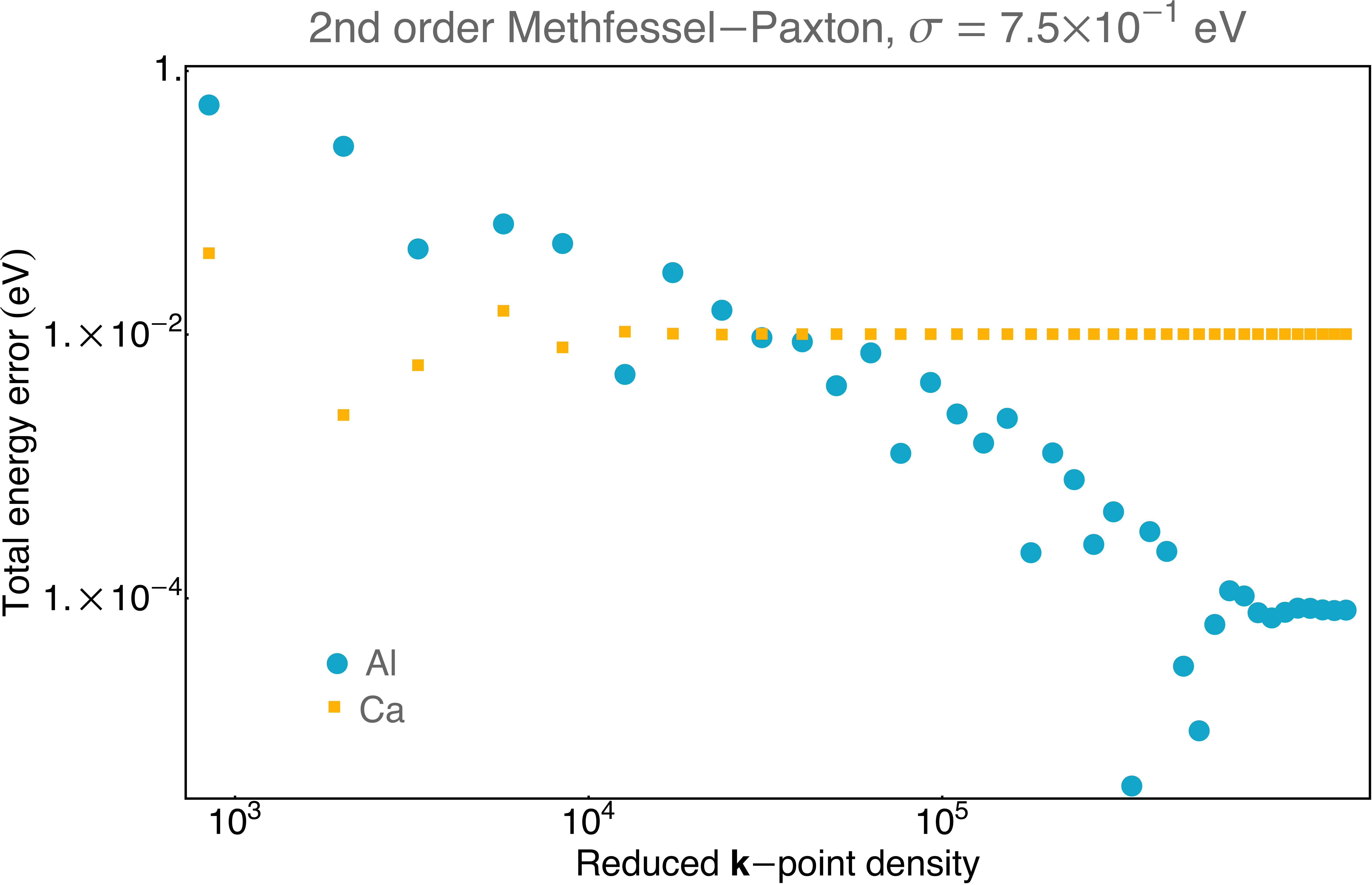}
\caption{The total energy error convergence for Al and Ca with second order Methfessel-Paxton smearing with a smearing parameter value of 0.75 eV. For Al, this is one of the few cases where Methfessel-Paxton smearing performs better than no smearing. The same smearing method and parameter performs very poorly for another metallic system, Ca, and demonstrates that the optimal smearing is dependent on the system.}
\label{fig:vasp_Al-Ca-comp}
\end{center}
\end{figure}

There are very few cases where Methfessel-Paxton smearing improves total energies: Al with $\sigma$ values of 1 eV and 0.75 eV are two examples. These same smearing parameters show very poor performance---leveling off of the error---for many other systems with Methfessel-Paxton smearing. There is no single, optimal value of the smearing parameter for all metals; the optimal smearing depends on the metal. One takeaway from the data is the smaller the smearing, the more accurate the total energies and forces. The size of the Fermi level plays a role in how much smearing occurs for a given smearing parameter, and greater caution is needed when selecting a single smearing parameter to study systems with widely varying Fermi levels or total energies. One cannot recommend a ``rule of thumb" as there is not one. This is demonstrated in Fig. \ref{fig:vasp_Al-Ca-comp}.

\begin{figure}[htbp]
\includegraphics[width=\linewidth,keepaspectratio]{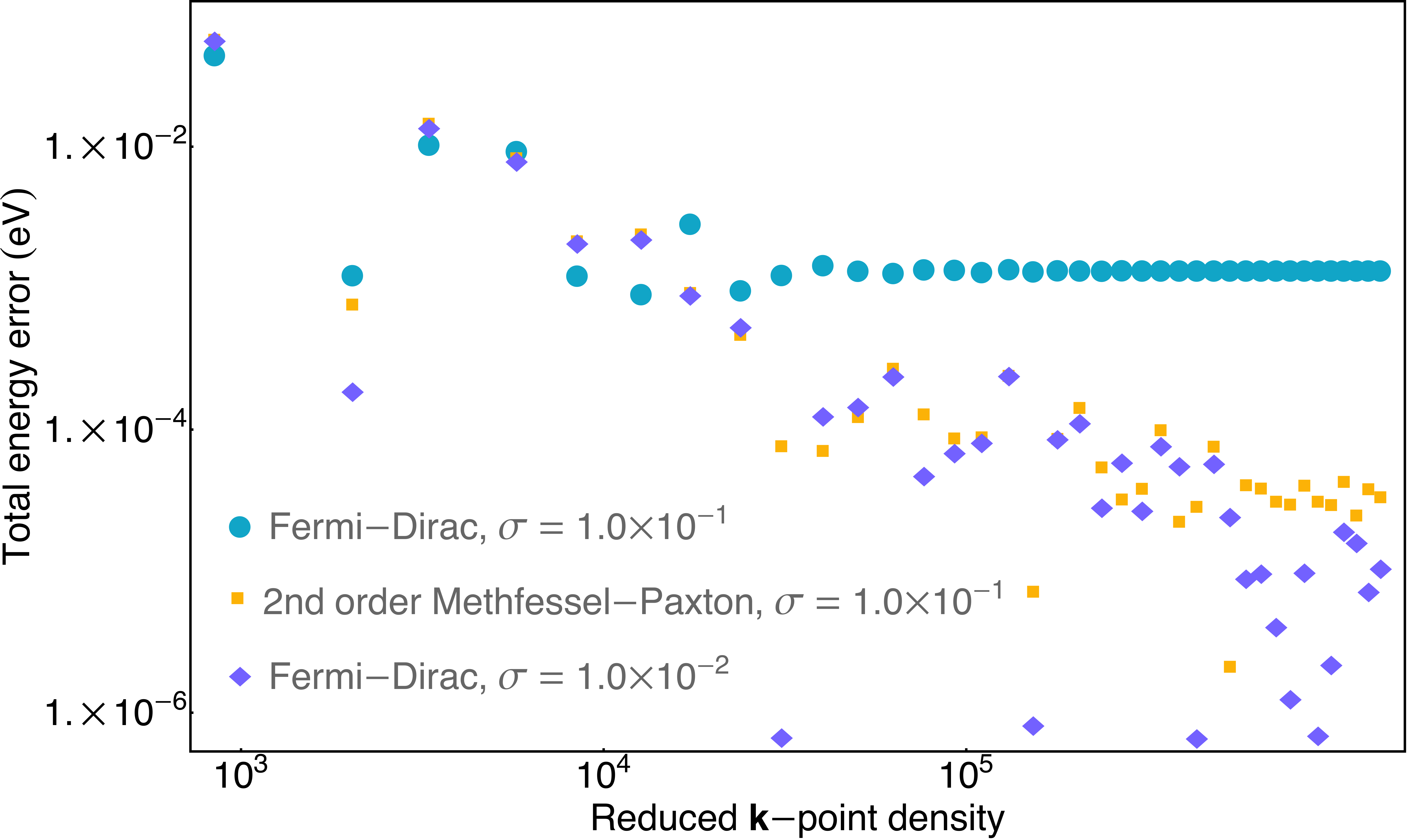}
\caption{The total energy error convergence for Ca. For the same value of the smearing parameter, Fermi-Dirac smearing results in more smearing than second order Methfessel-Paxton, which is apparent from the leveling off the convergence. It would appear that Methfessel-Paxton smearing performs better than Fermi-Dirac, but it is possible to select a smaller smearing parameter for Fermi-Dirac where the two smearing methods have comparable amounts of smearing and performance. In the limit the smearing parameter goes to zero, the performance of the two is identical.}
\label{fig:vasp_Ca-fd-mp}
\end{figure}

The level of smearing is not the same for the different smearing methods (Gaussian, Fermi Dirac, etc.). It is generally believed that Methfessel-Paxton smearing is superior to other smearing methods. However, Methfessel-Paxton smearing appears to perform better because it results in less smearing for the same smearing parameter than other smearing methods. For example, Fermi-Dirac and Methfessel-Paxton smearing perform the same in Fig. \ref{fig:vasp_Ca-fd-mp}.

\begin{figure}[htbp]
\includegraphics[width=\linewidth,keepaspectratio]{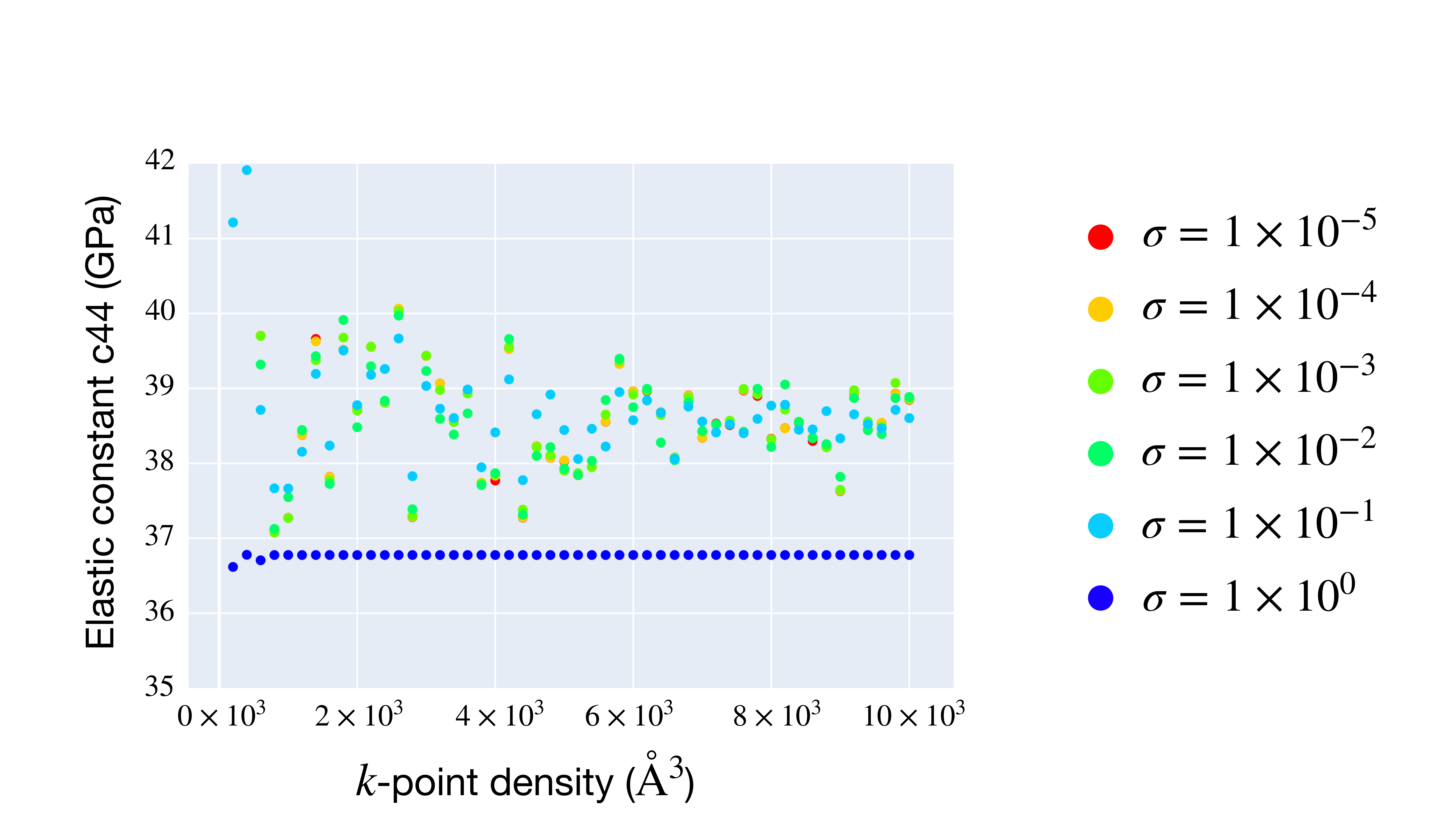}
\caption{Convergence of the elastic constant with respect to \kb-point density for varying smearing parameters in VASP. The system is HCP Zr and the smearing method is Gaussian. A large smearing parameter causes the elastic constant to converge quicker but converges to a value different than it would with no smearing. In some cases, the value of the elastic constant with smearing agrees better with experiment.}
\label{fig:elastic}
\end{figure}

Elastic constants converge slowly with respect to \kb-point density and require extreme \kb-point densities to converge without smearing. Smearing will cause the elastic constants to converge with a reasonable number of \kb-points but like other tests in this paper, what they converge to differs from the elastic constants without smearing (see Fig. \ref{fig:elastic}). Smearing may help the elastic constants agree with experimental values \cite{louail2004calculation}. In a comprehensive study, it seems likely that smearing would improve agreement with experiment for many systems but make some systems further from agreement. Broad testing of the effect of smearing on elastic constants is not in the scope of this work---the reader is encouraged to perform their own tests.

\begin{figure}[htbp]
\includegraphics[width=\textwidth,height=\textheight,keepaspectratio]{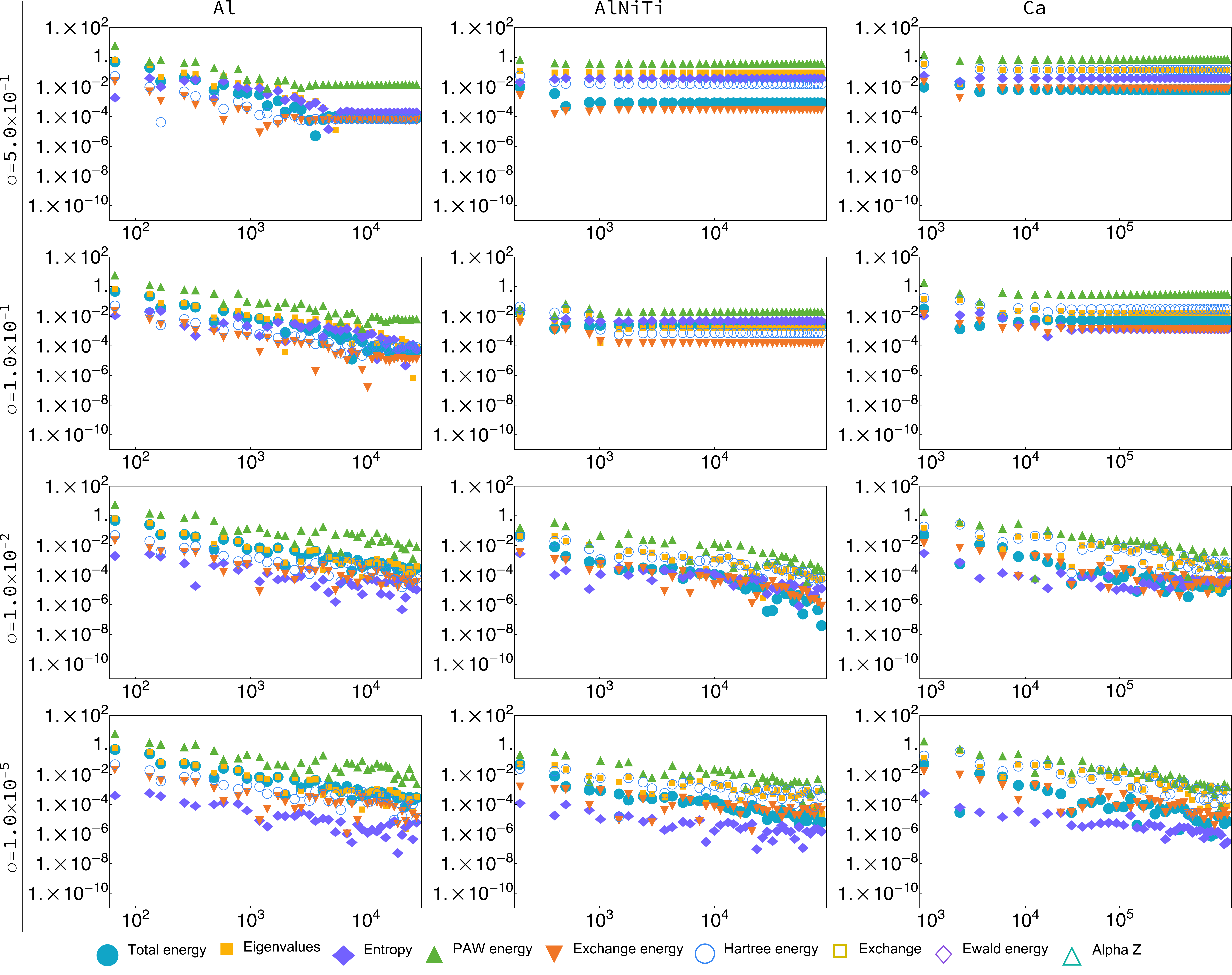}
\caption{The convergence of components of the total energy in VASP for the metals Al, AlNiTi, and Ca with 1st order Methfessel-Paxton smearing. The atomic energy contribution to the total energy is left out due to its lack of dependence on the amount smearing or the \kb-point density. For all plots, the $x$-axis is the reduced \kb-point density in units of cubic Angstroms.}
\label{fig:vasp_Al-AlNiTi-Ca-1st-order-Methfessel-Paxton}
\end{figure}

In the plots in Fig. \ref{fig:vasp_Al-AlNiTi-Ca-1st-order-Methfessel-Paxton}, the convergence of the component energies of the total energy for Al, Ca, and Cu with 1st order Methfessel-Paxton smearing in VASP is observed. The atomic energy contribution to the total energy is ignored due to its lack of dependence on smearing or \kb-point density. The errors in the component energies all decrease as the amount of smearing decreases. One exception is Cu with 2nd order Methfessel-Paxton smearing where there is improved convergence with $\sigma=2.5 \times 10^{-1}$ than $\sigma=1.0 \times 10^{-5}$, but, as previously discussed, this is an exception; this same smearing results in worse performance in other metals. Notice that some of the errors in the component energies are correlated.

\begin{figure}[htbp]
\includegraphics[width=\textwidth,height=\textheight,keepaspectratio]{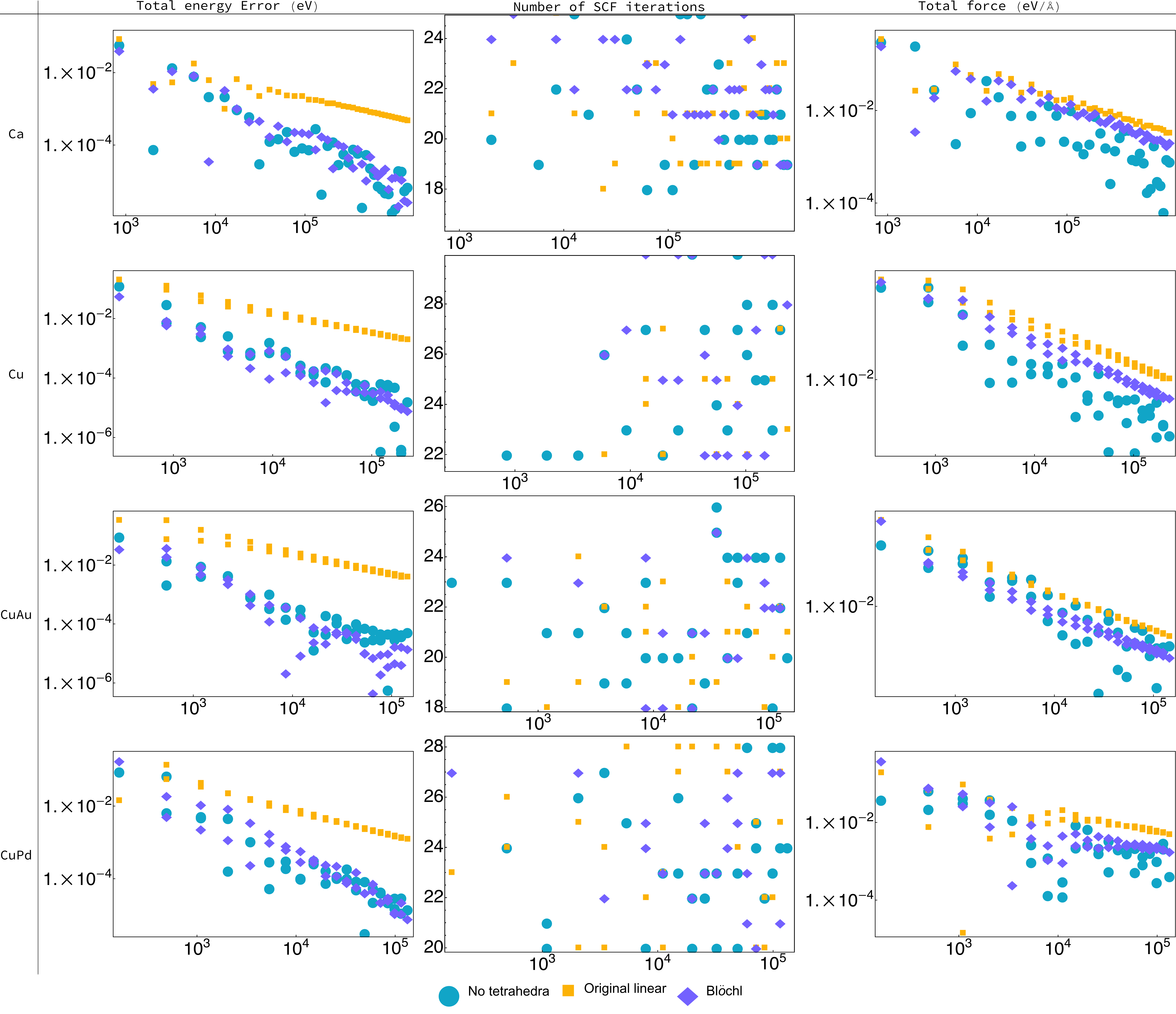}
\caption{The total energy convergence, number of SCF cycles, and force convergence of Ca, Cu, CuAu, and CuPd with tetrahedron methods in VASP. For all plots, the $x$-axis is the reduced \kb-point density in units of cubic Angstroms.}
\label{fig:vasp_Ca-Cu-CuAu-CuPd-comb-tet}
\end{figure}

For VASP (Fig. \ref{fig:vasp_Ca-Cu-CuAu-CuPd-comb-tet}), we look at the performance of tetrahedron method methods on Al, Ca, Cu, and CuAu. The performance of the uncorrected tetrahedron method also shows very consistent, poor convergence, but also consistent improvement in total energy error with higher \kb-point densities. The corrected tetrahedron method shows the same or better convergence of the total energy than calculations without tetrahedra and practically no smearing.

\begin{figure}[htbp]
\includegraphics[width=\linewidth]{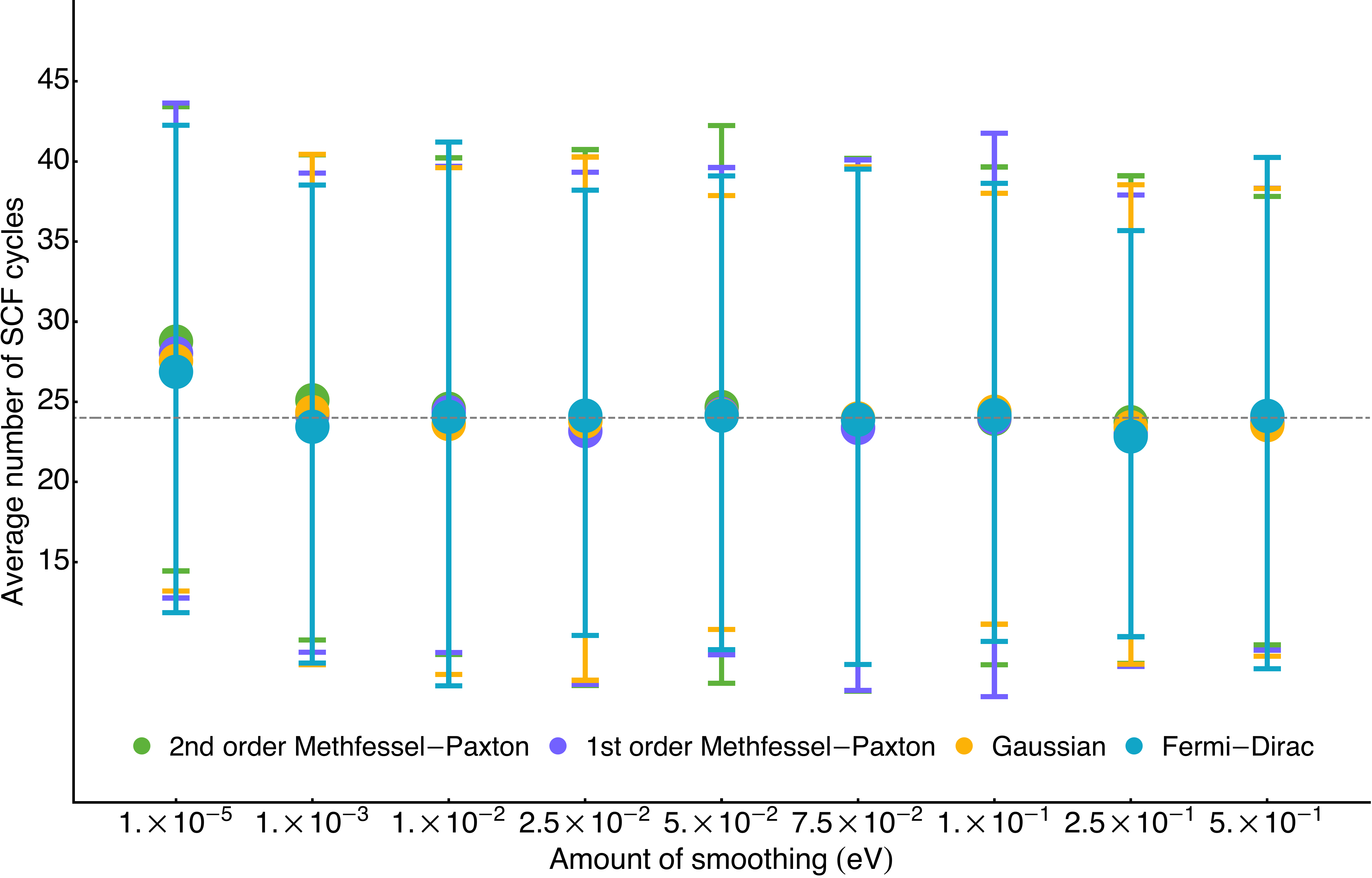}
\caption{The number of SCF cycles averaged over all 12 metals in VASP. The VASP calculations with practically no smearing require slightly more iterations. The horizontal black line is a guide for the eye.}
\label{fig:vasp-smearing-comparison}
\end{figure}

Finally, we look at the effect of smearing and tetrahedron methods on the number of SCF iterations. Smearing was developed to fix issues in the SCF cycle related to band sloshing, which should result in fewer SCF iterations. In QE, smearing and tetrahedron methods had little effect on the total number of SCF cycles needed to reach convergence. At higher \kb-point densities, the number of SCF iterations for all smearing and tetrahedron methods convergences to the same value to within a few iterations. In VASP, the number of SCF iterations to reach self-consistency is sporadic and random at all \kb-point densities and for all methods and amounts of smearing. Fig. \ref{fig:vasp-smearing-comparison} shows the average of the number of iterations in the SCF cycle for all systems of a given smearing method and value. The average of the standard deviations is represented with error bars. The smallest smearing value has an average number of SCF iterations that is 4 or 5 more iterations than all the others. All other smearing parameters, regardless of the method or amount, have approximately the same average and average standard deviation. Smearing has little effect on the number of SCF iterations in QE and reduces the number of iterations in VASP by around 5 iterations.

\section{Conclusion}

We ran about 40,000 DFT calculations on a suite of twelve metals to test the efficiency of smearing and tetrahedron methods. We used four smearing methods, three tetrahedron methods, and ten different smearing parameters in Quantum Espresso\cite{giannozzi2009quantum,giannozzi2017advanced}, and twelve metals, four smearing methods, two tetrahedron methods and eleven different smearing parameters in VASP \cite{kresse1993ab,kresse1996efficiency}. Smearing has significant, negative effects on the total energy when the smearing parameter is large, and little systematic, positive effect when small. Smearing has little effect on the number of SCF iterations in QE and decreases the number of SCF iterations in VASP by about 5 iterations on average. The observed reduction is \emph{independent} of the smearing parameter.

Tetrahedron methods have no effect on the number of SCF cycles in QE and significantly increase the number of SCF cycles in VASP. In QE, Bl\"{o}chl's tetrahedra slightly improved total energies and stresses. In VASP, Bl\"{o}chl's tetrahedra improved total energies on average. Small amounts of smearing showed slight improvement in the forces and stresses in VASP and QE.

Due to the risk of selecting a parameter that results in too much smearing, and the minimal effect of smearing on the number of SCF cycles, we recommend using a very small smearing parameter in DFT calculations, especially for high-throughput or machine learning applications. We recommend using Bl\"{o}chl's tetrahedron method in QE due to minor improvements to total energies and stresses. Bl\"{o}chl's tetrahedron method in VASP improved total energies but significantly increased the number of SCF cycles. We emphasize that, although smearing may lead to minor reductions in the number of SCF cycles and slightly more accurate forces for certain systems, smearing leads to inaccurate DFT calculations when the smearing parameter is large, and there does not exist an optimal smearing parameter; there is no reliable ``rule of thumb" to follow when choosing smearing parameters. The optimal smearing parameter is dependent on the system, smearing method, smearing parameter, and \kb-point density. As little smearing as possible is the safest option when treating DFT as a black box.

\section{Acknowledgements}

This work was supported by ONR (MURI N00014-13-1-0635).

\onecolumngrid

\newpage

\newpage
\twocolumngrid

\bibliography{/Users/jeremy/Papers/bib/master.bib}{}\bibliographystyle{unsrt}

\end{document}


\onecolumngrid

\FloatBarrier
\section{Quantum Espresso}

\FloatBarrier
\subsection{Misc. Plots}

\begin{figure}[H]
\begin{center}
\includegraphics[width=\textwidth,height=\textheight,keepaspectratio]{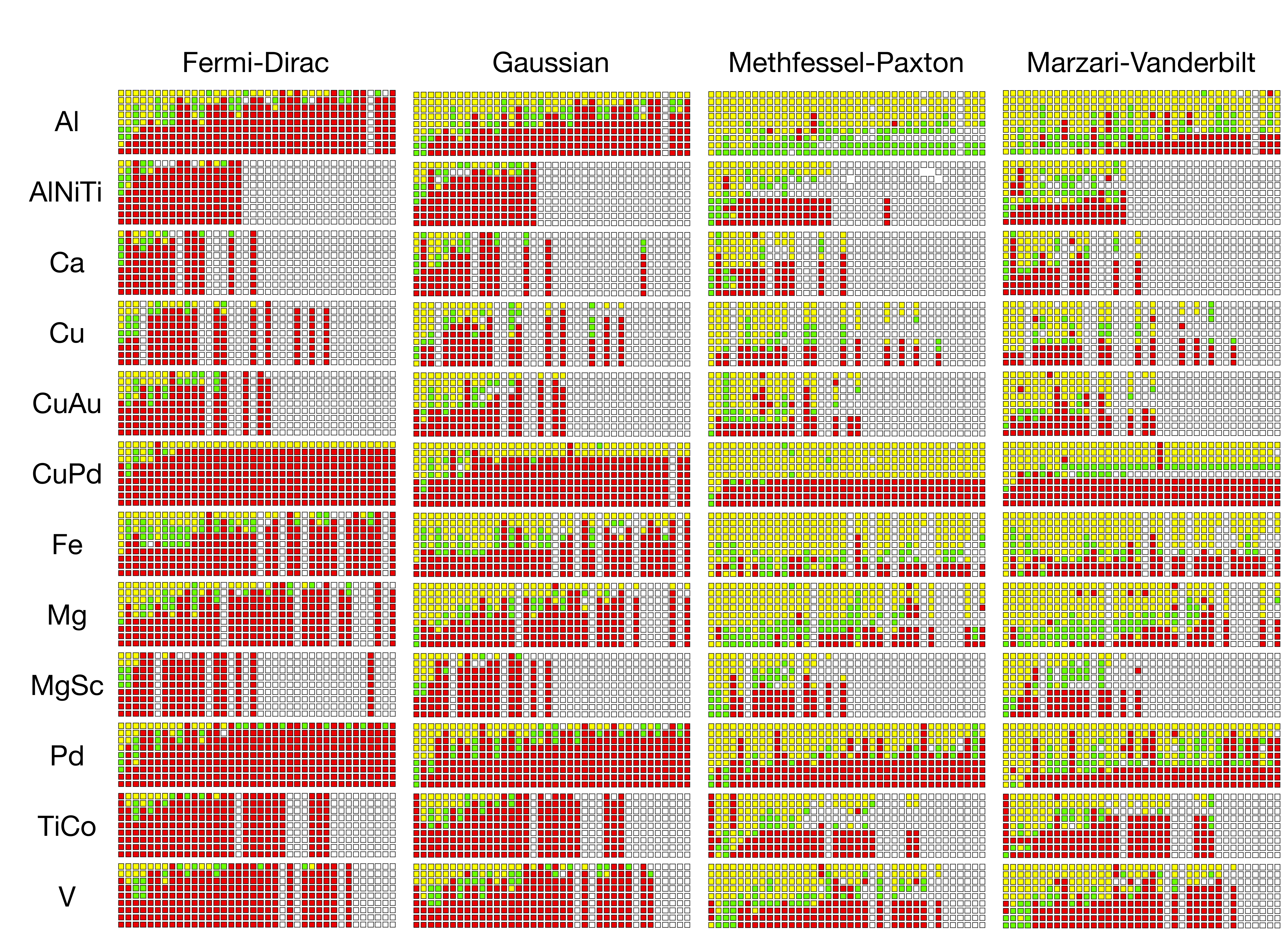}
\caption{Comparison of total energy errors in QE with and without smearing. The smearing parameter gets larger with each row (smearing parameter values in Ry are $1.0\times 10^{-3}$, $2.5\times10^{-3}$, $5.0\times10^{-3}$, $7.5\times10^{-3}$, $1.0\times10^{-2}$, $2.5\times10^{-2}$, $5.0\times10^{-2}$, $7.5\times10^{-2}$, and $1.0\times10^{-1}$) and the \kb-point density gets larger with each column (number of \kb-points is $3^3$, $4^3$, \dots, $40^3$). Smearing most often does not appreciably improve the accuracy of total energies in DFT calculations in QE. The optimal smearing is dependent on the smearing parameter, smearing type, \kb-point density, and metal.}
\end{center}
\end{figure}

\begin{figure}[t]
\begin{center}
\includegraphics[width=\textwidth,height=\textheight,keepaspectratio]{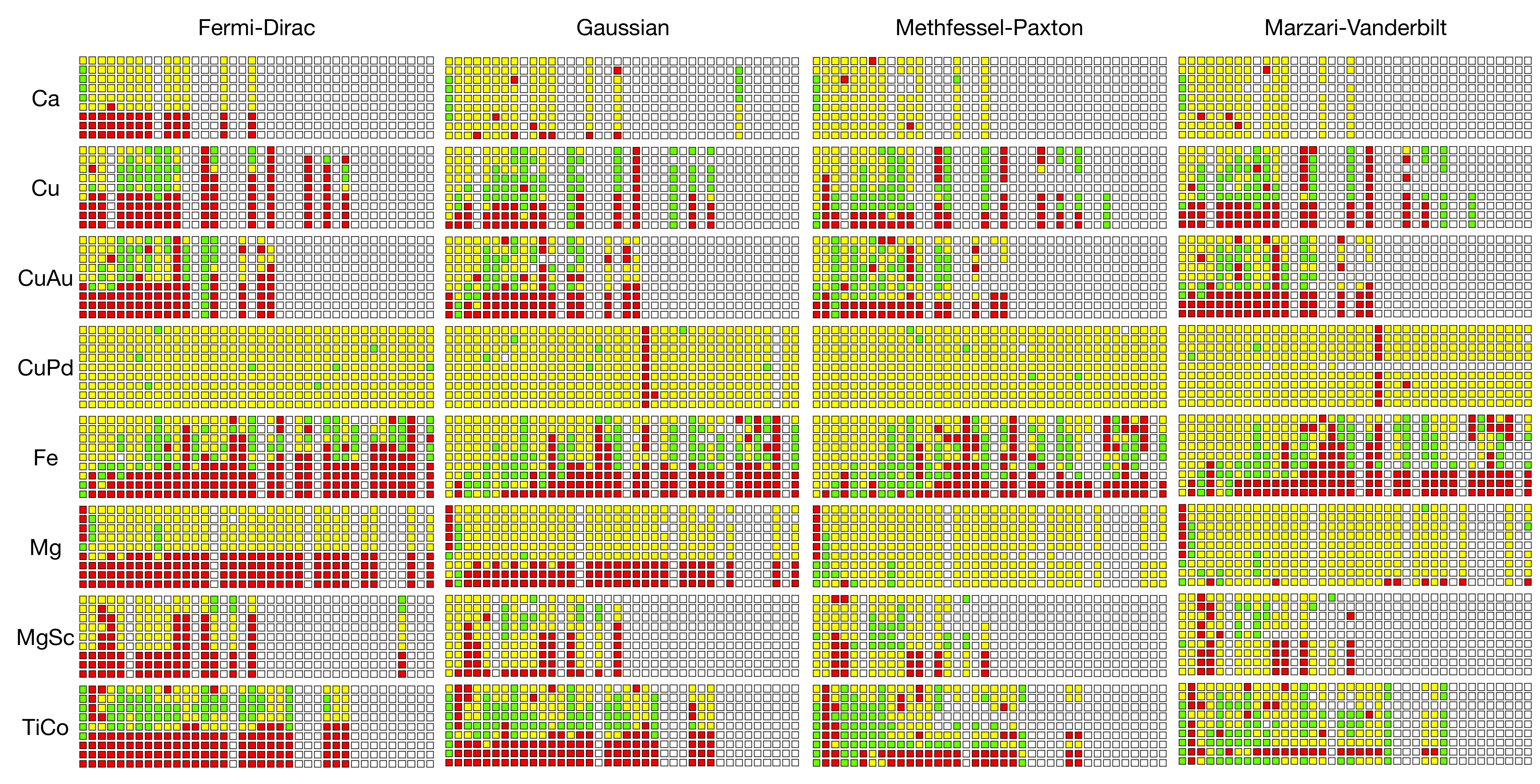}
\caption{Comparison of stress errors in QE with and without smearing. The smearing parameter gets larger with each row (smearing parameter values in Ry are $1.0\times 10^{-3}$, $2.5\times10^{-3}$, $5.0\times10^{-3}$, $7.5\times10^{-3}$, $1.0\times10^{-2}$, $2.5\times10^{-2}$, $5.0\times10^{-2}$, $7.5\times10^{-2}$, and $1.0\times10^{-1}$) and the \kb-point density gets larger with each column (number of \kb-points is $3^3$, $4^3$, \dots, $40^3$). Smearing most often does not appreciably improve the accuracy of stresses in DFT calculations in QE.}
\end{center}
\end{figure}

\begin{figure}[t]
\begin{center}
\includegraphics[width=\textwidth,height=\textheight,keepaspectratio]{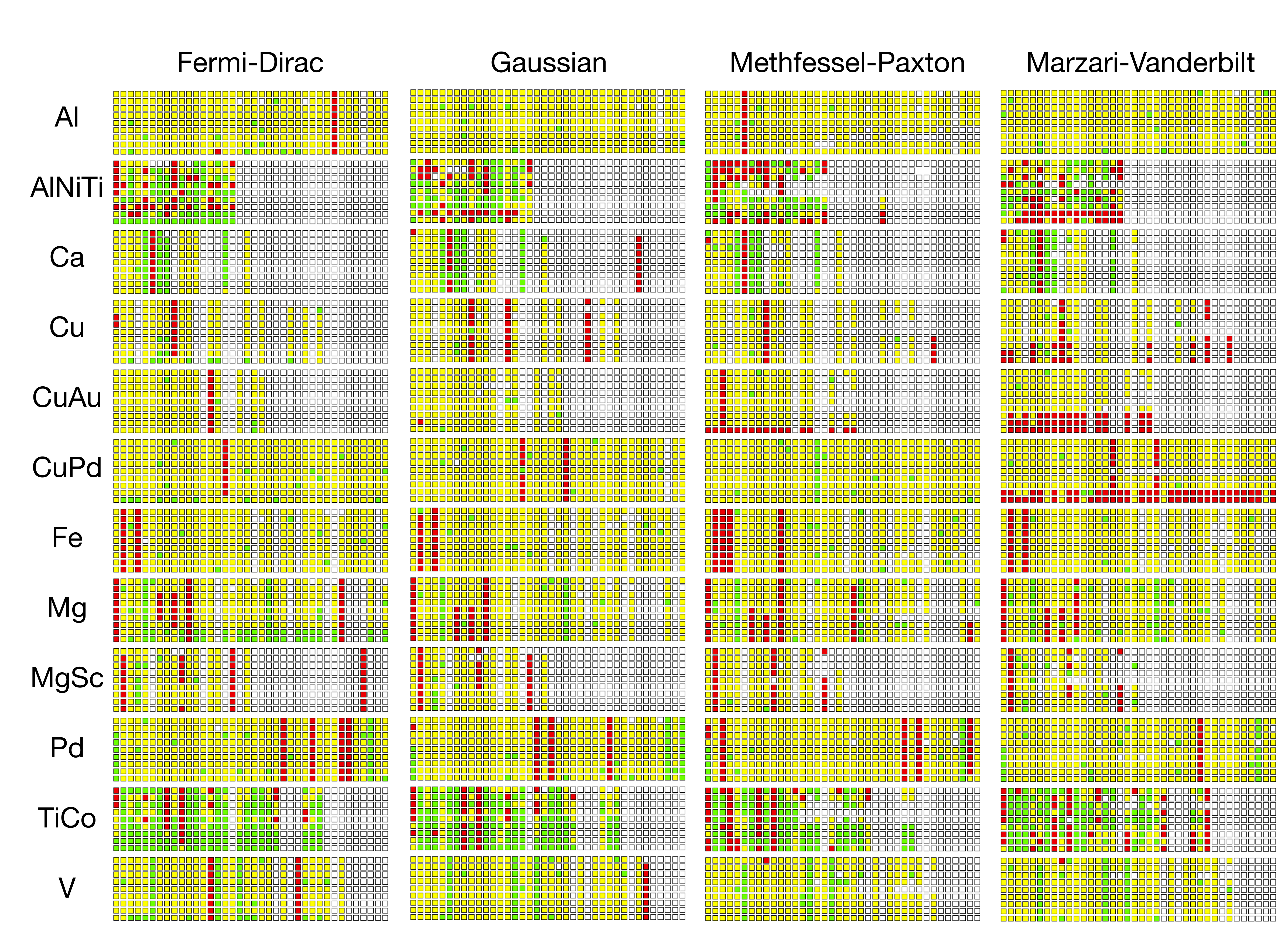}
\caption{Comparison of the number of SCF iterations in QE with and without smearing. The smearing parameter gets larger with each row (smearing parameter values in Ry are $1.0\times 10^{-3}$, $2.5\times10^{-3}$, $5.0\times10^{-3}$, $7.5\times10^{-3}$, $1.0\times10^{-2}$, $2.5\times10^{-2}$, $5.0\times10^{-2}$, $7.5\times10^{-2}$, and $1.0\times10^{-1}$) and the \kb-point density gets larger with each column (number of \kb-points is $3^3$, $4^3$, \dots, $40^3$). Smearing usually reduces the number of SCF cycles but the average reduction is less than 1 iteration.}
\end{center}
\end{figure}

\begin{figure}[t]
\begin{center}
\includegraphics[width=\textwidth,height=\textheight,keepaspectratio]{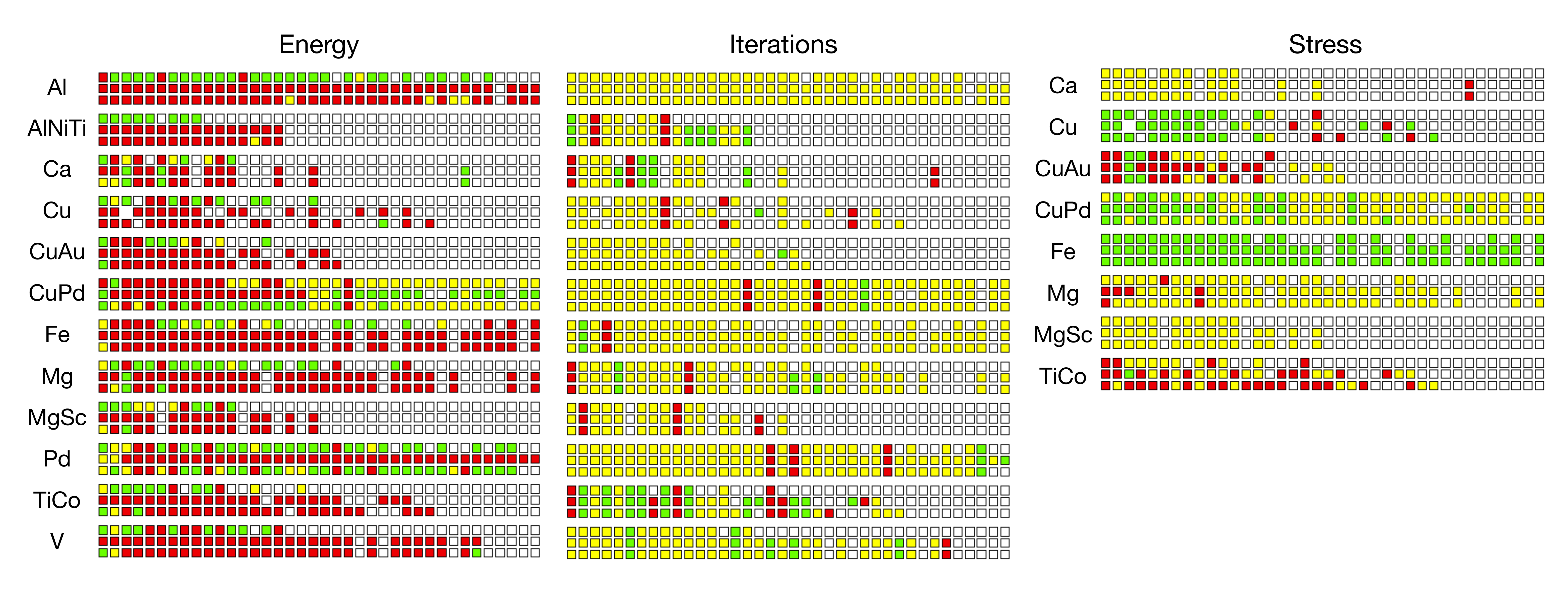}
\caption{The total energies, stresses and number of SCF iterations in QE with tetrahedron methods compared against the same values obtained without smearing or tetrahedra. The smearing parameter for all calculations was $1\times10^{-10}$ Ry. The tetrahedron methods were Bl\"ochl's, linear, and Kawamura's tetrahedra in order of row from top to bottom. The \kb-point density gets larger with each column (number of \kb-points is $3^3$, $4^3$, \dots, $40^3$). Bl\"ochl's tetrahedron method in QE usually improves total energies. All tetrahedron methods most often do not reduce the number of SCF iterations. Tetrahedra improve stress accuracy for certain metals.}
\end{center}
\end{figure}

\begin{figure}[t]
\begin{center}
\includegraphics[width=5in,keepaspectratio]{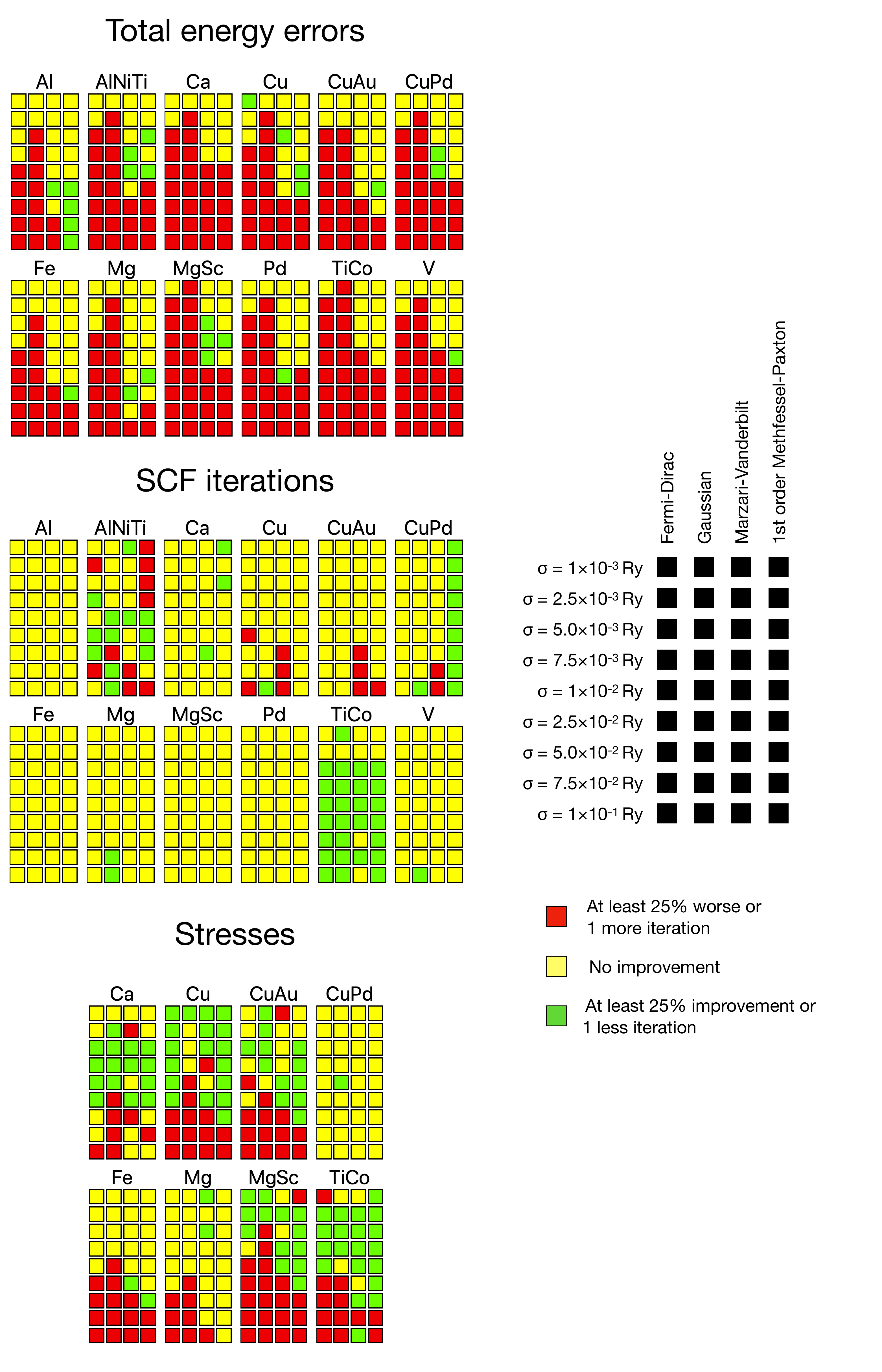}
\caption{QE total energy error, SCF iteration differences, and stress errors with smearing compared to calculations without smearing. Each colored square is an average over \kb-point densities. We excluded calculations where the total energy error was less than 0.1 meV. In each block with a system label, the columns from left to right are smearing types Gaussian, Fermi-Dirac, Marzari-Vanderbilt, and order 1 Methfessel-Paxton. The rows from top to bottom are different smearing parameters starting with 0.001 Ry at the top and increasing to 0.1 Ry at the bottom. For total energy and stress errors, green squares indicate smearing decreases error on average by more than 25\%,  red squares indicate smearing increases error on average by 25\%, and yellows squares are located where errors are between the two. For SCF iterations, green squares are located where smearing reduces the number of iterations on average by at least 1, red squares are where smearing increases the number of iterations on average by at least 1, and yellow squares are located where the iteration difference is in between.}
\end{center}
\end{figure}

\begin{figure}[t]
\begin{center}
\includegraphics[width=\textwidth,height=\textheight,keepaspectratio]{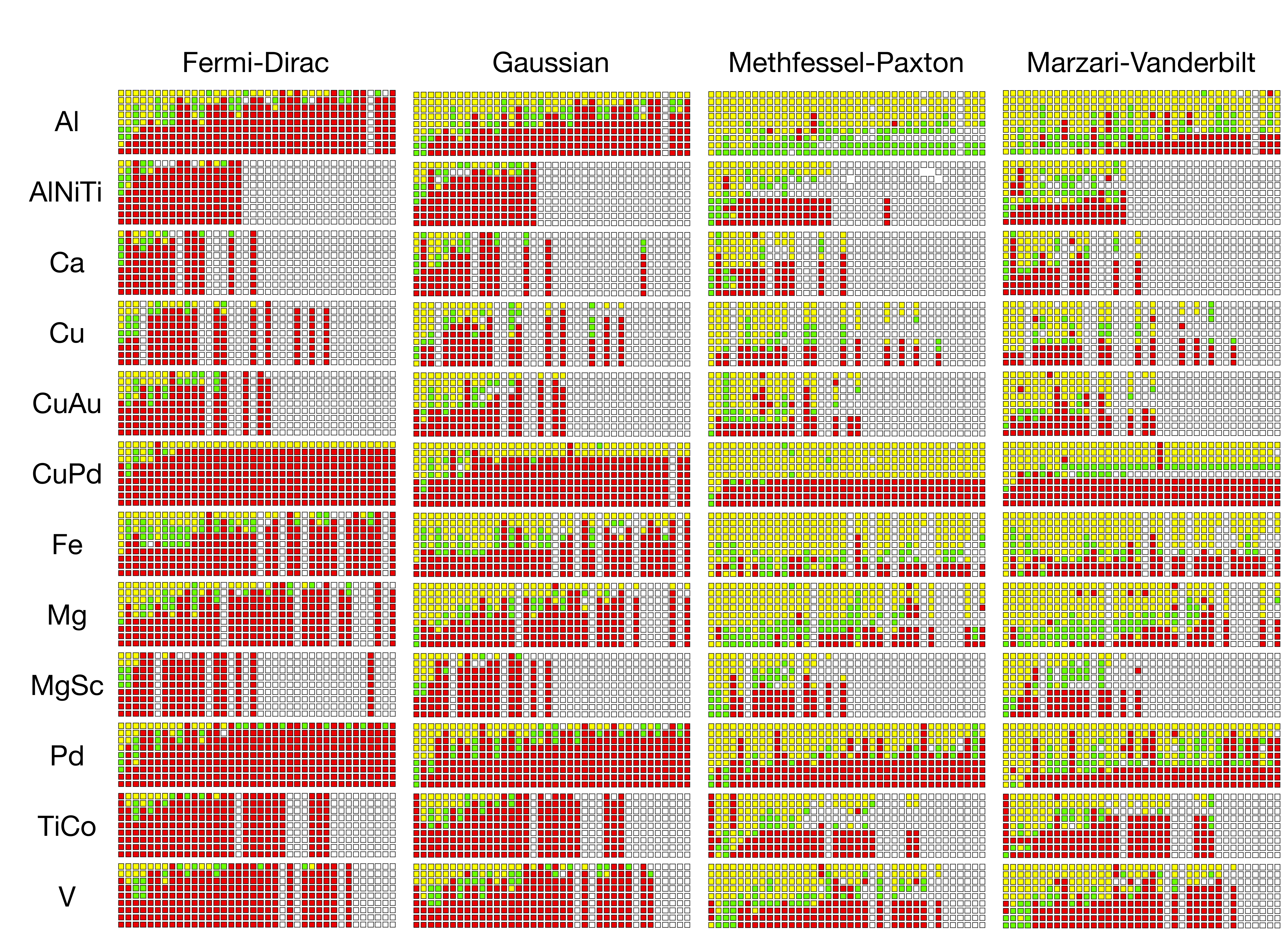}
\caption{Mean and mean deviation of energy ratios, stress ratios, and iteration differences in QE. Tetrahedron methods were only tested with the smallest smearing parameter.}
\end{center}
\end{figure}

\FloatBarrier
\subsection{Smearing tests in Quantum Espresso}

\begin{figure}[H]
\includegraphics[width=\textwidth,height=\textheight,keepaspectratio]{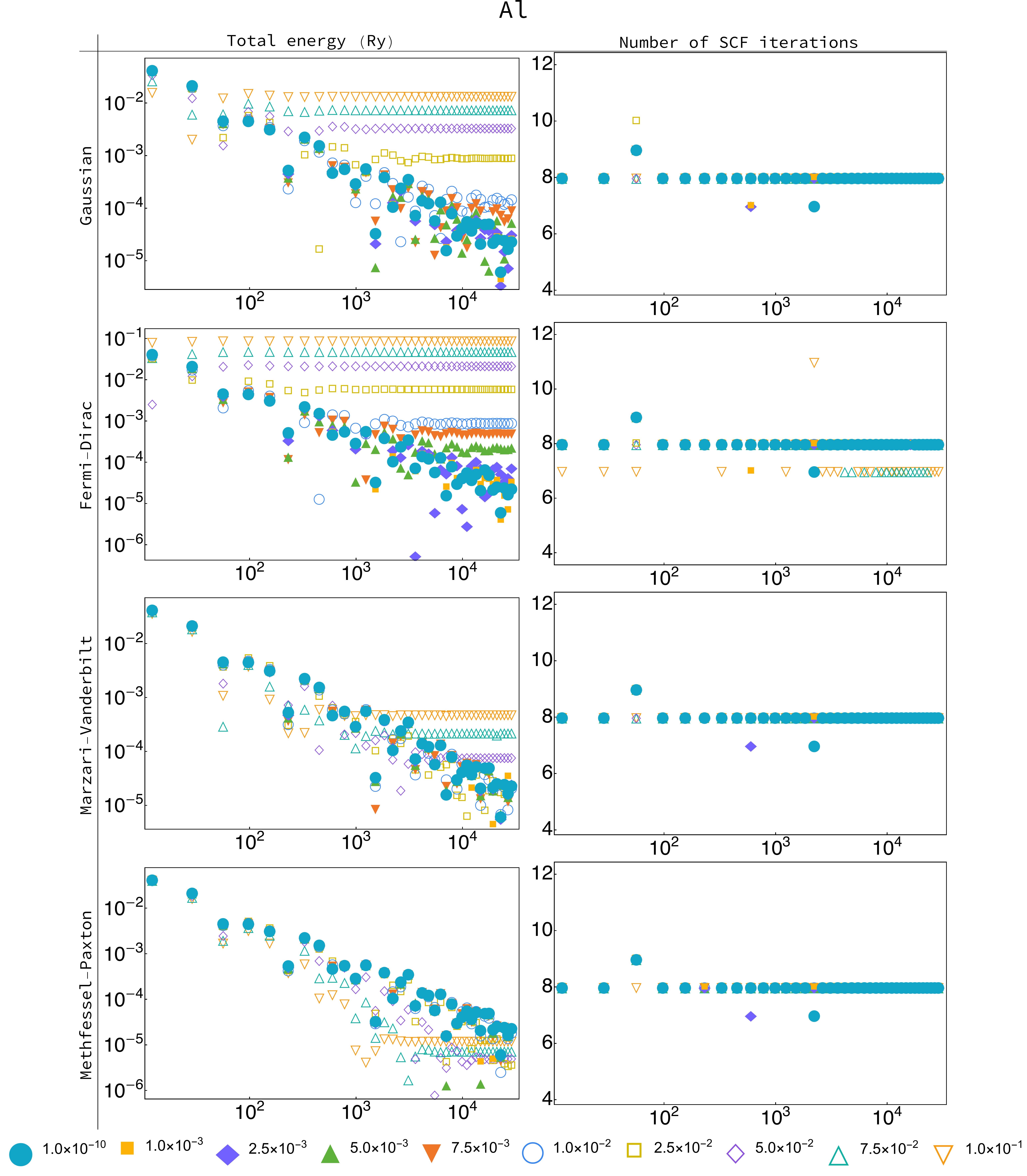}
\caption{The total energy convergence and number of SCF cycles Al in Quantum Espresso as a function of \kb-point density. For all plots, the $x$-axis is the reduced \kb-point density in units of cubic Bohrs. The legend at the bottom gives the amount of smearing in Rydbergs.}
\label{fig:qe_Al-smooth}
\end{figure}

\begin{figure}[h]
\includegraphics[width=6in]{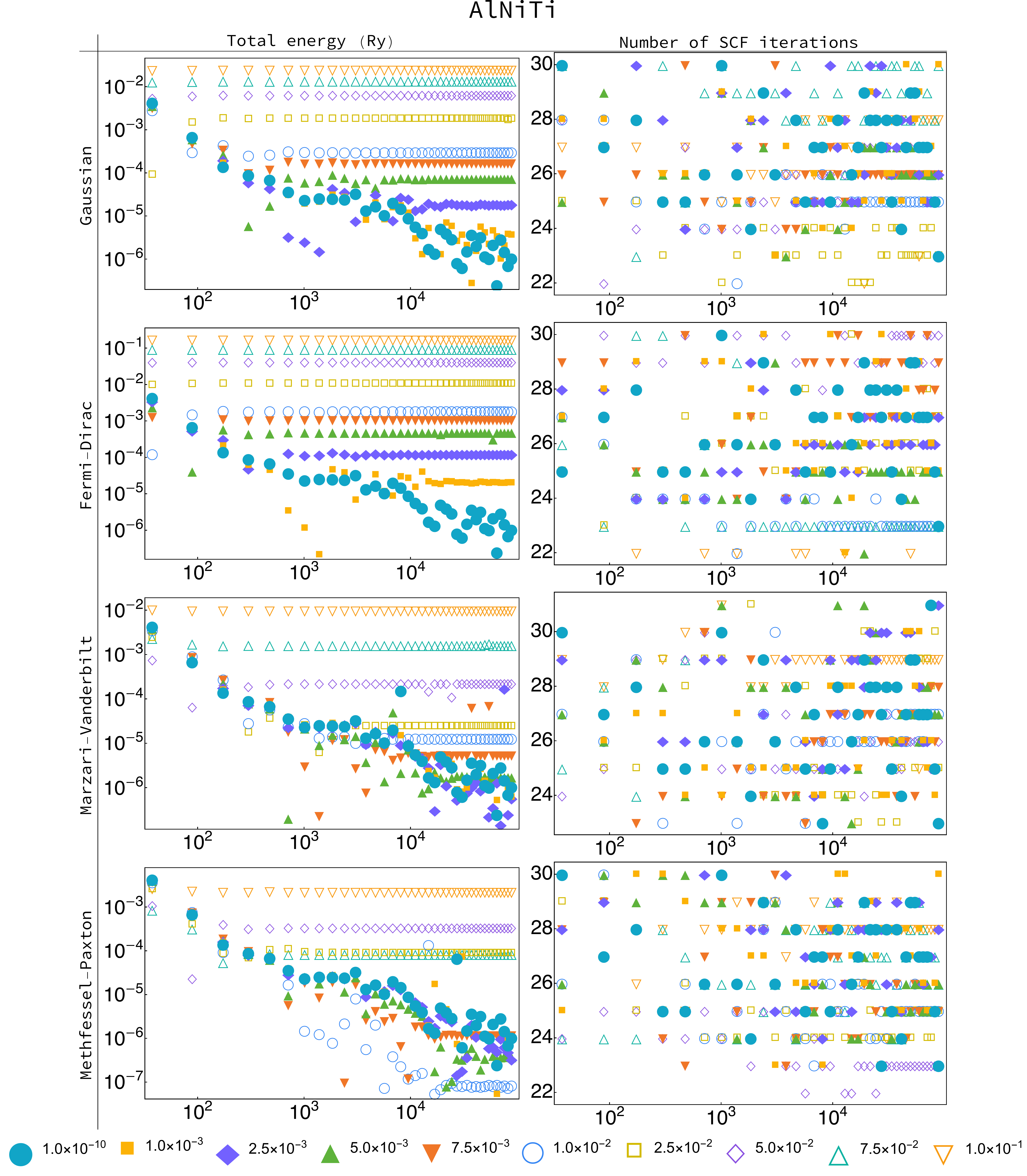}
\caption{The total energy convergence and number of SCF cycles for AlNiTi in Quantum Espresso as a function of \kb-point density. For all plots, the $x$-axis is the reduced \kb-point density in units of cubic Bohrs. The legend at the bottom gives the amount of smearing in Rydbergs.}
\label{fig:qe_AlNiTi-smooth}
\end{figure}

\begin{figure}[h]
\includegraphics[width=\textwidth]{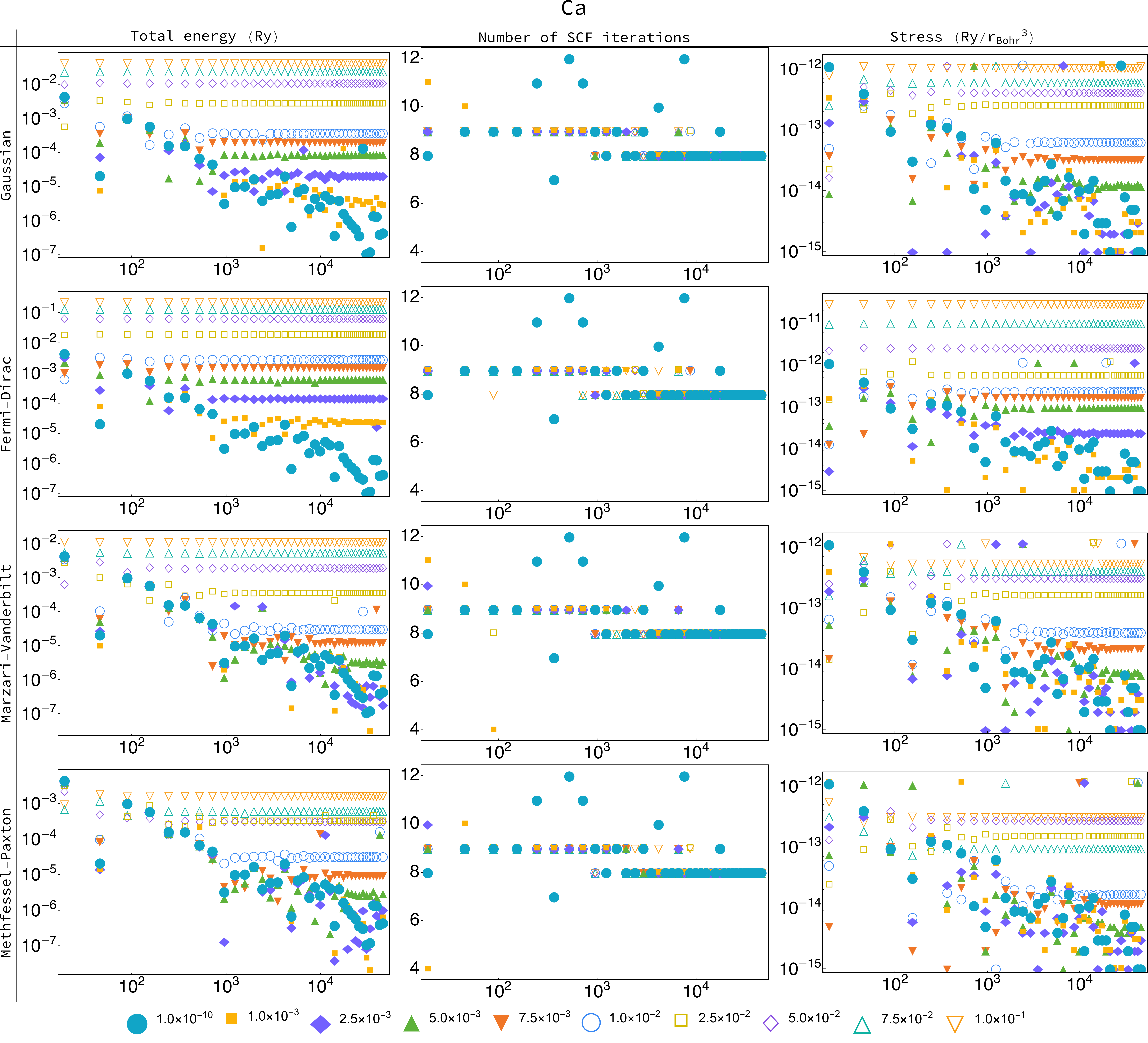}
\caption{The total energy convergence, number of SCF cycles, and stress convergence for Ca in Quantum Espresso. For all plots, the $x$-axis is the reduced \kb-point density in units of cubic Bohrs. The legend at the bottom gives the amount of smearing in Rydbergs.}
\label{fig:qe_Ca-smooth}
\end{figure}

\begin{figure}[p]
\includegraphics[width=\textwidth,height=\textheight,keepaspectratio]{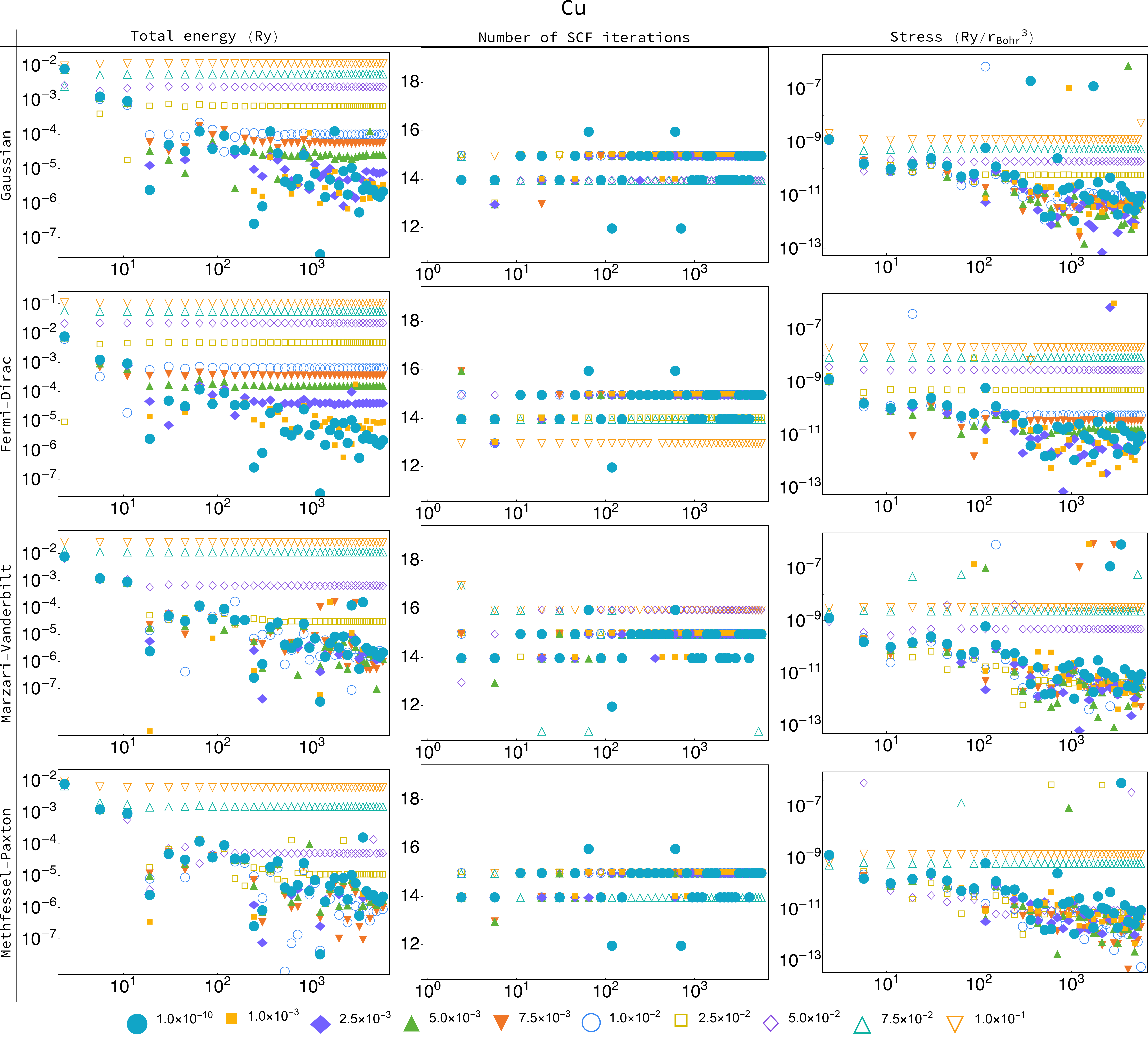}
\caption{The total energy convergence, number of SCF cycles, and stress convergence for Cu in Quantum Espresso. For all plots, the $x$-axis is the reduced \kb-point density in units of cubic Bohrs. The legend at the bottom gives the amount of smearing in Rydbergs.}
\label{fig:qe_Cu-smooth}
\end{figure}
\FloatBarrier

\begin{figure}[p]
\includegraphics[width=\textwidth,height=\textheight,keepaspectratio]{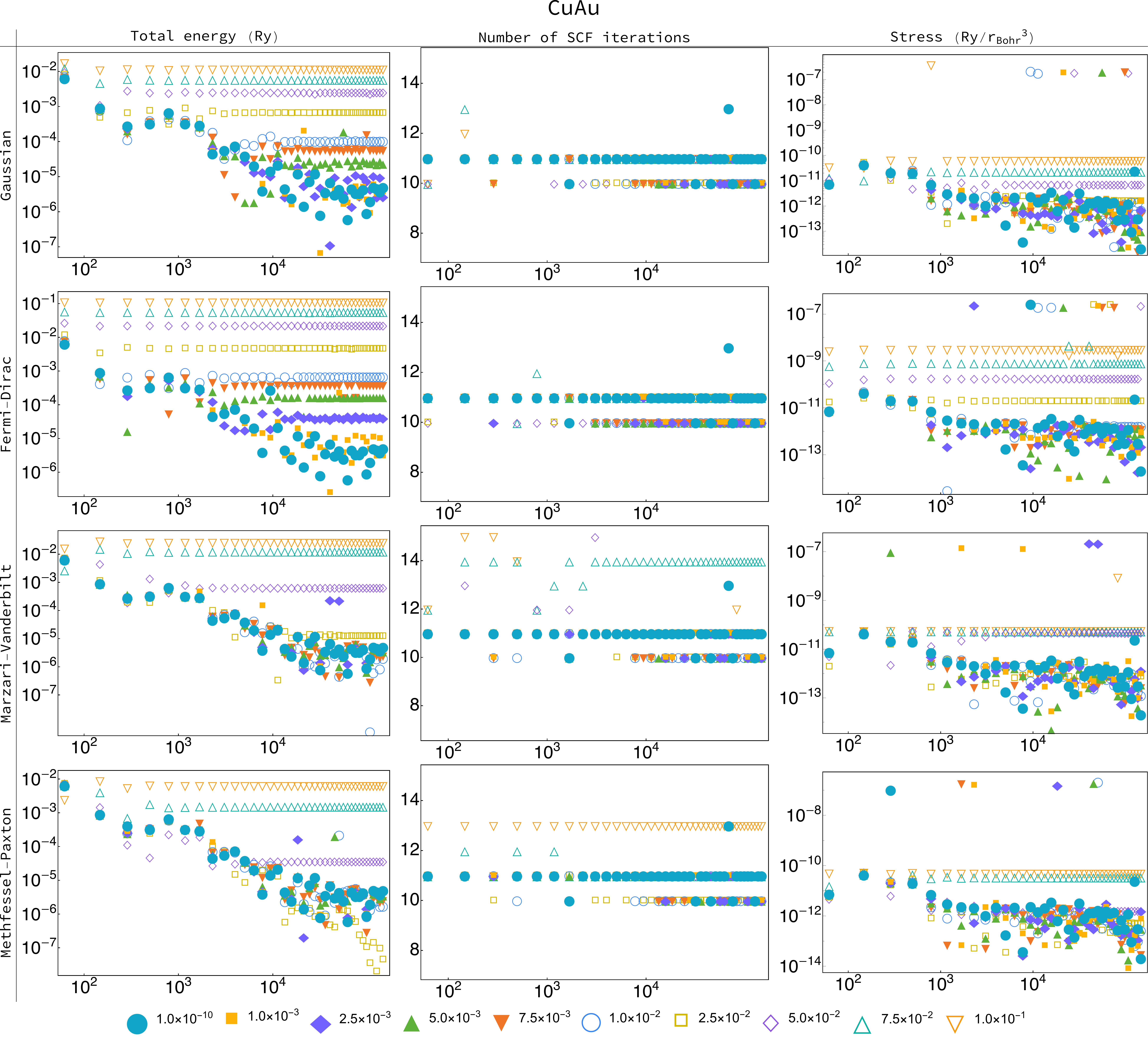}
\caption{The total energy convergence, number of SCF cycles, and stress convergence for CuAu in Quantum Espresso. For all plots, the $x$-axis is the reduced \kb-point density in units of cubic Bohrs. The legend at the bottom gives the amount of smearing in Rydbergs.}
\label{fig:qe_CuAu-smooth}
\end{figure}
\FloatBarrier

\begin{figure}[p]
\includegraphics[width=\textwidth,height=\textheight,keepaspectratio]{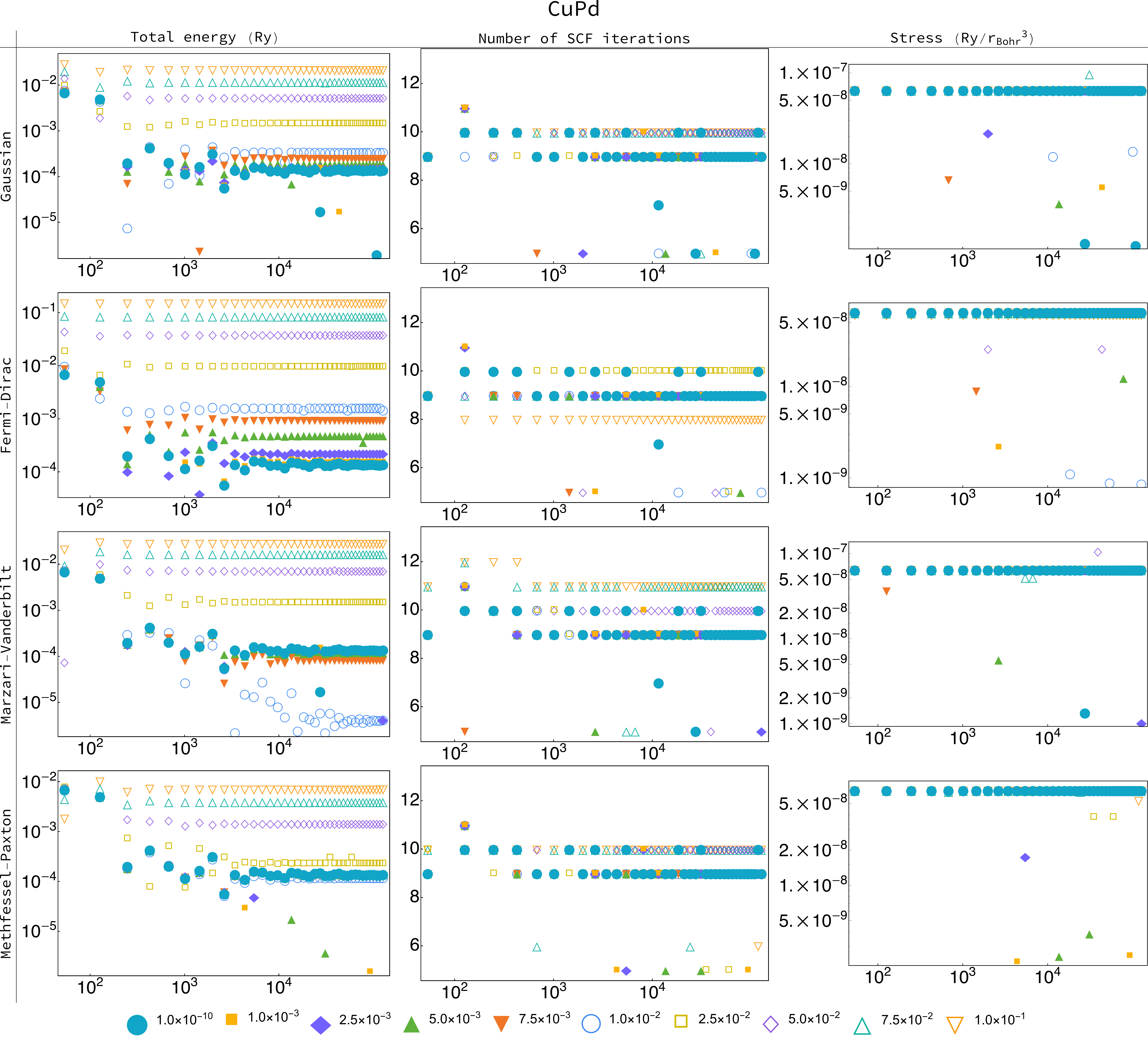}
\caption{The total energy convergence, number of SCF cycles, and stress convergence for CuPd in Quantum Espresso. For all plots, the $x$-axis is the reduced \kb-point density in units of cubic Bohrs. The legend at the bottom gives the amount of smearing in Rydbergs.}
\label{fig:qe_CuPd-smooth}
\end{figure}
\FloatBarrier

\begin{figure}[p]
\includegraphics[width=\textwidth,height=\textheight,keepaspectratio]{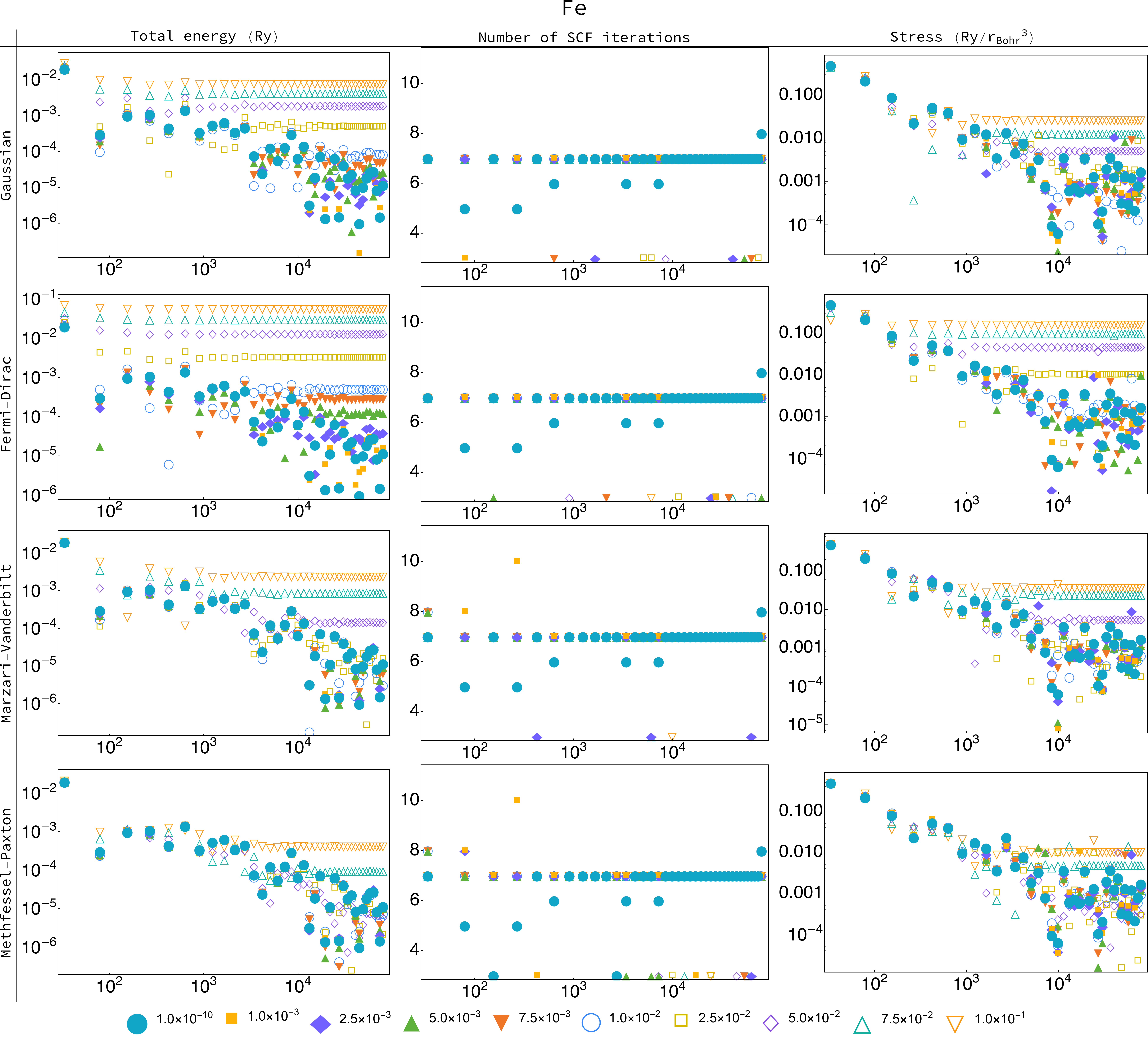}
\caption{The total energy convergence, number of SCF cycles, and stress convergence for Fe in Quantum Espresso. For all plots, the $x$-axis is the reduced \kb-point density in units of cubic Bohrs. The legend at the bottom gives the amount of smearing in Rydbergs.}
\label{fig:qe_Fe-smooth}
\end{figure}
\FloatBarrier

\begin{figure}[p]
\includegraphics[width=\textwidth,height=\textheight,keepaspectratio]{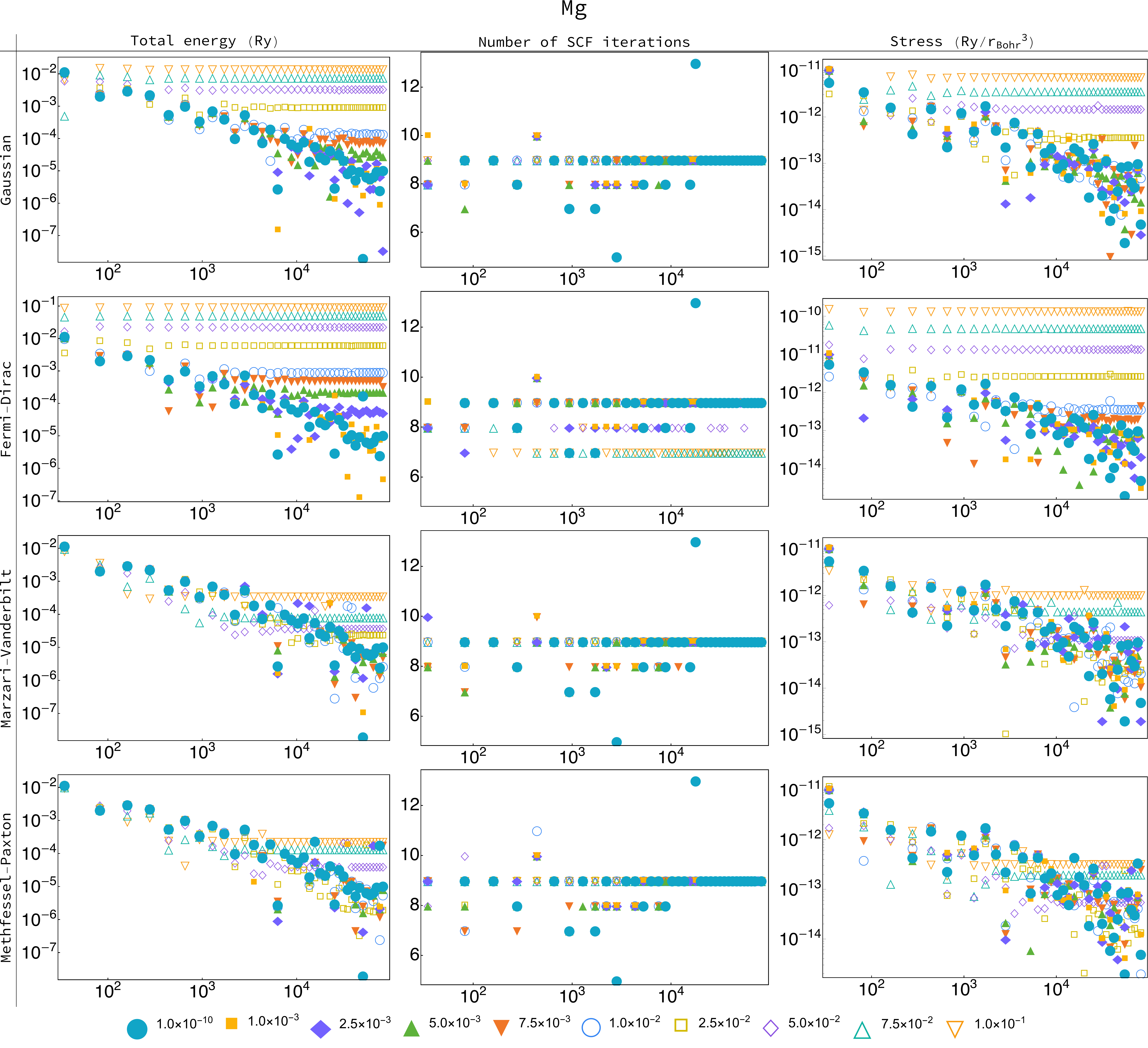}
\caption{The total energy convergence, number of SCF cycles, and stress convergence for Mg in Quantum Espresso. For all plots, the $x$-axis is the reduced \kb-point density in units of cubic Bohrs. The legend at the bottom gives the amount of smearing in Rydbergs.}
\label{fig:qe_Mg-smooth}
\end{figure}
\FloatBarrier

\begin{figure}[p]
\includegraphics[width=\textwidth,height=\textheight,keepaspectratio]{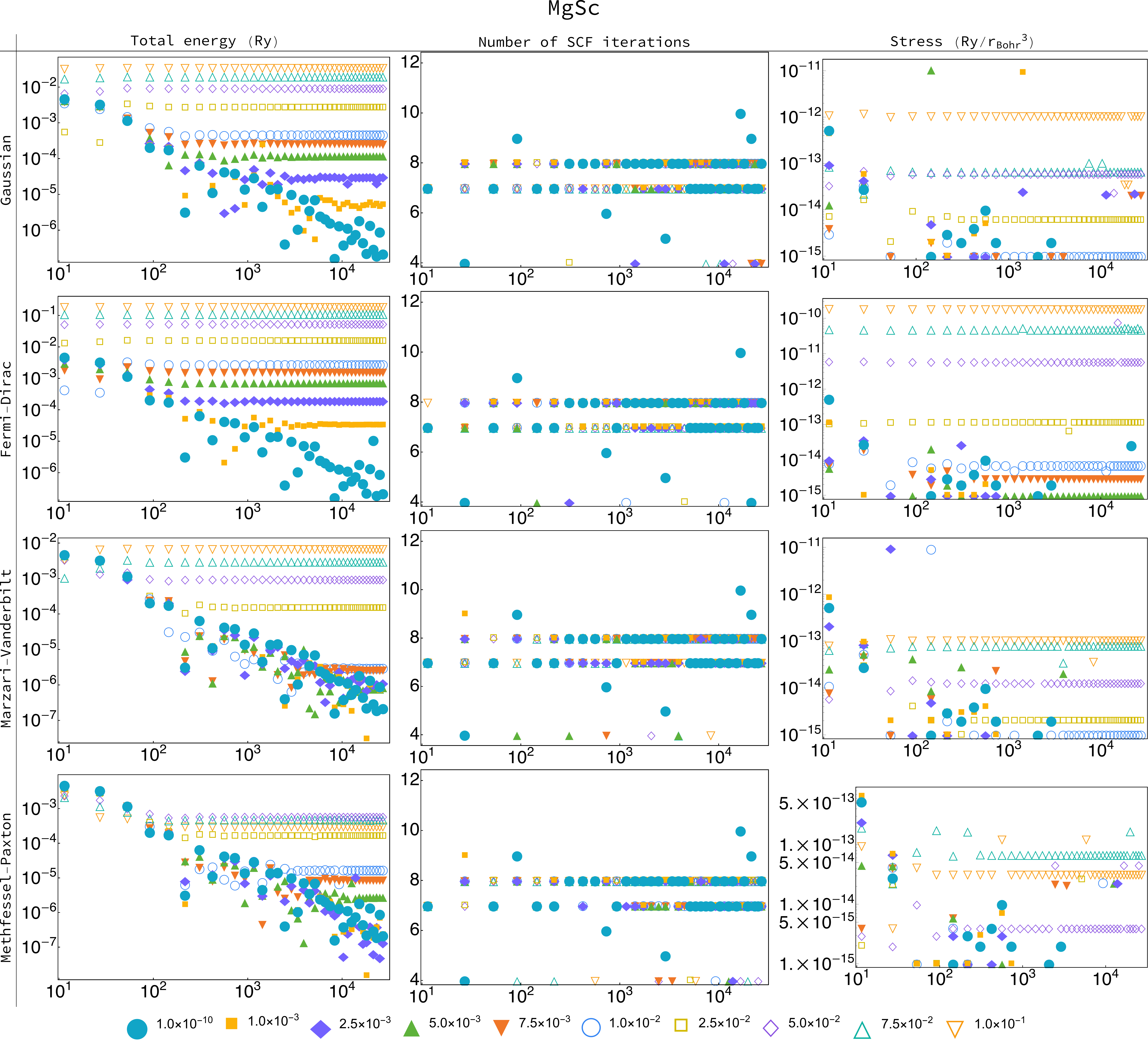}
\caption{The total energy convergence, number of SCF cycles, and stress convergence for MgSc in Quantum Espresso. For all plots, the $x$-axis is the reduced \kb-point density in units of cubic Bohrs. The legend at the bottom gives the amount of smearing in Rydbergs.}
\label{fig:qe_MgSc-smooth}
\end{figure}
\FloatBarrier

\begin{figure}[p]
\includegraphics[width=6in,height=\textheight,keepaspectratio]{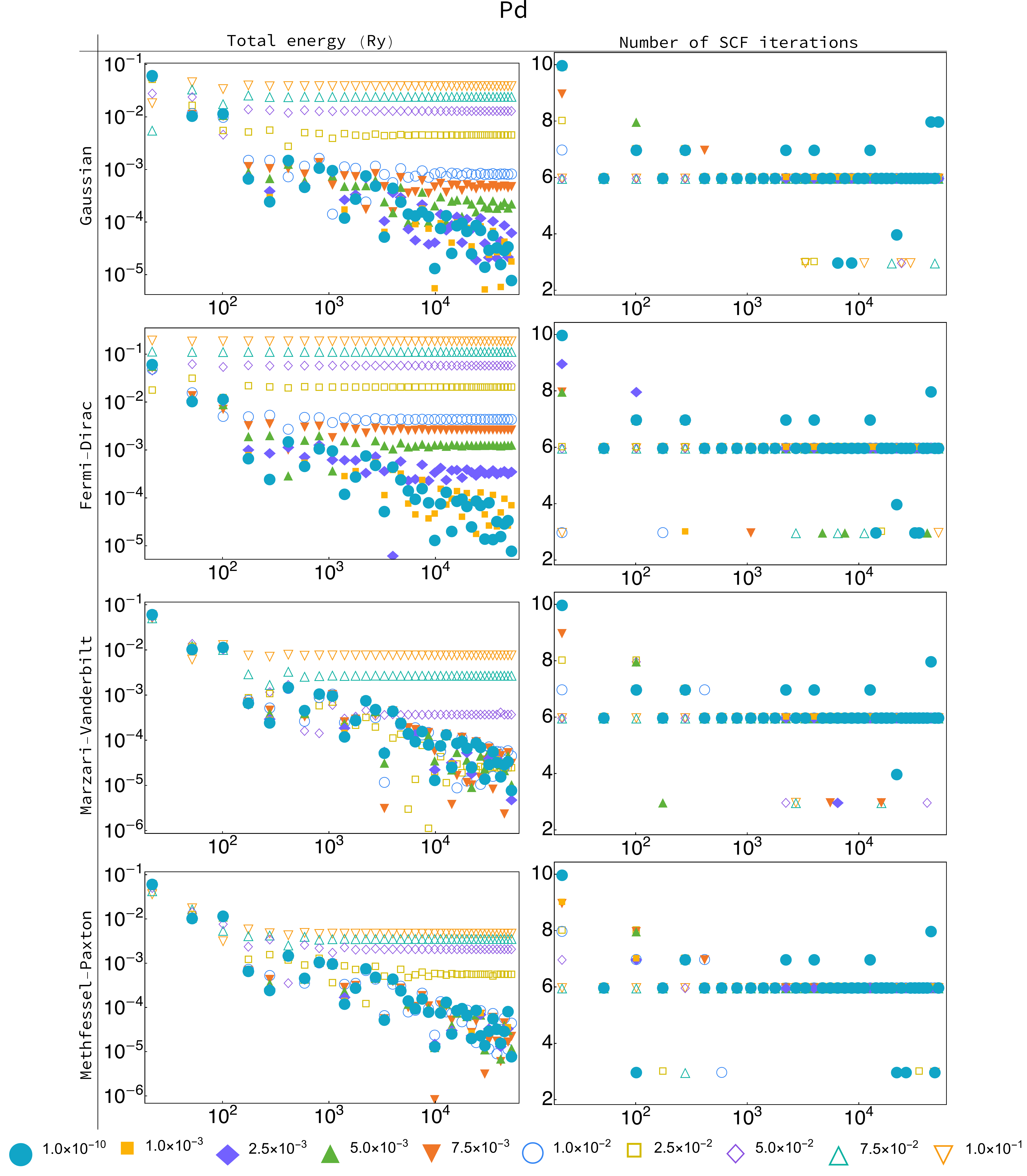}
\caption{The total energy convergence and number of SCF cycles for Pd in Quantum Espresso as a function of \kb-point density. For all plots, the $x$-axis is the reduced \kb-point density in units of cubic Bohrs. The legend at the bottom gives the amount of smearing in Rydbergs.}
\label{fig:qe_Pd-smooth}
\end{figure}
\FloatBarrier

\begin{figure}[p]
\includegraphics[width=6in,height=\textheight,keepaspectratio]{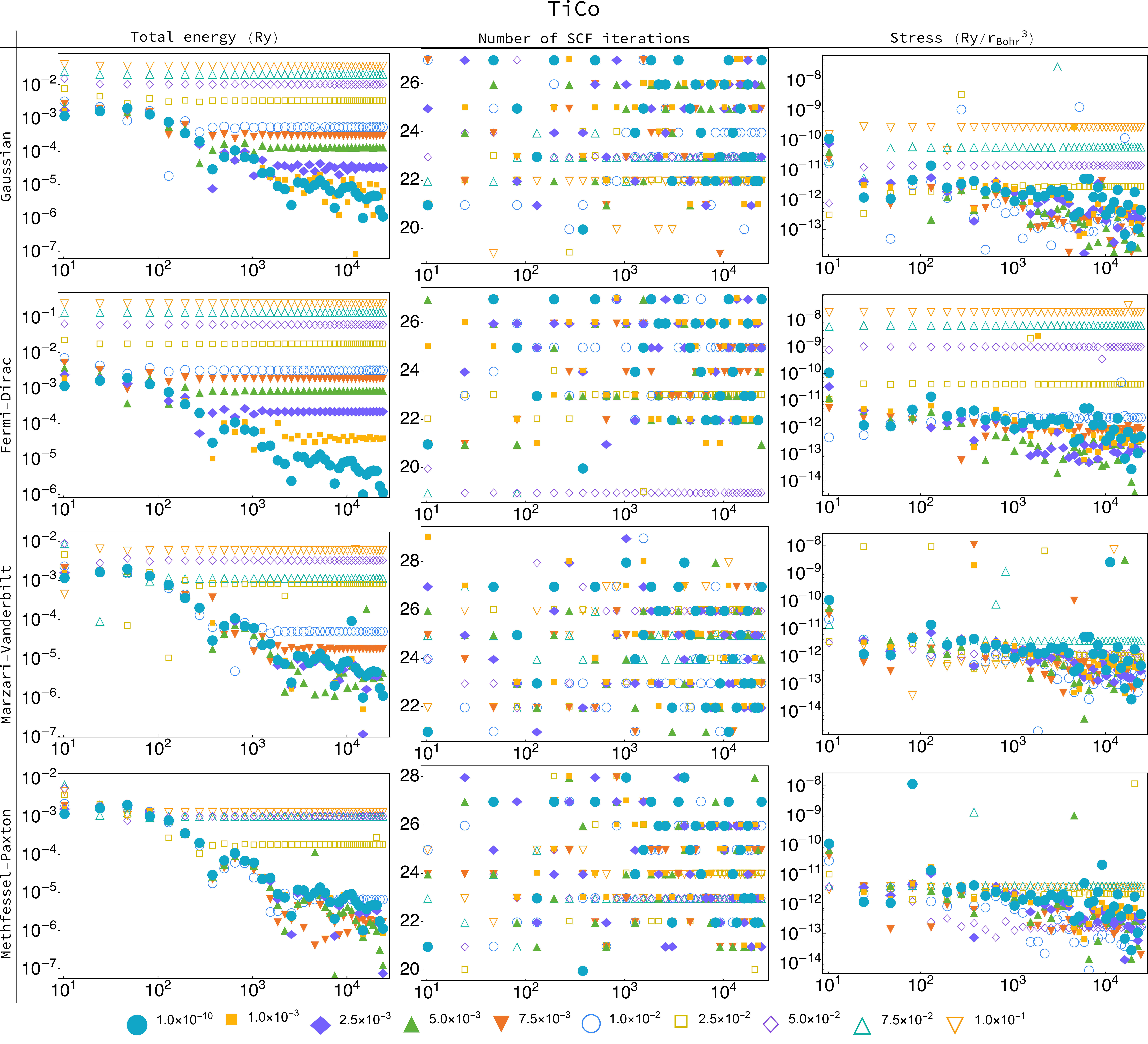}
\caption{The total energy convergence and number of SCF cycles for TiCo in Quantum Espresso as a function of \kb-point density. For all plots, the $x$-axis is the reduced \kb-point density in units of cubic Bohrs. The legend at the bottom gives the amount of smearing in Rydbergs.}
\label{fig:qe_TiCo-smooth}
\end{figure}
\FloatBarrier

\begin{figure}[p]
\includegraphics[width=6in,height=\textheight,keepaspectratio]{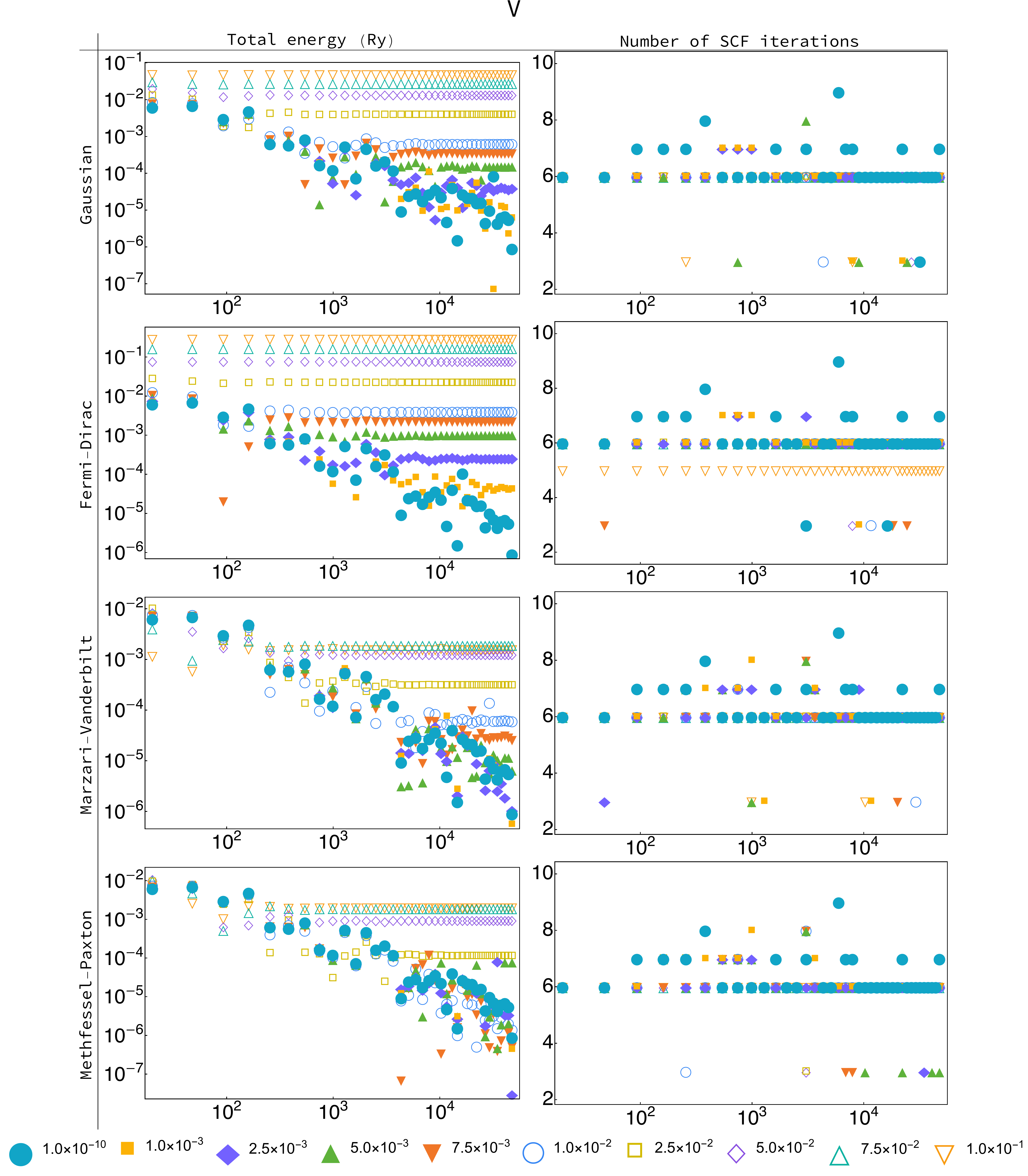}
\caption{The total energy convergence number of SCF cycles for V in Quantum Espresso. For all plots, the $x$-axis is the reduced \kb-point density in units of cubic Bohrs. The legend at the bottom gives the amount of smearing in Rydbergs.}
\label{fig:qe_V-smooth}
\end{figure}

\FloatBarrier
\subsection{Tetrahedra tests in Quantum Espresso}

\begin{figure}[H]
\begin{center}
\includegraphics[width=\textwidth,height=\textheight,keepaspectratio]{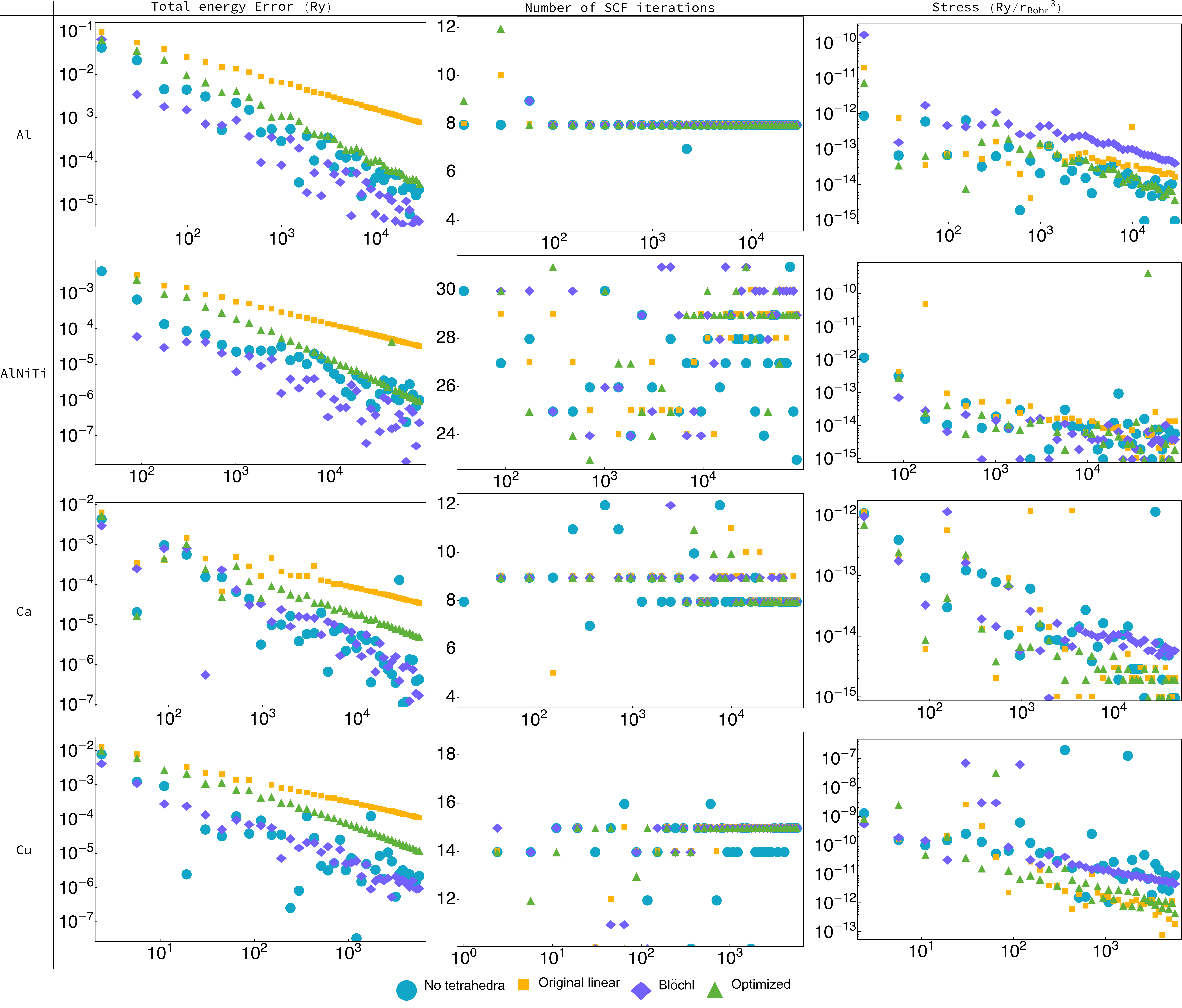}
\caption{The total energy convergence, number of SCF cycles, and stress convergence of Al, AlNiTi, Ca, and Cu with tetrahedron methods in Quantum Espresso. For all plots, the $x$-axis is the reduced \kb-point density in units of cubic Bohrs.}
\end{center}
\end{figure}

\begin{figure}[h]
\includegraphics[width=\textwidth,height=\textheight,keepaspectratio]{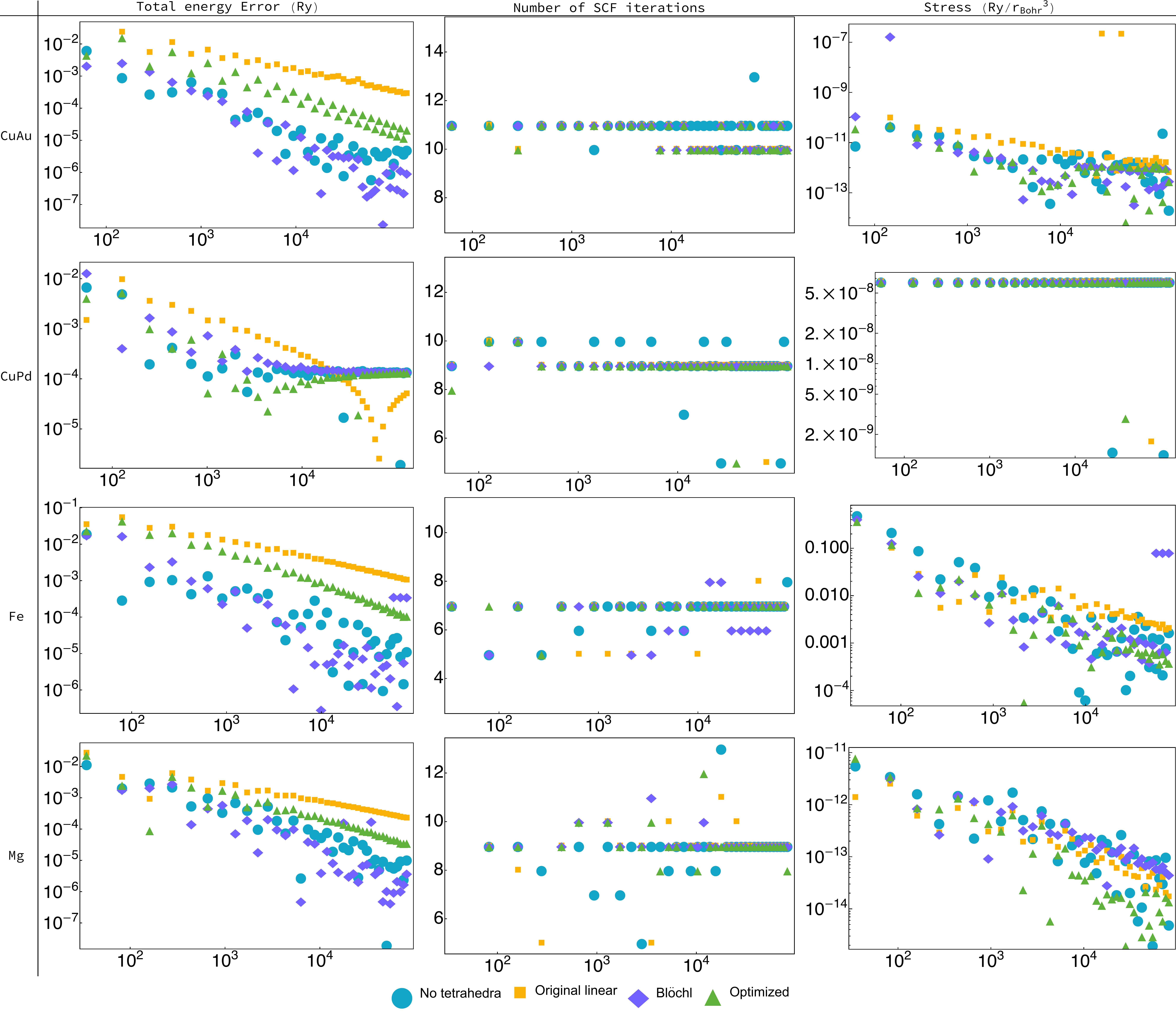}
\caption{The total energy convergence, number of SCF cycles, and stress convergence of CuAu, CuPd, Fe, and Mg with tetrahedron methods in Quantum Espresso. For all plots, the $x$-axis is the reduced \kb-point density in units of cubic Bohrs.}
\label{fig:qe_CuPd-Fe-K-Mg-comb-tet}
\end{figure}
\FloatBarrier

\begin{figure}[p]
\includegraphics[width=\textwidth,height=\textheight,keepaspectratio]{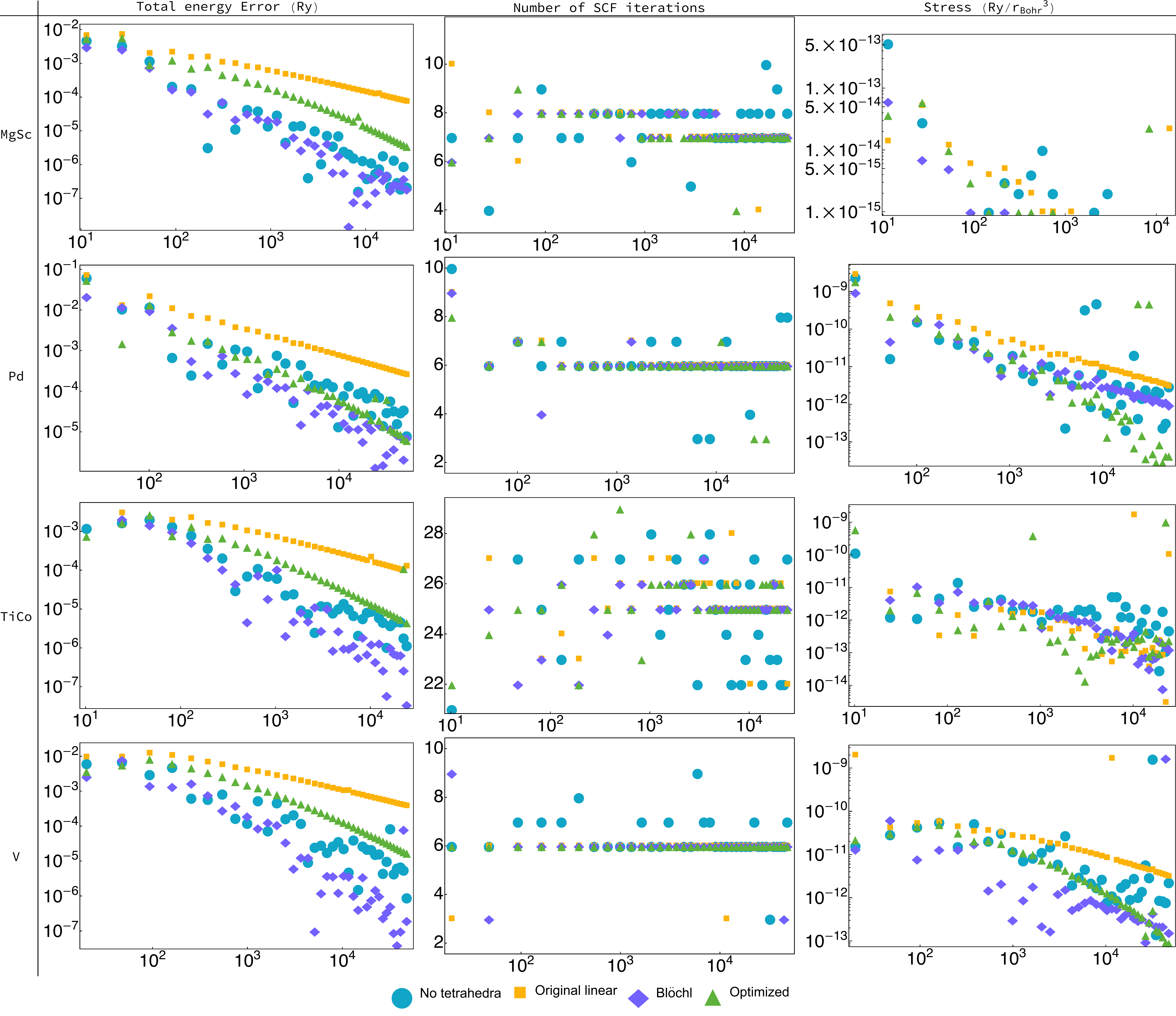}
\caption{The total energy convergence, number of SCF cycles, and stress convergence of MgSc, Pd, TiCo, and V with tetrahedron methods in Quantum Espresso. For all plots, the $x$-axis is the reduced \kb-point density in units of cubic Bohrs.}
\label{fig:qe_MgSc-Pd-TiCo-V-comb-tet}
\end{figure}

\FloatBarrier
\subsection{Energy component tests in Quantum Espresso}

\begin{figure}[H]
\includegraphics[width=\textwidth,height=\textheight,keepaspectratio]{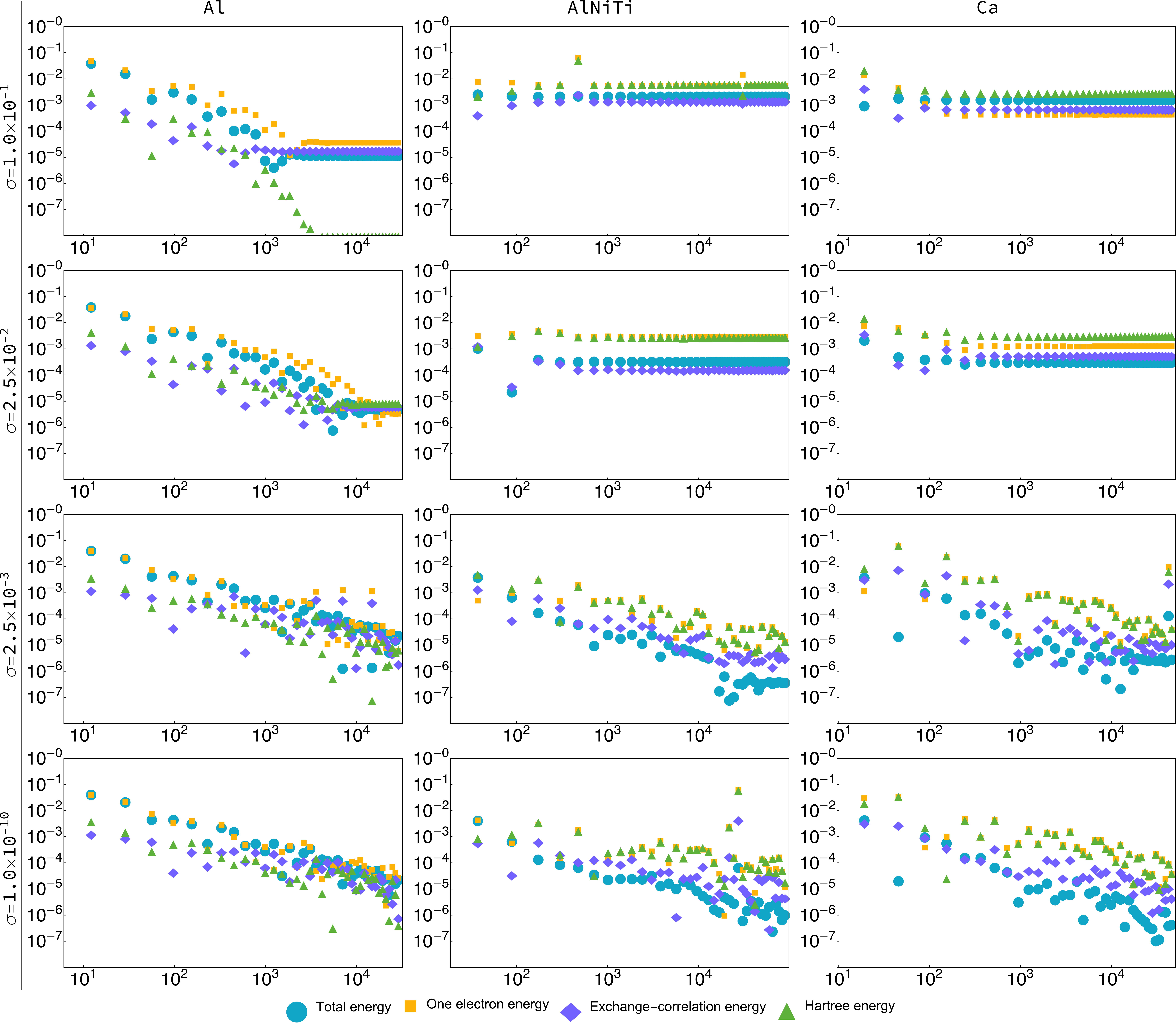}
\caption{The convergence of components of the total energy in Quantum Espresso for the metals Al, AlNiTi, and Ca with Methfessel-Paxton smearing. The Ewald contribution to the total energy is left out due to its lack of dependence on the amount smearing or the \kb-point density. For all plots, the $x$-axis is the reduced \kb-point density in units of cubic Bohrs.}
\label{fig:qe_Al-AlNiTi-Ca-methfessel-paxton}
\end{figure}
\FloatBarrier

\begin{figure}[h]
\includegraphics[width=\textwidth,height=\textheight,keepaspectratio]{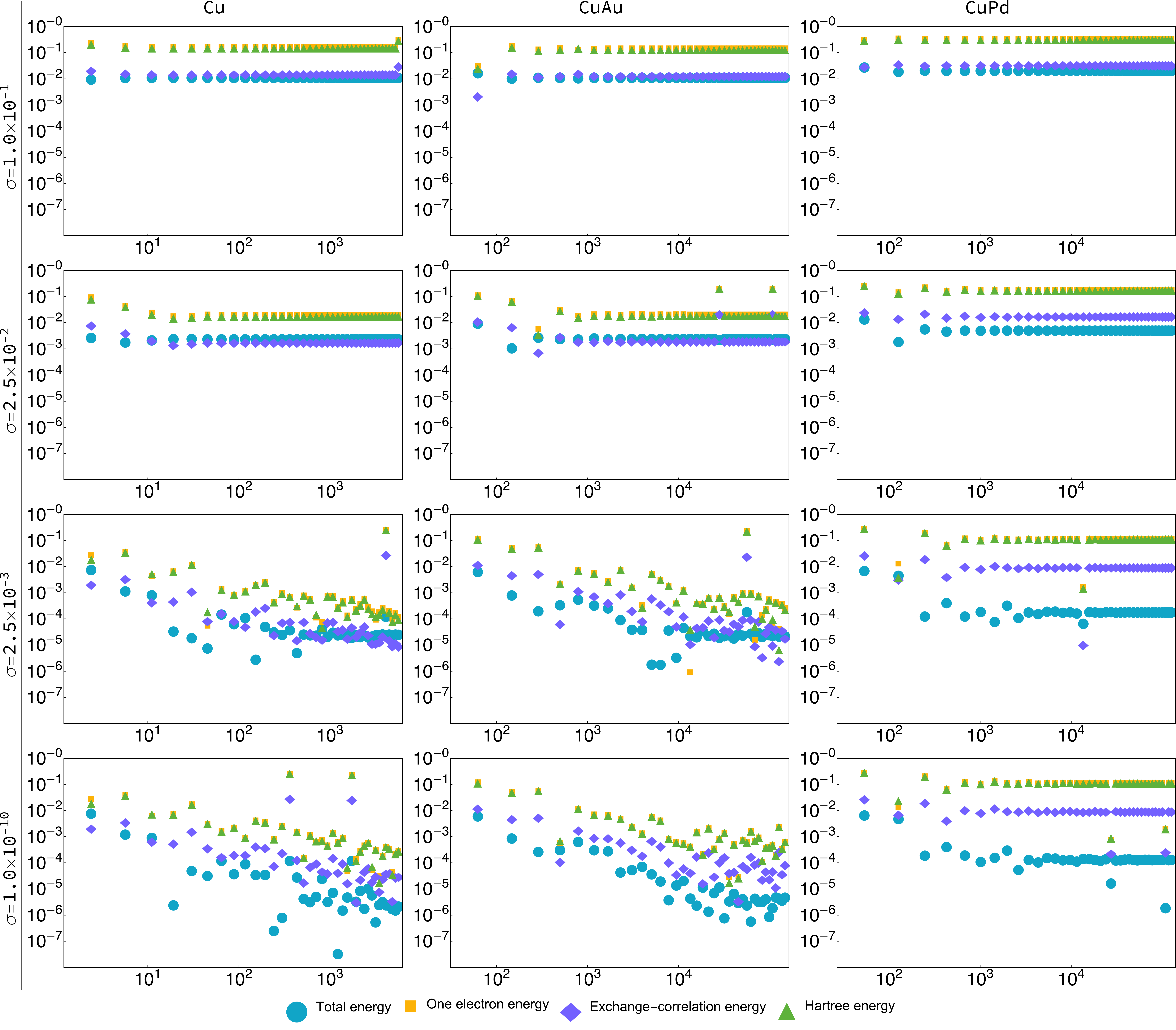}
\caption{The convergence of components of the total energy in Quantum Espresso for the metals Cu, CuAu, and CuPd with Gaussian smearing. The Ewald contribution to the total energy is left out due to its lack of dependence on the amount smearing or the \kb-point density. For all plots, the $x$-axis is the reduced \kb-point density in units of cubic Bohrs.}
\label{fig:qe_Cu-CuAu-CuPd-gaussian}
\end{figure}
\FloatBarrier

\begin{figure}[p]
\includegraphics[width=\textwidth,height=\textheight,keepaspectratio]{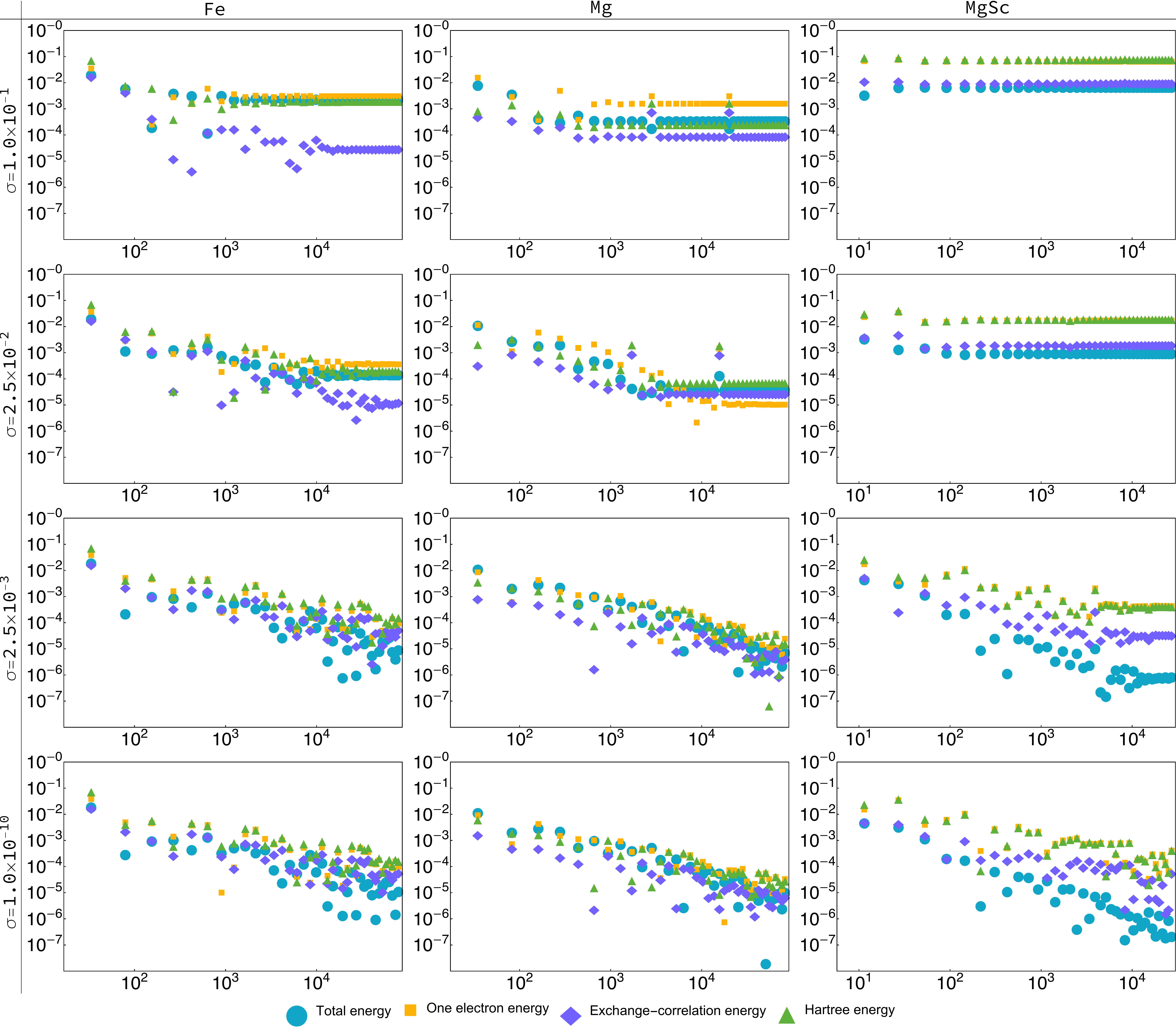}
\caption{The convergence of components of the total energy in Quantum Espresso for the metals Fe, Mg, and MgSc with Marzari-Vanderbilt smearing. The Ewald contribution to the total energy is left out due to its lack of dependence on the amount smearing or the \kb-point density. For all plots, the $x$-axis is the reduced \kb-point density in units of cubic Bohrs.}
\label{fig:qe_Fe-Mg-MgSc-marzari-vanderbilt}
\end{figure}
\FloatBarrier

\newpage
\begin{figure}[h]
\begin{center}
\includegraphics[width=\textwidth,height=\textheight,keepaspectratio]{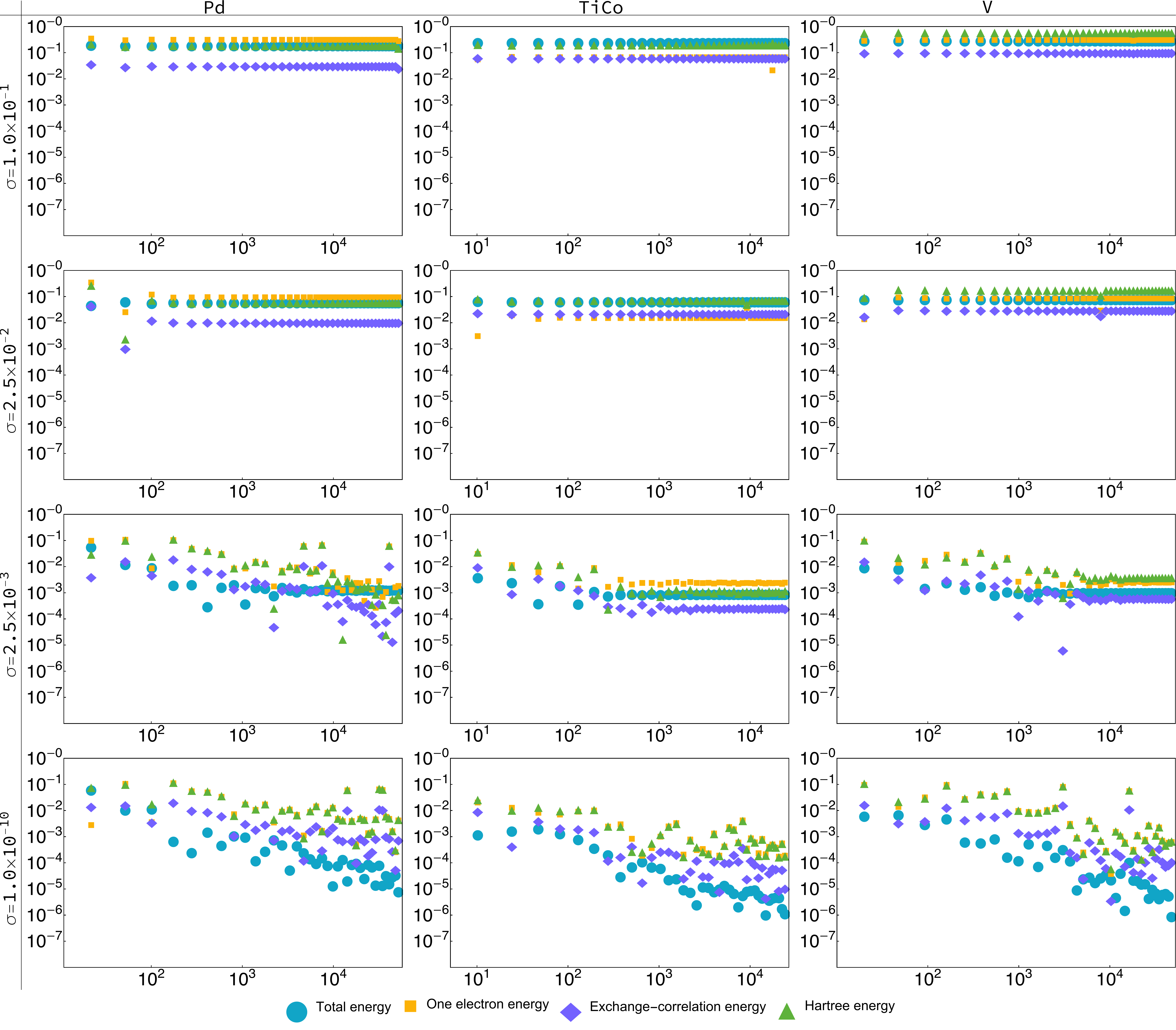}
\caption{The convergence of components of the total energy in Quantum Espresso for the metals Pd, TiCo, and V with Fermi-Dirac smearing. The Ewald contribution to the total energy is left out due to its lack of dependence on the amount smearing or the \kb-point density. For all plots, the $x$-axis is the reduced \kb-point density in units of cubic Bohrs.}
\end{center}
\end{figure}

\newpage
\section{VASP}
\subsection{Misc. Plots}
\begin{figure}[b]
\begin{center}
\includegraphics[width=4in,keepaspectratio]{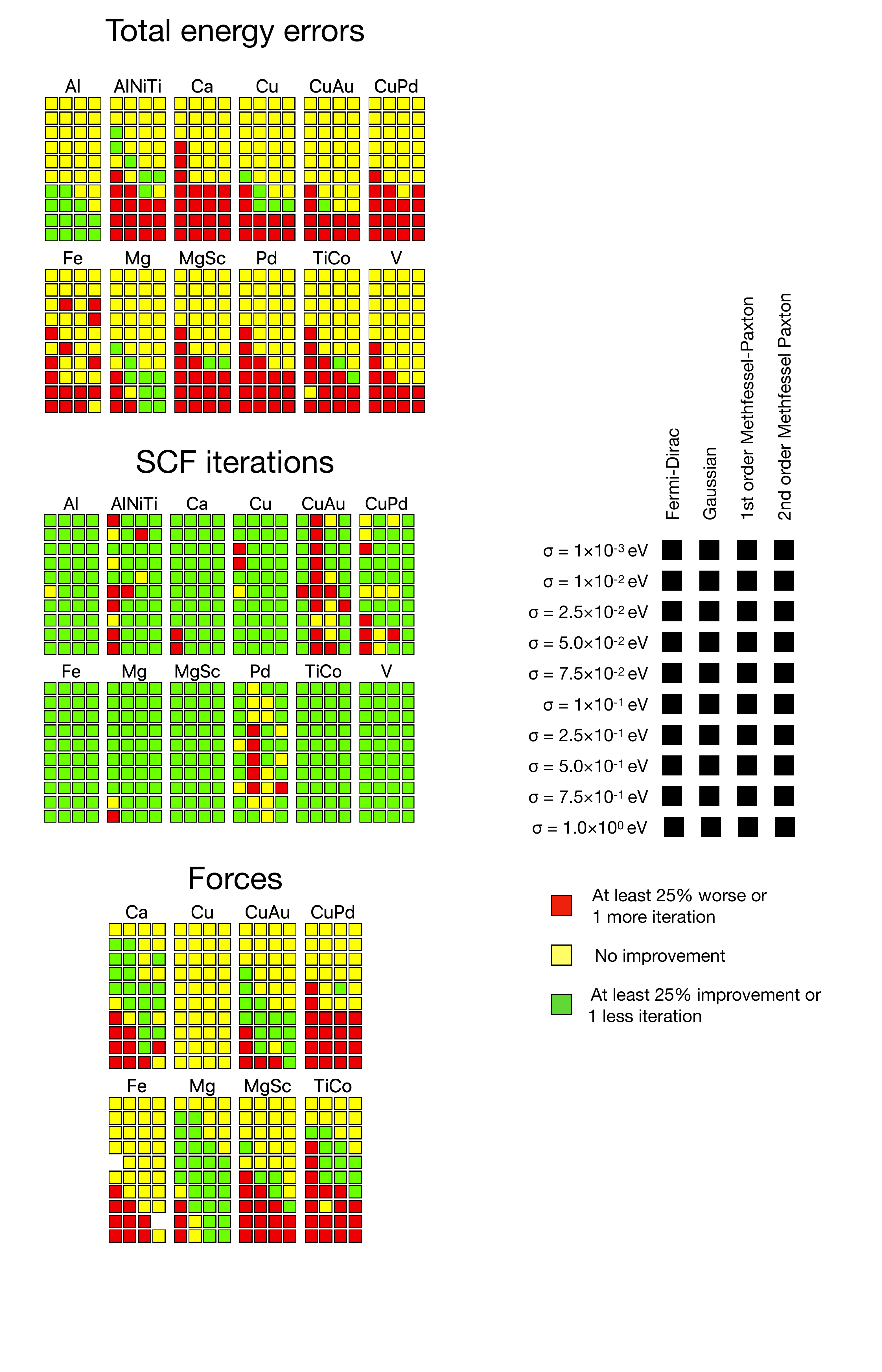}
\caption{VASP total energy errors, SCF iterations, and force error calculations with smearing compared to those without. Each colored square is an average over \kb-point densities. We exclude calculations where the total energy error is less than 0.1 meV. In each block with a system label, the columns from left to right are smearing types Fermi-Dirac, Gaussian, order 1 Methfessel-Paxton, and order 2 Methfessel-Paxton. The rows from top to bottom are different smearing parameters starting with 0.001 eV at the top and increasing to 1 eV at the bottom. For total energy and force errors, green squares are located where smearing decreases errors by more than 25\%, yellow where smearing errors are within 25\% of errors without smearing, and red where smearing errors are more than 25\% more than the errors without smearing. For SCF iterations, green squares are located where smearing reduces the number of iterations by at least 1, yellow squares are where smearing and no smearing iterations are within 1 iteration of each other, and red squares are where smearing has one average iteration more than no smearing.}
\end{center}
\end{figure}

\FloatBarrier
\subsection{Smearing tests in VASP}

\begin{figure}[H]
\includegraphics[width=6in,keepaspectratio]{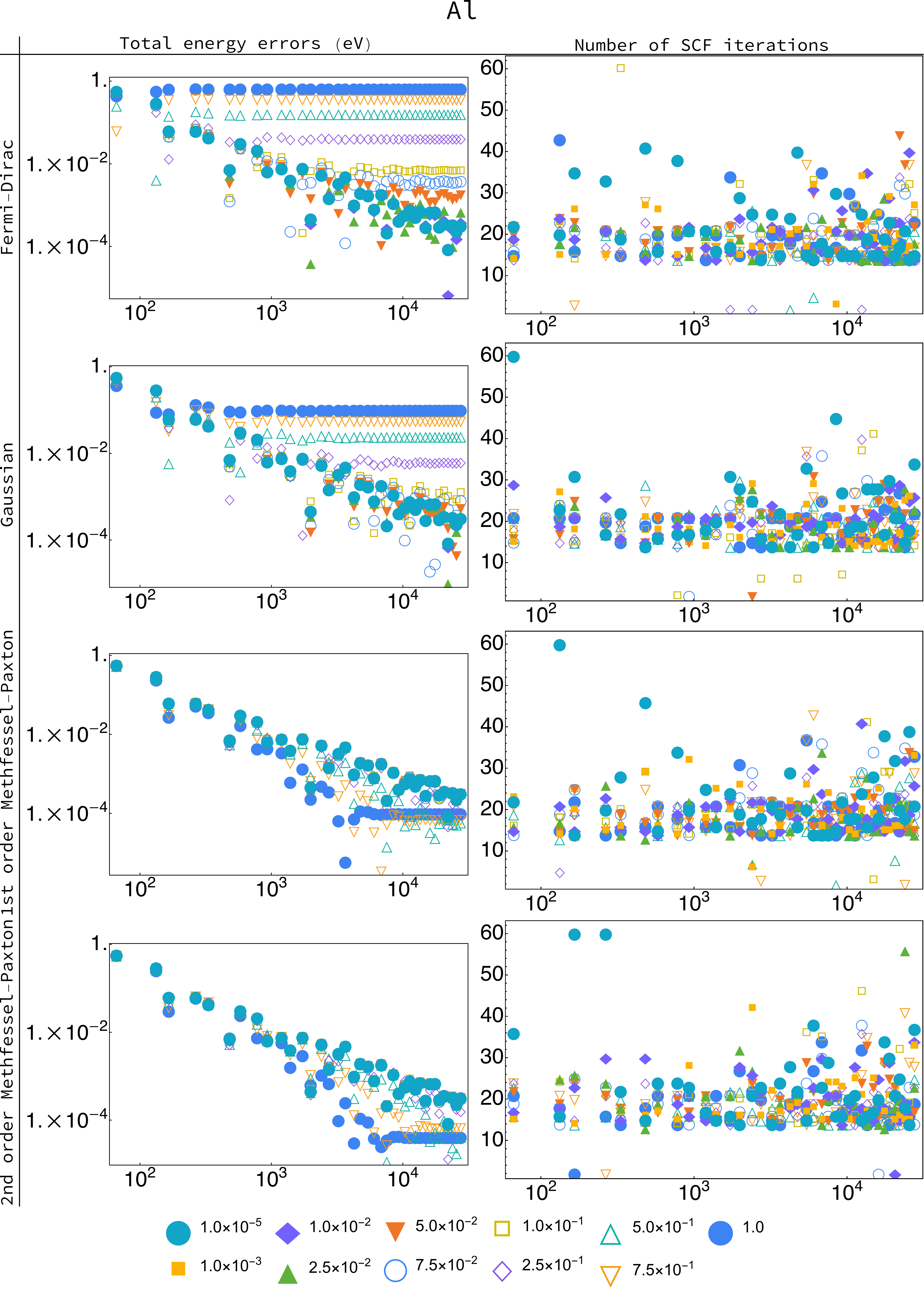}
\caption{The total energy convergence and number of SCF cycles for Al in VASP as a function of \kb-point density. For all plots, the $x$-axis is the reduced \kb-point density in units of cubic Angstroms. The legend at the bottom gives the amount of smearing in electron volts.}
\label{fig:vasp_Al-smooth}
\end{figure}
\FloatBarrier

\begin{figure}[h]
\centering
\includegraphics[width=6in,height=\textheight,keepaspectratio]{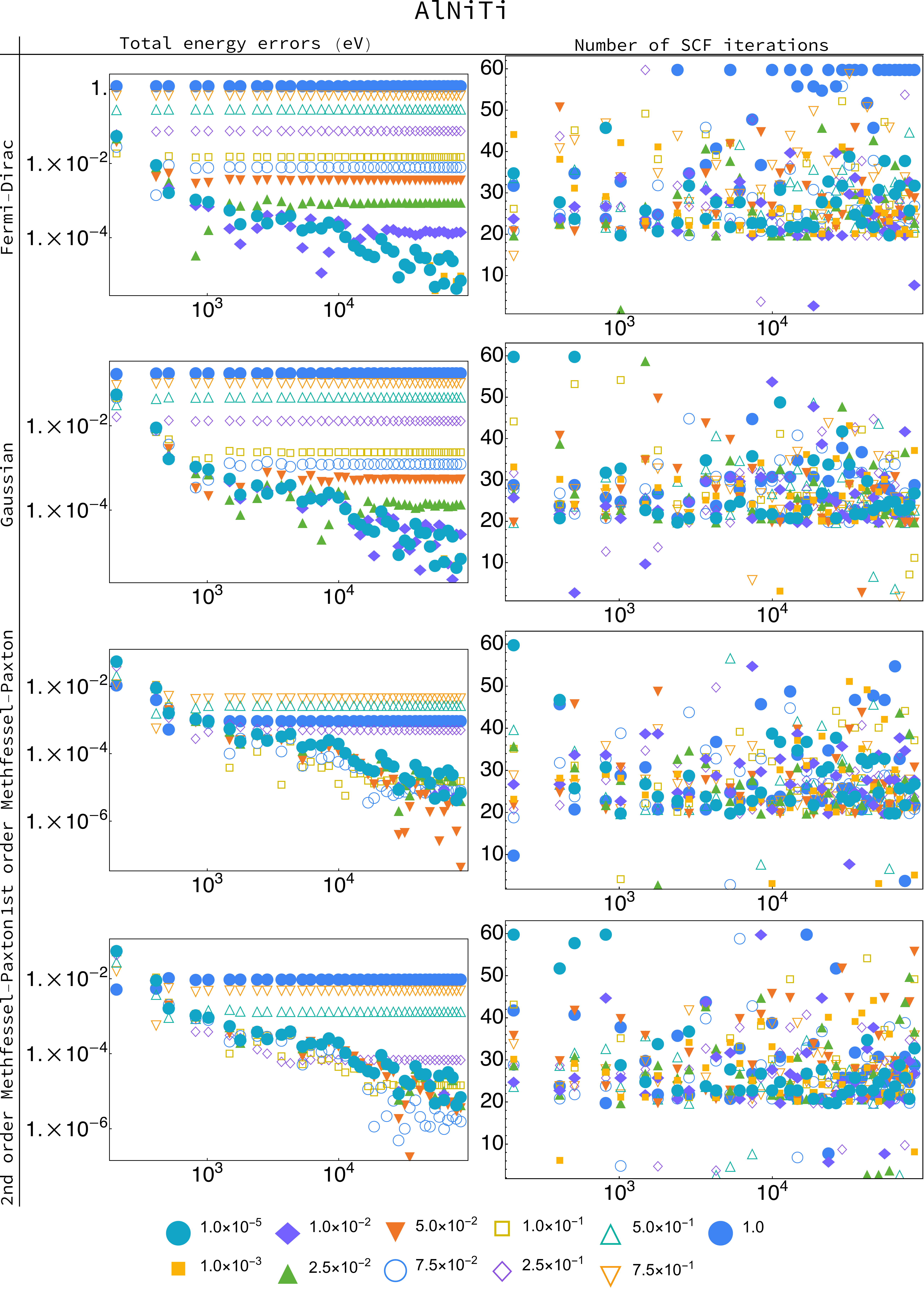}
\caption{The total energy convergence and number of SCF cycles AlNiTi in VASP as a function of \kb-point density. For all plots, the $x$-axis is the reduced \kb-point density in units of cubic Angstroms. The legend at the bottom gives the amount of smearing in electron volts. The empty plot in the lower right corner is due to stresses equal to zero for almost all runs.}
\label{fig:vasp_AlNiTi-smooth}
\end{figure}
\FloatBarrier

\begin{figure}[h]
\includegraphics[width=\textwidth,height=\textheight,keepaspectratio]{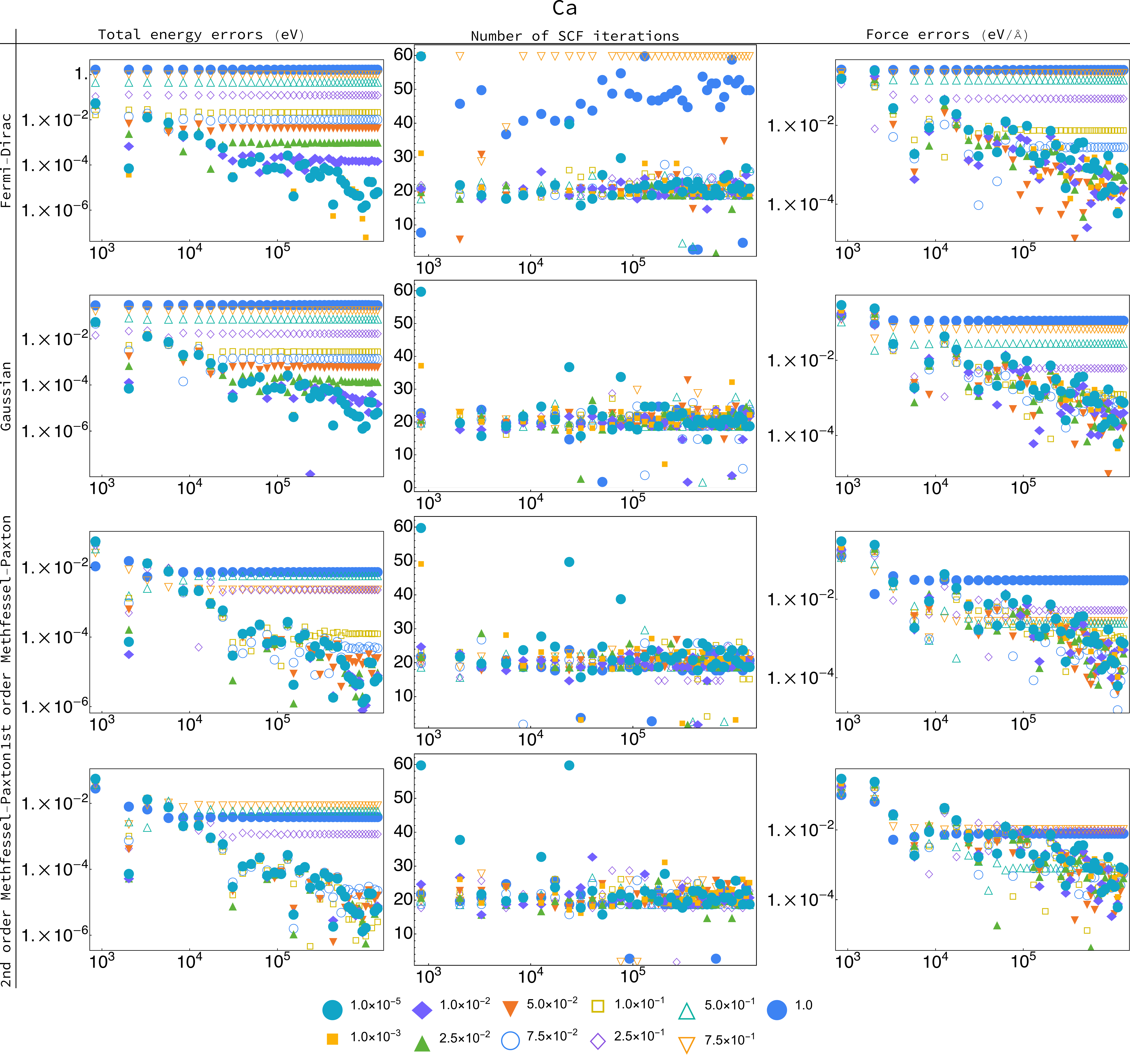}
\caption{The total energy convergence, number of SCF cycles, and force convergence for Ca in VASP. For all plots, the $x$-axis is the reduced \kb-point density in units of cubic Angstroms. The legend at the bottom gives the amount of smearing in electron volts.}
\label{fig:vasp_Ca-smooth}
\end{figure}
\FloatBarrier

\begin{figure}[p]
\includegraphics[width=\textwidth,height=\textheight,keepaspectratio]{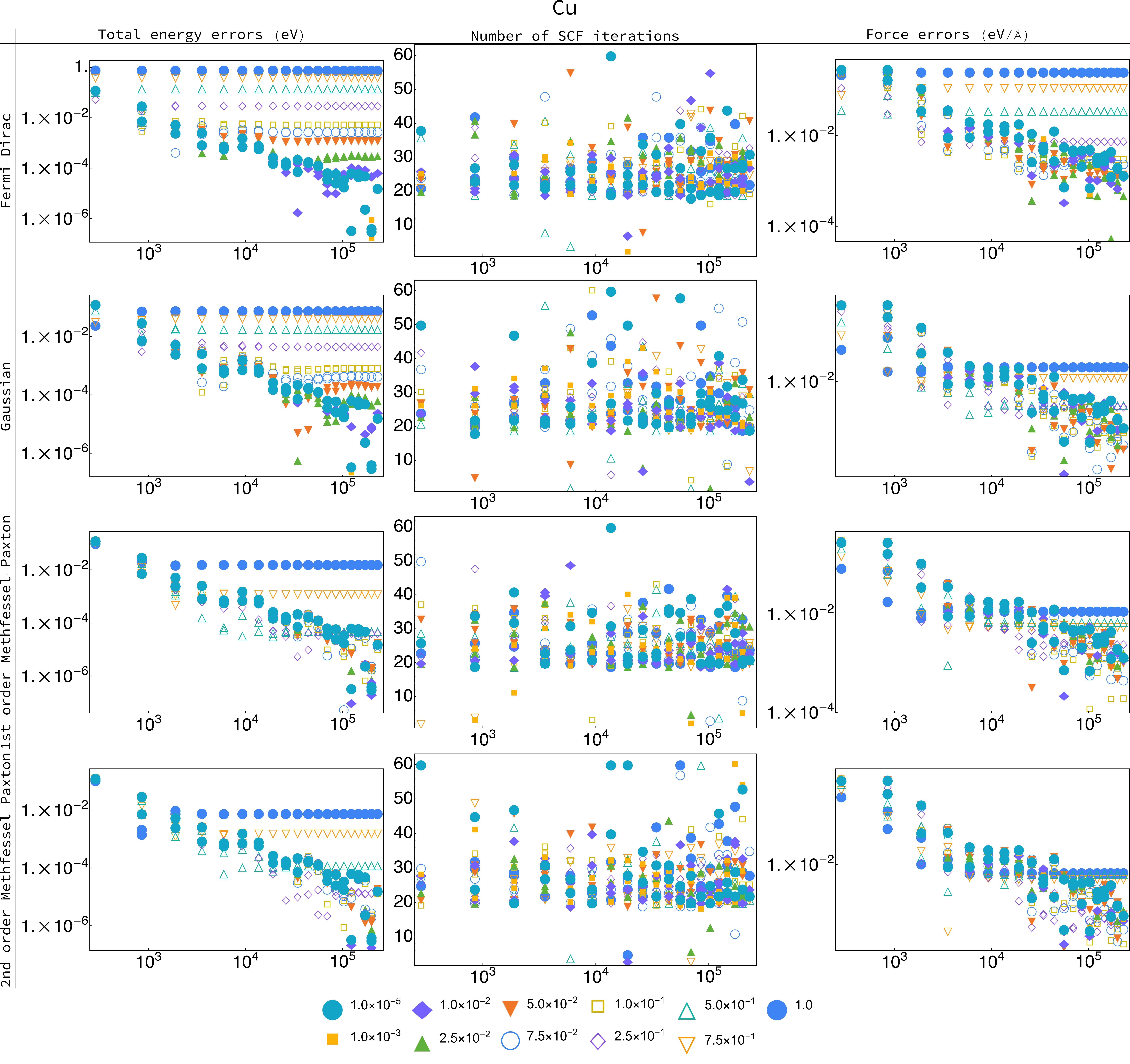}
\caption{The total energy convergence, number of SCF cycles, and force convergence for Cu in VASP. For all plots, the $x$-axis is the reduced \kb-point density in units of cubic Angstroms. The legend at the bottom gives the amount of smearing in electron volts.}
\label{fig:vasp_Cu-smooth}
\end{figure}
\FloatBarrier

\begin{figure}[p]
\includegraphics[width=\textwidth,height=\textheight,keepaspectratio]{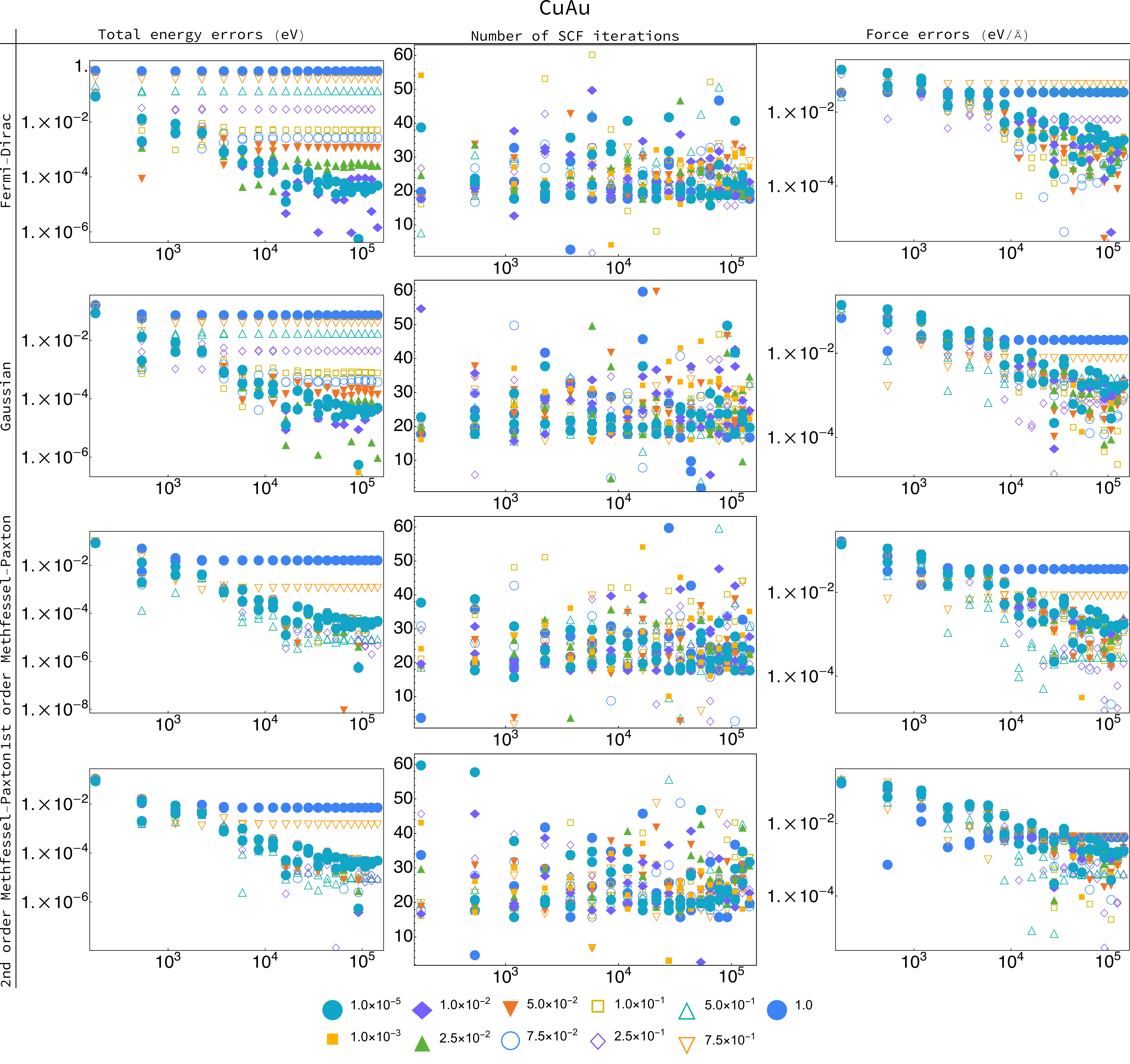}
\caption{The total energy convergence, number of SCF cycles, and force convergence for CuAu in VASP. For all plots, the $x$-axis is the reduced \kb-point density in units of cubic Angstroms. The legend at the bottom gives the amount of smearing in electron volts.}
\label{fig:vasp_CuAu-smooth}
\end{figure}
\FloatBarrier

\begin{figure}[p]
\includegraphics[width=\textwidth,height=\textheight,keepaspectratio]{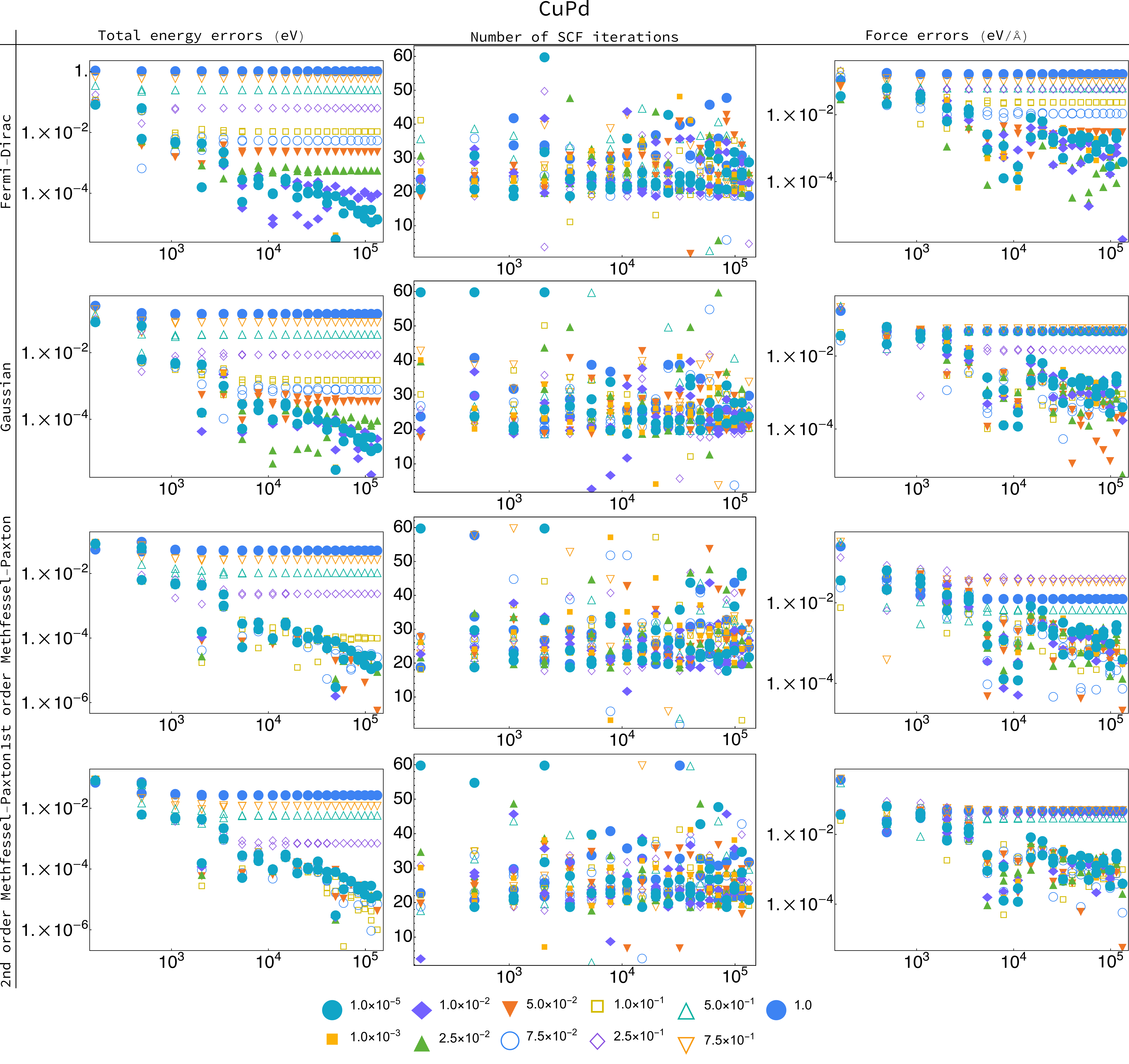}
\caption{The total energy convergence, number of SCF cycles, and force convergence for CuPd in VASP. For all plots, the $x$-axis is the reduced \kb-point density in units of cubic Angstroms. The legend at the bottom gives the amount of smearing in electron volts.}
\label{fig:vasp_CuPd-smooth}
\end{figure}
\FloatBarrier

\begin{figure}[p]
\includegraphics[width=\textwidth,height=\textheight,keepaspectratio]{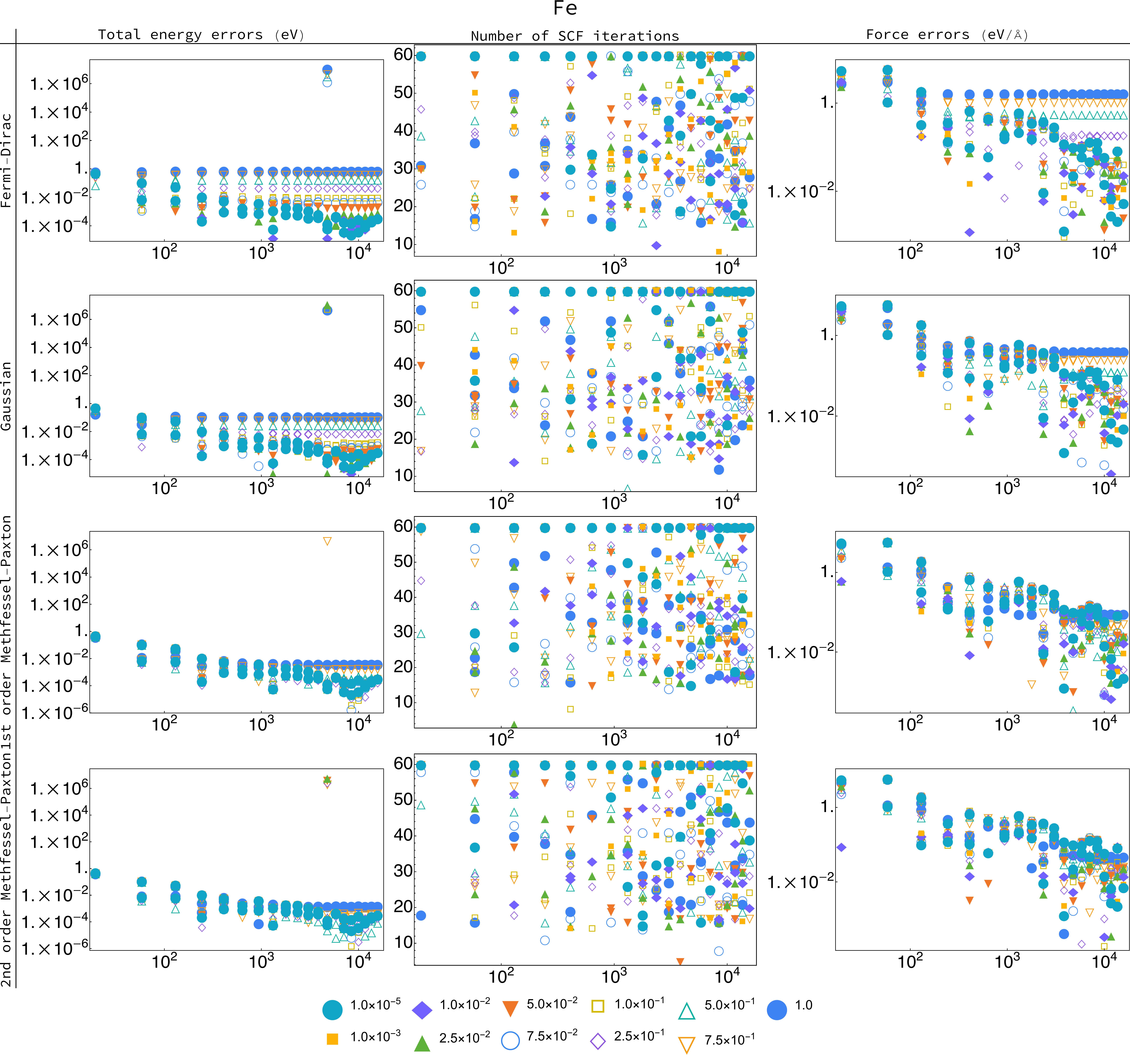}
\caption{The total energy convergence, number of SCF cycles, and force convergence for Fe in VASP. For all plots, the $x$-axis is the reduced \kb-point density in units of cubic Angstroms. The legend at the bottom gives the amount of smearing in electron volts.}
\label{fig:vasp_Fe-smooth}
\end{figure}
\FloatBarrier

\begin{figure}[p]
\includegraphics[width=\textwidth,height=\textheight,keepaspectratio]{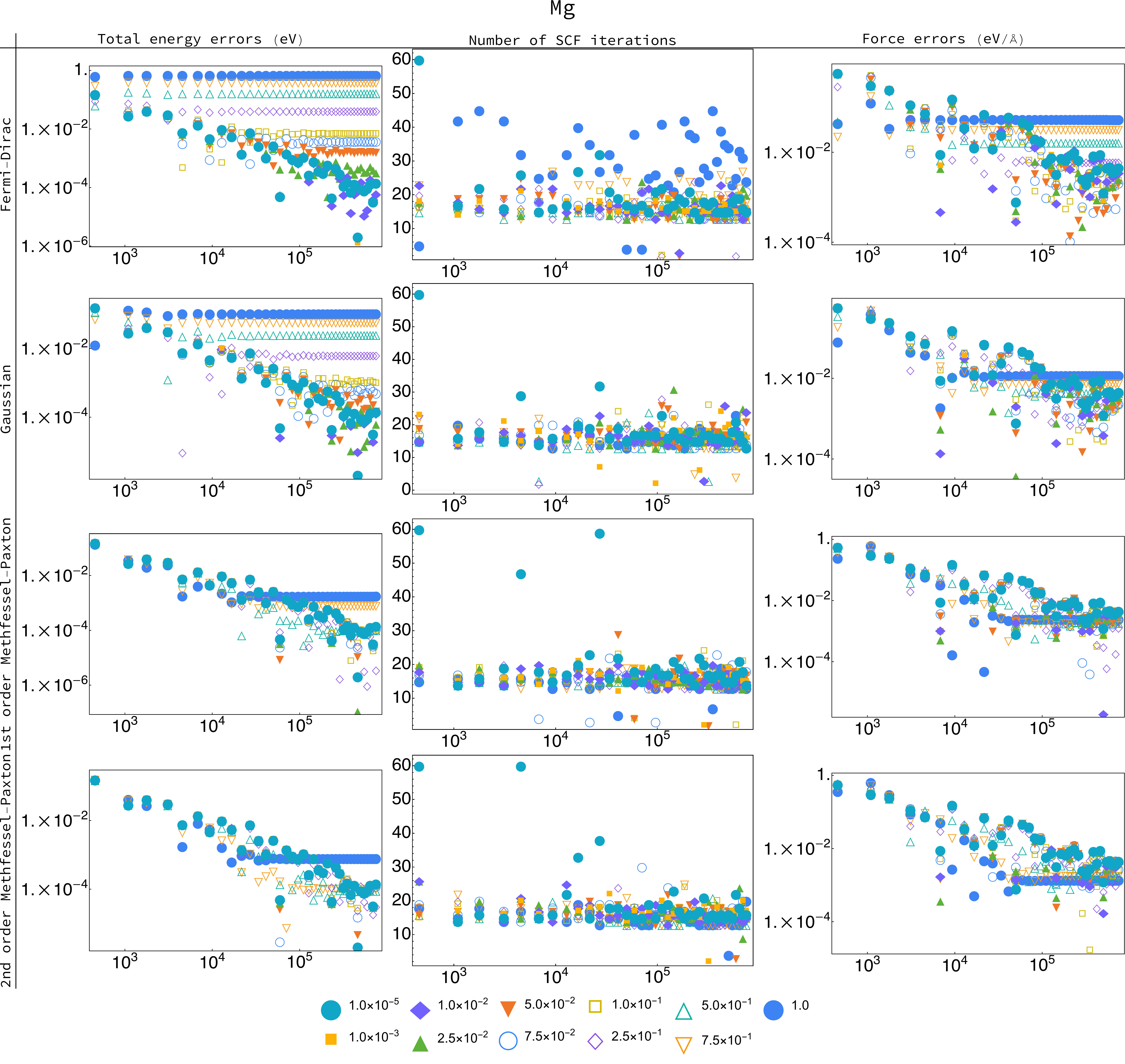}
\caption{The total energy convergence, number of SCF cycles, and force convergence for Mg in VASP. For all plots, the $x$-axis is the reduced \kb-point density in units of cubic Angstroms. The legend at the bottom gives the amount of smearing in electron volts.}
\label{fig:vasp_Mg-smooth}
\end{figure}
\FloatBarrier

\begin{figure}[p]
\includegraphics[width=\textwidth,height=\textheight,keepaspectratio]{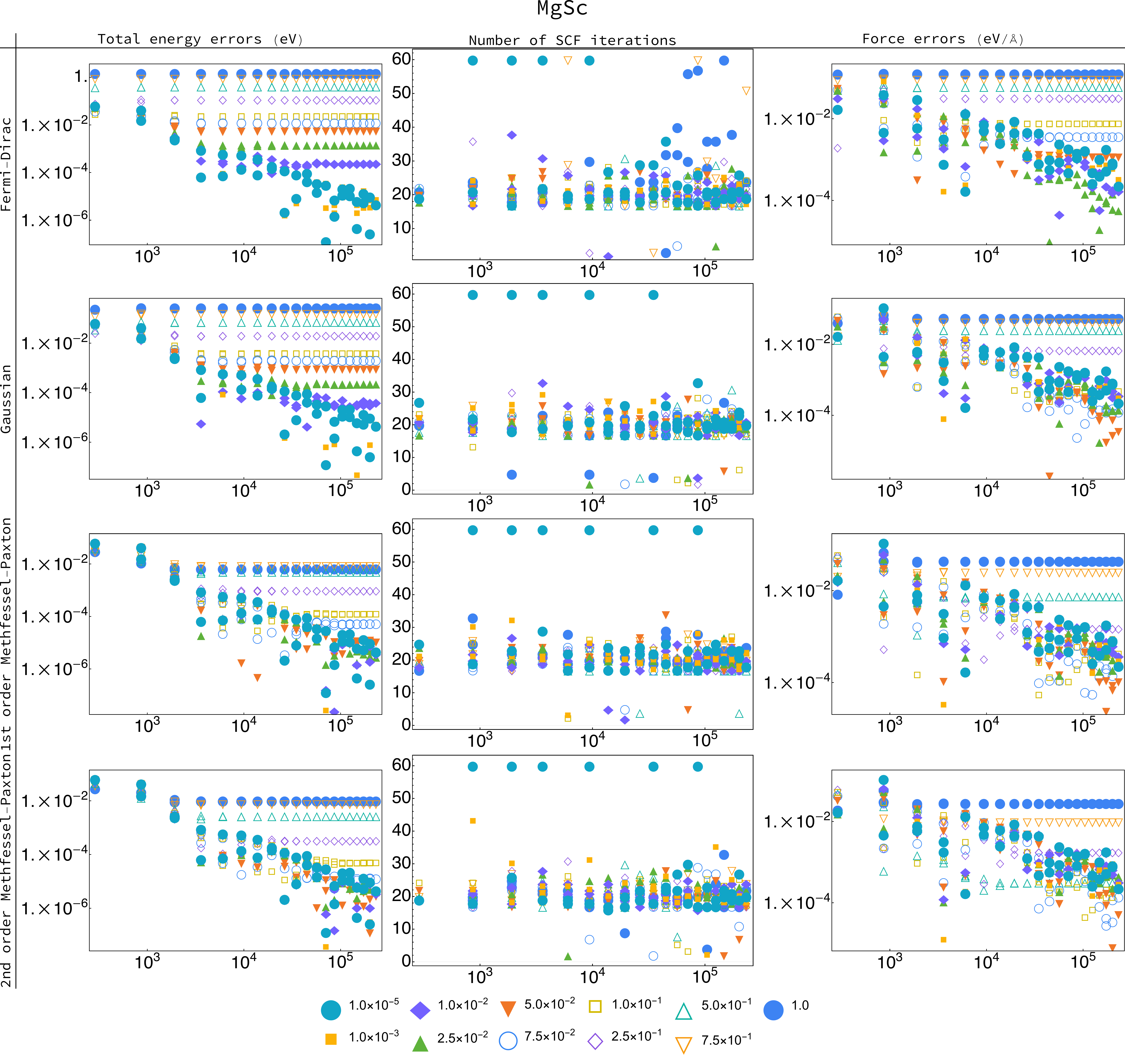}
\caption{The total energy convergence, number of SCF cycles, and force convergence for MgSc in VASP. For all plots, the $x$-axis is the reduced \kb-point density in units of cubic Angstroms. The legend at the bottom gives the amount of smearing in electron volts.}
\label{fig:vasp_MgSc-smooth}
\end{figure}
\FloatBarrier

\begin{figure}[p]
\includegraphics[width=6in,height=\textheight,keepaspectratio]{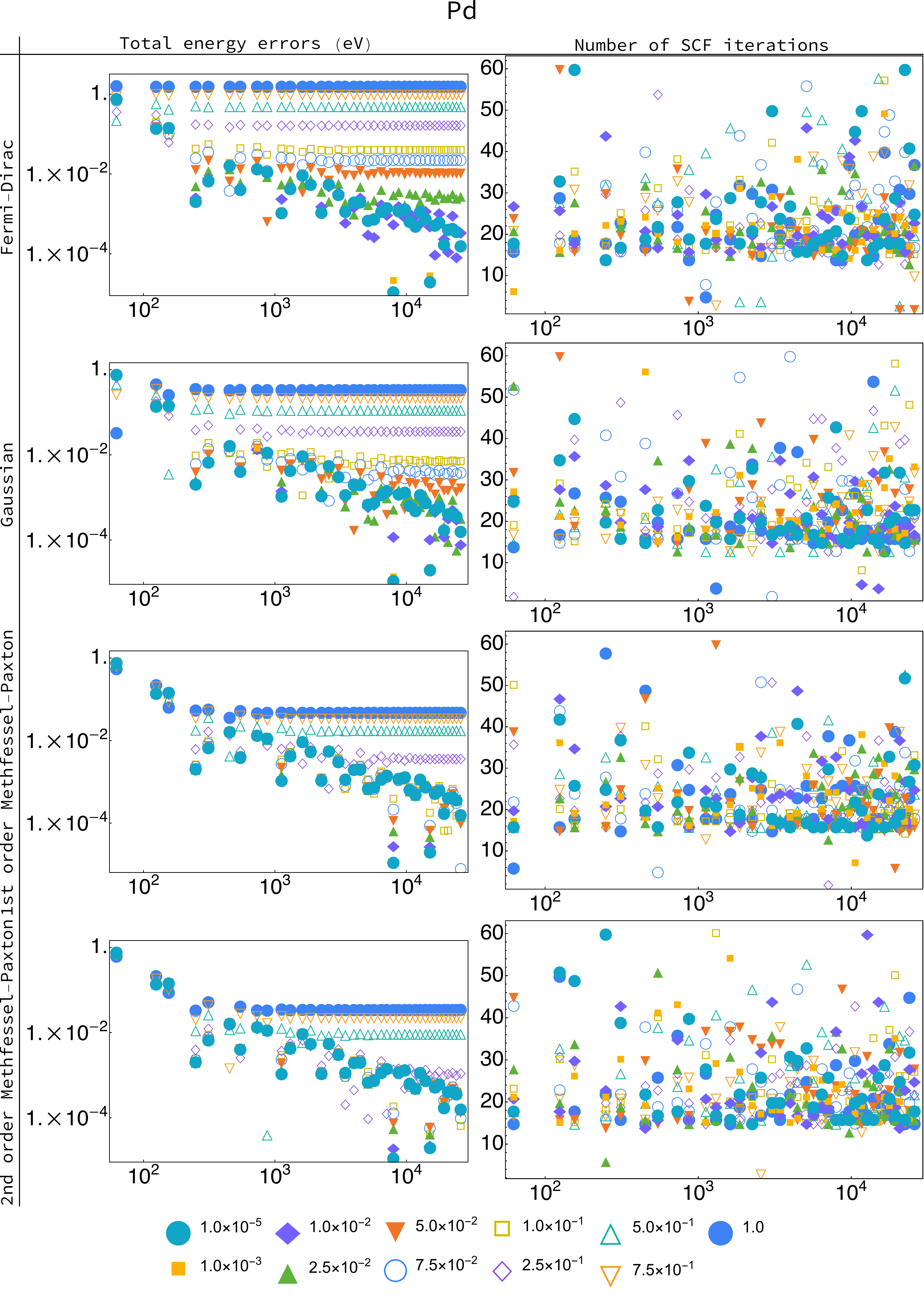}
\caption{The total energy convergence and number of SCF cycles for Pd in VASP as a function of \kb-point density. For all plots, the $x$-axis is the reduced \kb-point density in units of cubic Angstroms. The legend at the bottom gives the amount of smearing in electron volts.}
\label{fig:vasp_Pd-smooth}
\end{figure}
\FloatBarrier

\begin{figure}[p]
\includegraphics[width=\textwidth,height=\textheight,keepaspectratio]{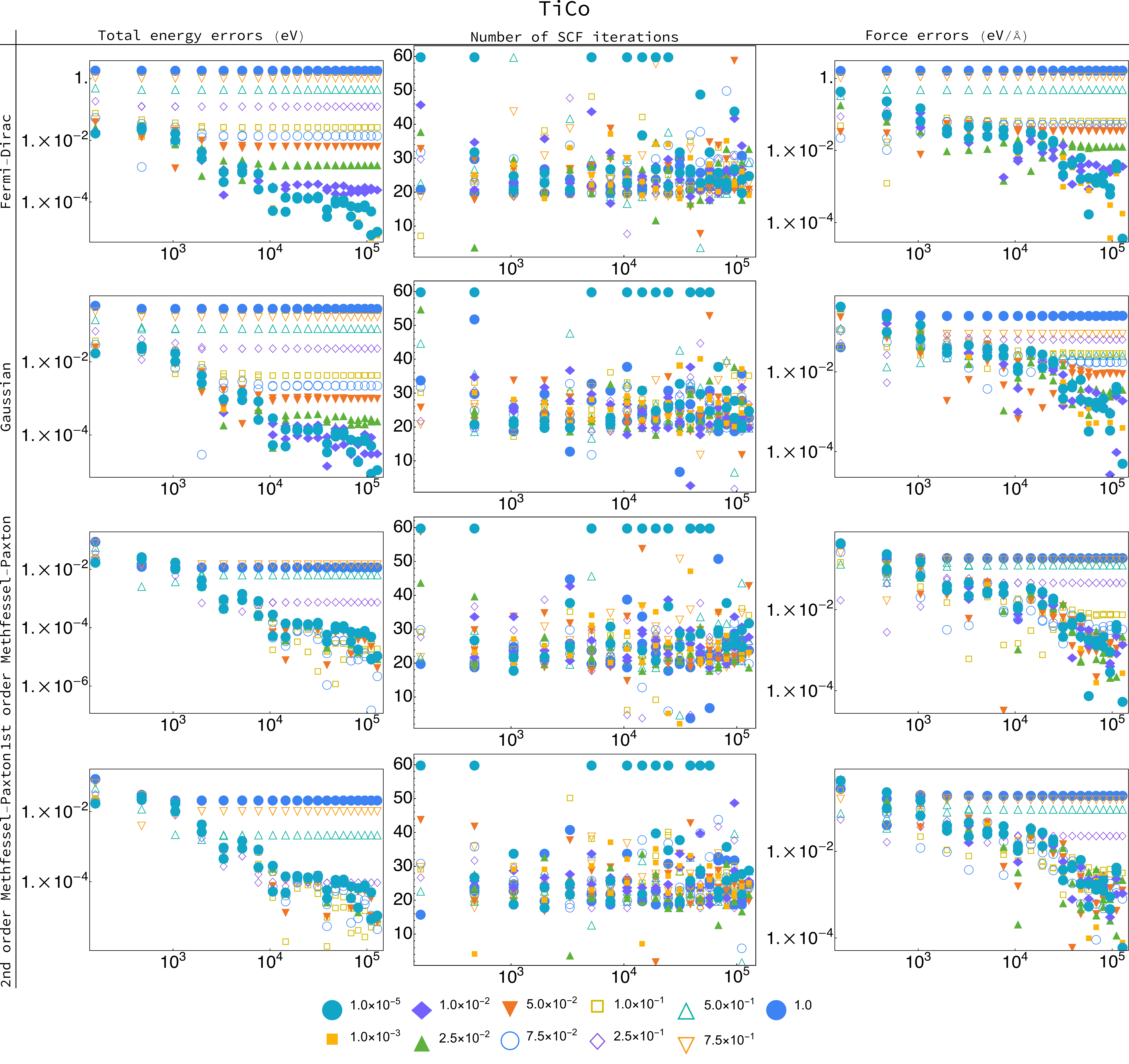}
\caption{The total energy convergence, number of SCF cycles, and force convergence for TiCo in VASP. For all plots, the $x$-axis is the reduced \kb-point density in units of cubic Angstroms. The legend at the bottom gives the amount of smearing in electron volts.}
\label{fig:vasp_TiCo-smooth}
\end{figure}
\FloatBarrier

\begin{figure}[p]
\includegraphics[width=6in,height=\textheight,keepaspectratio]{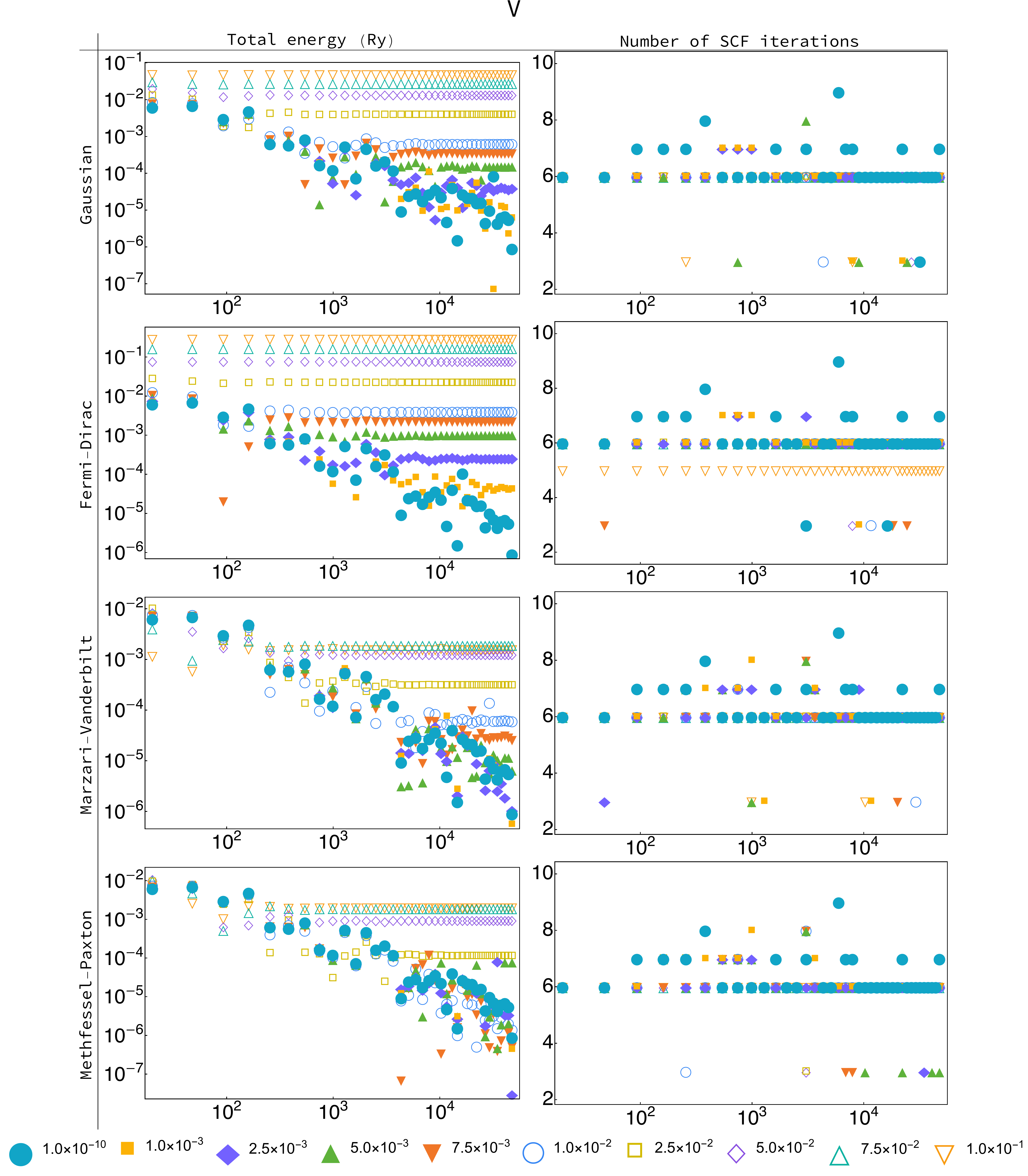}
\caption{The total energy convergence and number of SCF cycles for V in VASP as a function of \kb-point density. For all plots, the $x$-axis is the reduced \kb-point density in units of cubic Angstroms. The legend at the bottom gives the amount of smearing in electron volts.}
\label{fig:vasp_V-smooth}
\end{figure}

\FloatBarrier
\subsection{Tetrahedra tests in VASP}

\begin{figure}[h]
\includegraphics[width=6in,height=\textheight,keepaspectratio]{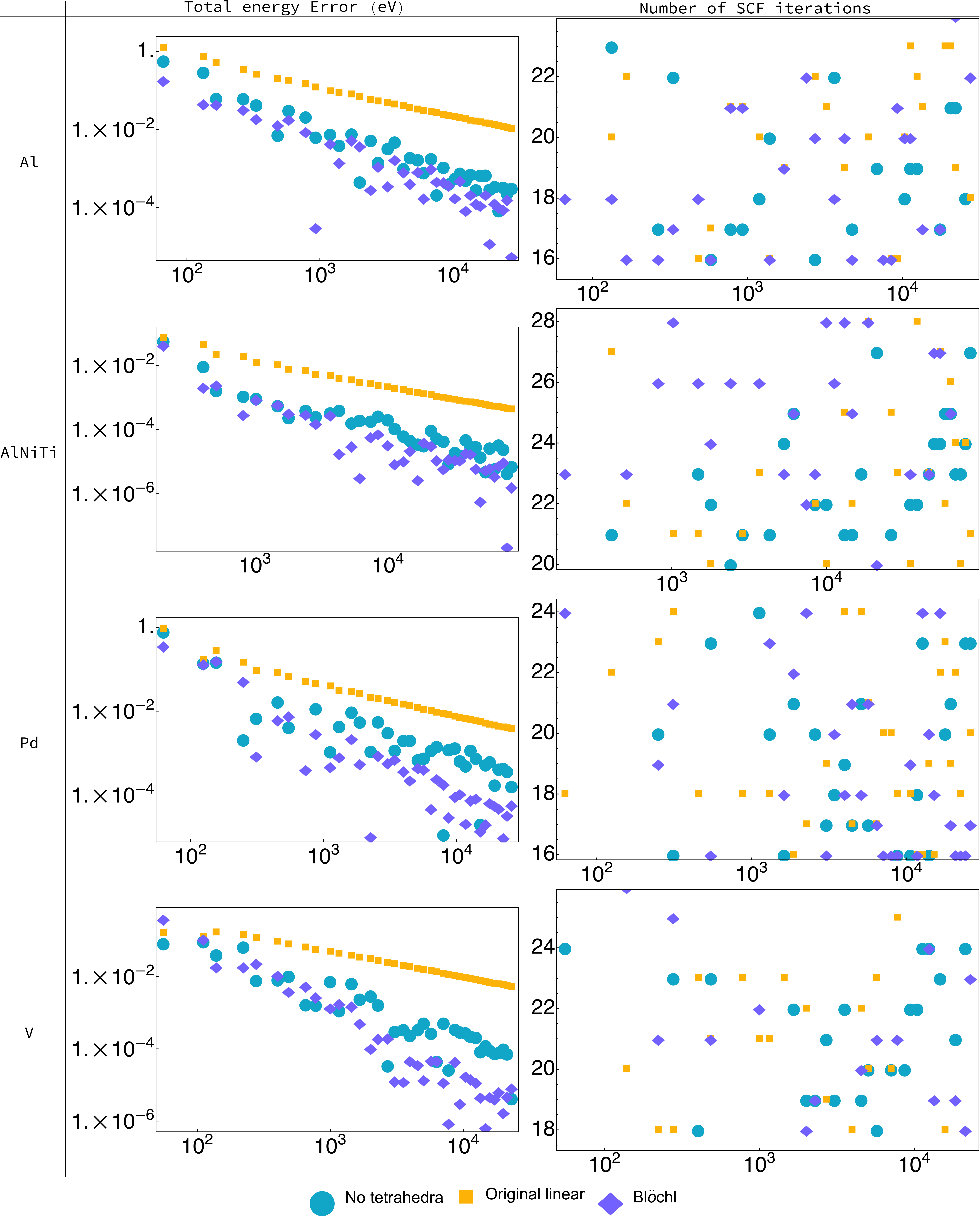}
\caption{The total energy convergence and number of SCF cycles of Al, AlNiTi, Pd, and V with tetrahedron methods in VASP as a function of \kb-point density. For all plots, the $x$-axis is the reduced \kb-point density in units of cubic Angstroms.}
\label{fig:vasp_Al-AlNiTi-Pd-V-comb-tet}
\end{figure}
\FloatBarrier

\begin{figure}[p]
\includegraphics[width=\textwidth,height=\textheight,keepaspectratio]{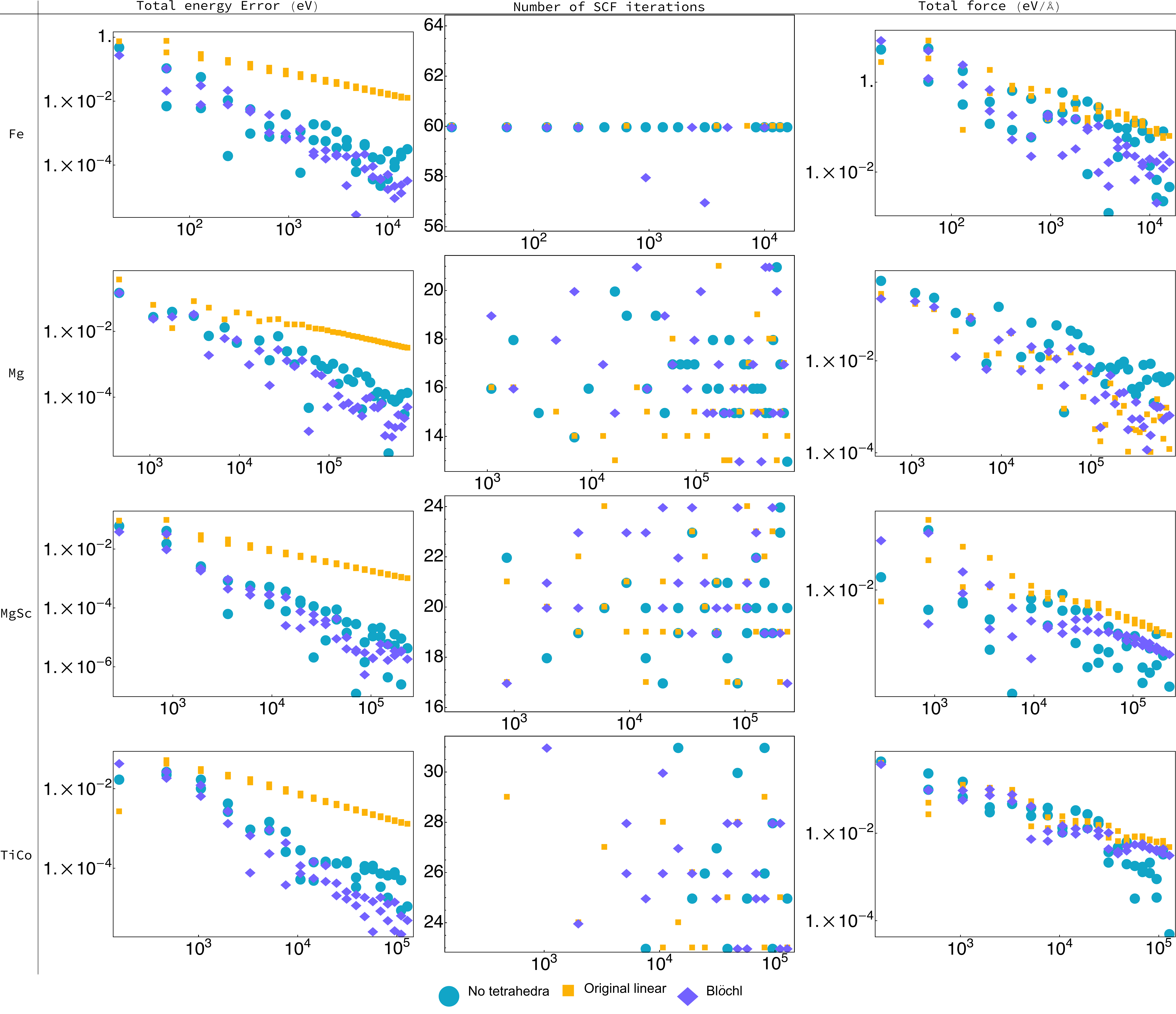}
\caption{The total energy convergence and number of SCF cycles Fe, Mg, MgSc, and TiCo with tetrahedron methods in VASP as a function of \kb-point density. For all plots, the $x$-axis is the reduced \kb-point density in units of cubic Angstroms.}
\label{fig:vasp_Fe-Mg-MgSc-TiCo-comb-tet}
\end{figure}

\FloatBarrier
\subsection{Energy component tests in VASP}

\begin{figure}[h]
\includegraphics[width=\textwidth,height=\textheight,keepaspectratio]{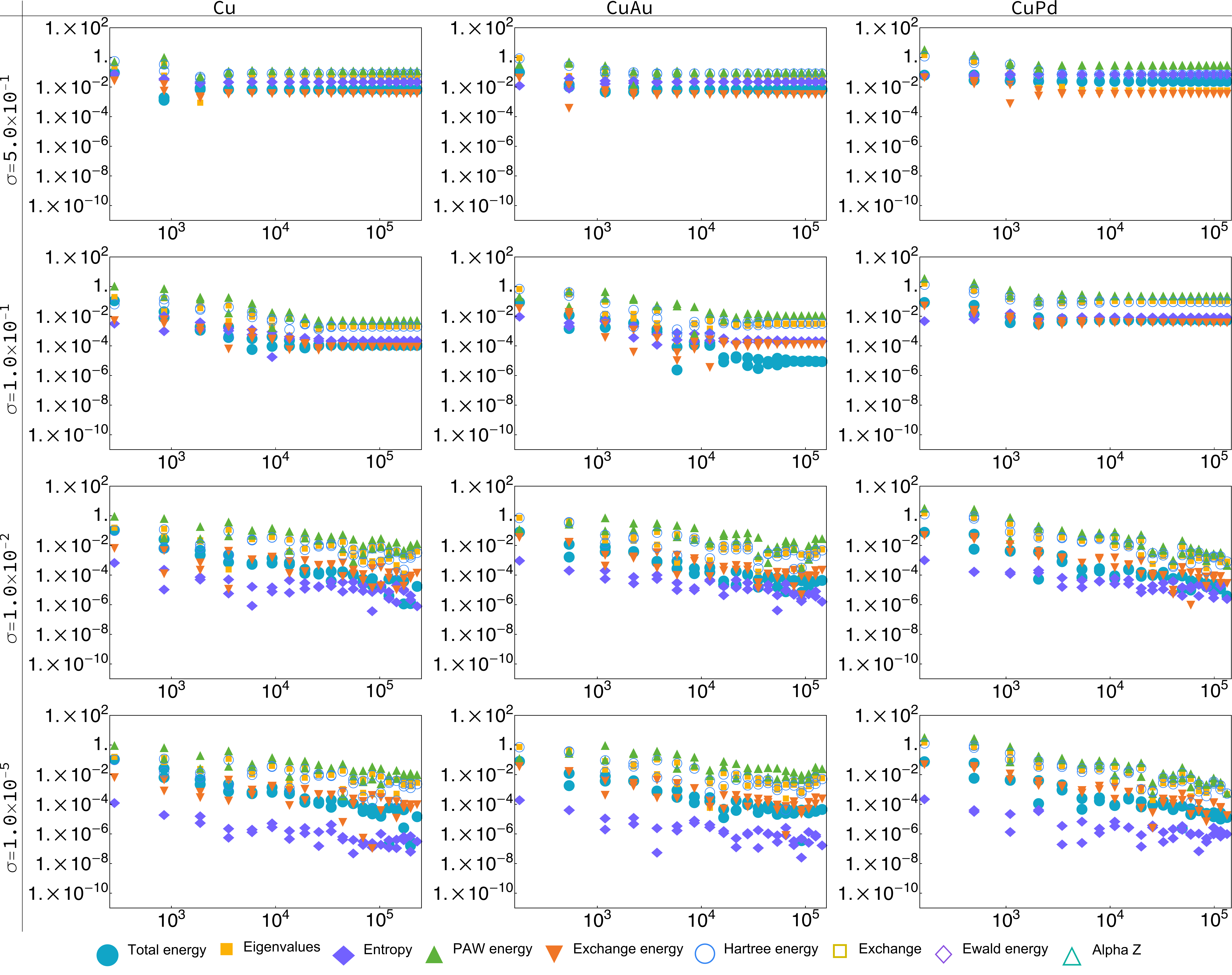}
\caption{The convergence of components of the total energy in VASP for the metals CuAu, CuPd, and Fe with 2nd order Methfessel-Paxton smearing. The atomic energy contribution to the total energy is left out due to its lack of dependence on the amount smearing or the \kb-point density. For all plots, the $x$-axis is the reduced \kb-point density in units of cubic Angstroms.}
\label{fig:vasp_Cu-CuAu-CuPd-2nd-order-Methfessel-Paxton}
\end{figure}
\FloatBarrier

\begin{figure}[p]
\includegraphics[width=\textwidth,height=\textheight,keepaspectratio]{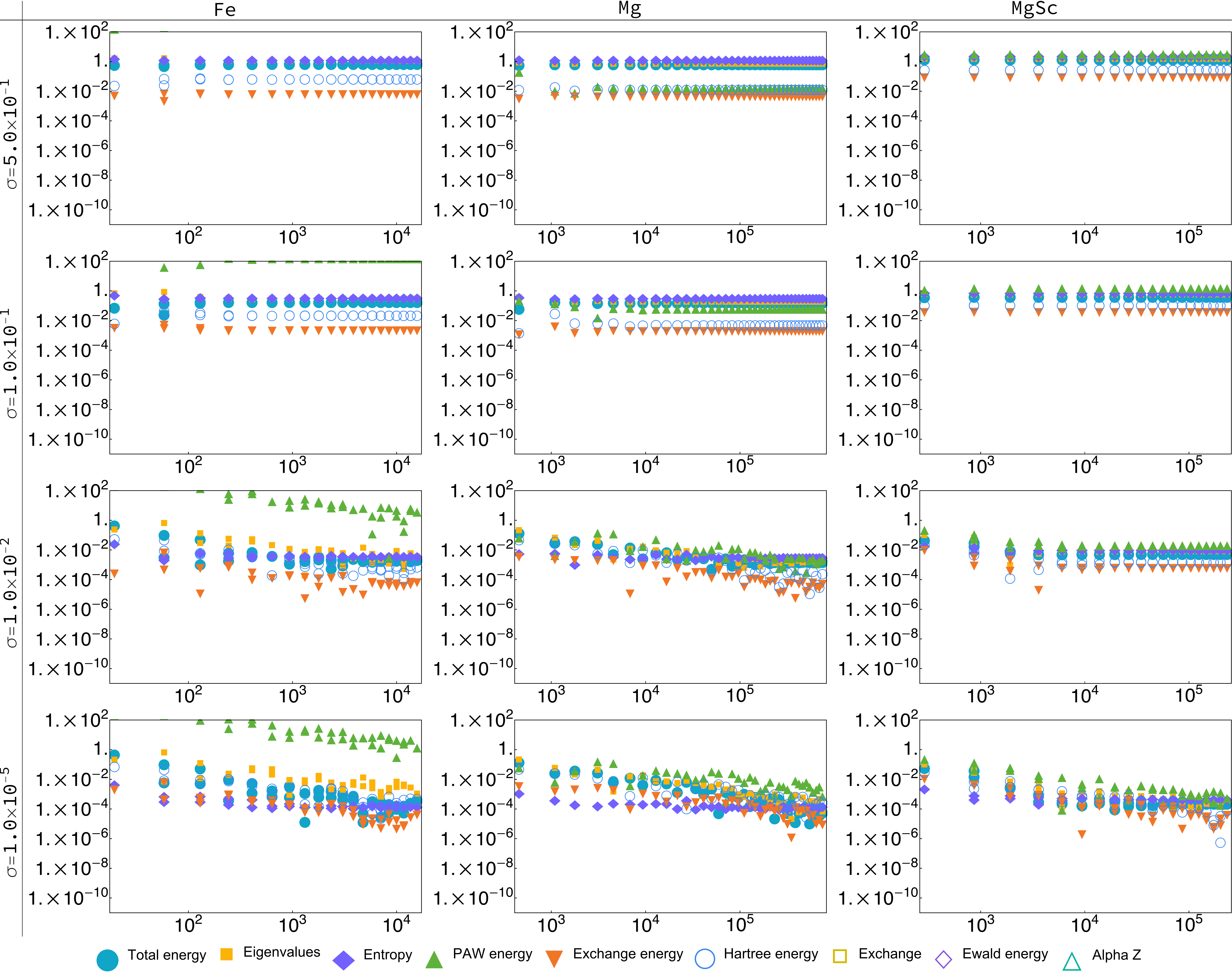}
\caption{The convergence of components of the total energy in VASP for the metals Fe, Mg, and MgSc with Fermi-Dirac smearing. The atomic energy contribution to the total energy is left out due to its lack of dependence on the amount smearing or the \kb-point density. For all plots, the $x$-axis is the reduced \kb-point density in units of cubic Angstroms.}
\label{fig:vasp_K-Mg-MgSc-Fermi-Dirac}
\end{figure}
\FloatBarrier

\begin{figure}[p]
\includegraphics[width=\textwidth,height=\textheight,keepaspectratio]{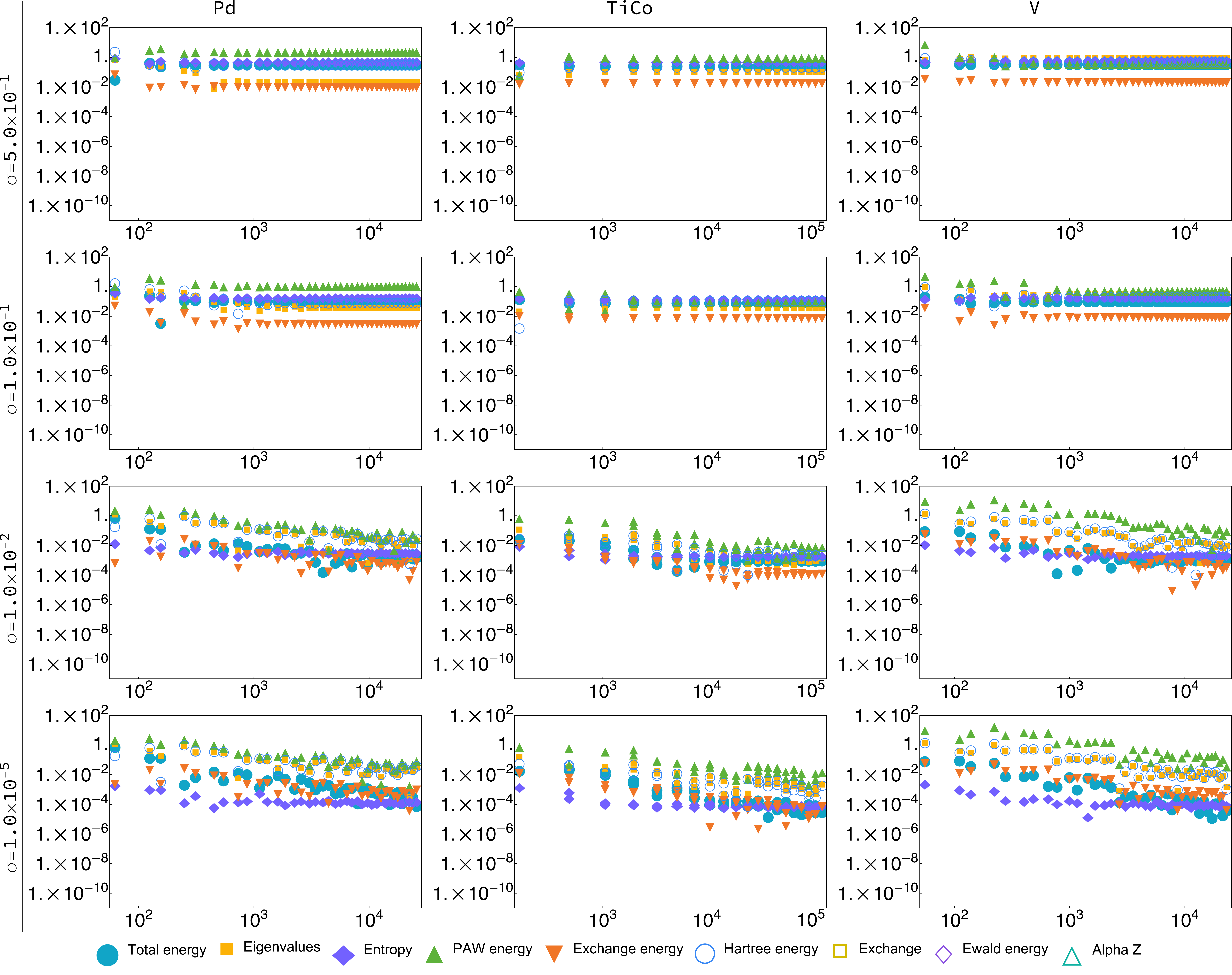}
\caption{The convergence of components of the total energy in VASP for the metals Pd, TiCo, and V with Gaussian smearing. The atomic energy contribution to the total energy is left out due to its lack of dependence on the amount smearing or the \kb-point density. For all plots, the $x$-axis is the reduced \kb-point density in units of cubic Angstroms.}
\label{fig:vasp_Pd-TiCo-V-Gaussian}
\end{figure}